\documentclass[
reprint,
superscriptaddress,
amsmath,
amssymb,
aps,
prb,
floatfix
]{revtex4-2}
\usepackage{graphicx}
\usepackage{hyperref}
\usepackage{xcolor}
\usepackage{multirow}

\begin{document}

\title{Electronic structure, phase stability, and transport properties of the\\AlTiVCr lightweight high-entropy alloy: A computational study}

\author{Christopher D. Woodgate}
\thanks{Present address: Yusuf Hamied Department of Chemistry, University of Cambridge, Lensfield Road, Cambridge, CB2 1EW, United Kingdom}
\email{cw2016@cam.ac.uk}
\affiliation{H.H. Wills Physics Laboratory, University of Bristol, Royal Fort, Bristol, BS8 1TL, United Kingdom}
\author{Hubert J. Naguszewski}
\affiliation{Department of Physics, University of Warwick, Coventry, CV4 7AL, United Kingdom}
\author{Nicolas F. Piwek}
\affiliation{Lasercenter HM, Munich University of Applied Sciences HM, Lothstra{\ss}e 34, 80335 Munich, Germany}
\author{David Redka}
\affiliation{Lasercenter HM, Munich University of Applied Sciences HM, Lothstra{\ss}e 34, 80335 Munich, Germany}

\begin{abstract}
We investigate the thermodynamics and phase stability of the AlTiVCr lightweight high-entropy alloy using a combination of \textit{ab initio} electronic structure calculations, a concentration wave analysis, and atomistic Monte Carlo simulations. 
In alignment both with experimental data and with results obtained using other computational approaches, we predict a B2 (CsCl) chemical ordering emerging in this alloy at comparatively high temperatures, which is driven by Al and Ti moving to separate sublattices, while V and Cr express weaker site preferences. 
The impact of this B2 chemical ordering on the electronic transport properties of the alloy is investigated within a Kubo--Greenwood linear response framework and it is found that, counter-intuitively, the alloy's residual resistivity increases as the material transitions from the A2 (disordered bcc) phase to our predicted B2 (partially) ordered structure. 
Our results give fresh insight into the atomic-scale structure and consequent physical properties of this well-studied, technologically relevant material.
\end{abstract}

\date{November 16, 2026}

\maketitle

\noindent\textbf{Keywords:} High-entropy alloy, Alloy, Thermodynamics, Simulation, Electronic structure, Electrical properties

\section{Introduction}
\label{sec:introduction}

Though first reported more than twenty years ago by Yeh \textit{et al.}~\cite{yeh_nanostructured_2004} and Cantor \textit{et al.}~\cite{cantor_microstructural_2004}, high-entropy alloys (HEAs)---single-phase alloys combining four or more elements in near-equal ratios---continue to attract significant attention in the field of materials science~\cite{miracle_critical_2017, george_high-entropy_2019}. These materials, also sometimes referred to as `complex concentrated' or `multi-principal element' alloys, are of interest for a range of advanced engineering applications because they frequently exhibit superior physical properties as compared to traditional binary/ternary alloys, including impressive fracture resistance~\cite{gludovatz_fracture-resistant_2014, gludovatz_exceptional_2016, liu_exceptional_2022} and corrosion resistance~\cite{qiu_corrosion_2017}, good radiation damage tolerance~\cite{el-atwani_outstanding_2019, el_atwani_quinary_2023}, and excellent structural properties at elevated temperatures~\cite{praveen_highentropy_2018, chen_review_2018}. From a more fundamental perspective, these materials also exhibit a range of interesting physical behaviours. For example, various HEAs and related medium-entropy systems have previously been shown to exhibit superconductivity~\cite{kozelj_discovery_2014}, complex magnetism~\cite{billington_bulk_2020}, quantum critical behaviour~\cite{sales_quantum_2016}, and high levels of Fermi surface smearing~\cite{robarts_extreme_2020}.

\begin{figure}[t]
    \centering
    \includegraphics[width=\linewidth]{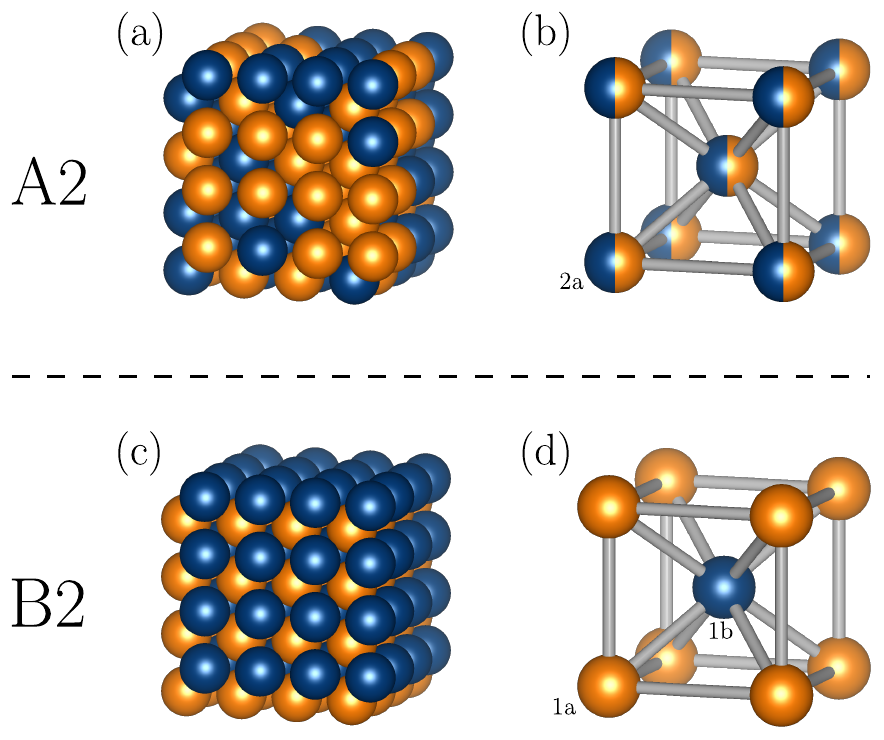}
    \caption{Illustrations---for an equiatomic binary alloy---of the disordered bcc structure, Strukturbericht designation A2, and the ordered CsCl structure, Strukturbericht designation B2. Panel (a) shows a supercell representation of the A2 structure, while panel (b) shows a representation of the unit cell where half-coloured spheres indicate partial lattice site occupancies. Panel (c) then shows a supercell representation of the B2 structure, while panel (d) shows a representation of the unit cell. Such chemically ordered structures can emerge in an alloy if it is annealed below its disorder--order transition temperature. In panels (b) and (d), lattice sites are given their Wyckoff labels. Images generated using \textsc{Vesta}~\cite{momma_vesta_2011}.}
    \label{fig:a2_b2}
\end{figure}

One composition of particular interest is the well-studied, equiatomic AlTiVCr HEA, which has a low density compared to many other bcc HEAs (because it contains only Al and early $3d$ transition metals) but also retains good mechanical properties~\cite{qiu_lightweight_2017, qiu_microstructure_2018, huang_lightweight_2019, huang_order-disorder_2022}. In contrast to many HEAs containing only transition-metal elements and forming disordered solid solutions, the addition of Al to HEAs is understood to promote atomic short- and long-range ordering tendencies~\cite{senkov_effect_2014, senkov_microstructure_2014}. In the case of AlTiVCr, it is understood that the material undergoes a phase transition from an A2 (disordered bcc) phase to an ordered B2 (CsCl) phase which subsequently remains stable across a wide temperature range~\cite{huang_order-disorder_2022}. (The B2 structure, conceptually illustrated in Fig.~\ref{fig:a2_b2}, is a chemically ordered structure imposed on the bcc lattice.) AlTiVCr is therefore perhaps best referred to as a refractory high-entropy superalloy (RSA), on account of the analogy with Ni--Al superalloys which also form ordered intermetallic phases with exceptional mechanical properties~\cite{miracle_refractory_2020}.

Computational studies have a key role to play in elucidating the nature of atomic short- and long-range order in HEAs, and can also provide key insights into how the atomic-scale structure of these complex materials influences their physical properties~\cite{widom_modeling_2018, eisenbach_first-principles_2019, ferrari_frontiers_2020, ferrari_simulating_2023}. The materials modeller's `toolkit' for studying the phase stability of HEAs includes simulation approaches based on density functional theory (DFT) calculations~\cite{widom_hybrid_2014, tamm_atomic-scale_2015, widom_first-principles_2024}; interatomic potentials including both semi-empirical and machine-learning models~\cite{kostiuchenko_impact_2019, rosenbrock_machine-learned_2021, kormann_b2_2021, zhang_roadmap_2025, cao_capturing_2025}; and lattice-based models such as cluster expansions~\cite{fernandez-caballero_short-range_2017, sobieraj_chemical_2020, kim_interaction_2023, vazquez_deciphering_2024}. Specific to the study of alloy thermodynamics, a range of conventional and enhanced sampling techniques (\textit{i.e.}, Monte Carlo and related methods) can be used to study phase transitions and phase equilibria~\cite{widom_modeling_2018, eisenbach_first-principles_2019, ferrari_frontiers_2020, ferrari_simulating_2023}, as can semi-empirical thermodynamic approaches such as CALPHAD~\cite{zhang_calphad_2022, li_calphad-aided_2023}. There are also approaches available based on effective medium theories such as the coherent potential approximation (CPA)~\cite{singh_atomic_2015, kormann_long-ranged_2017, singh_ta-nb-mo-w_2018}.

In this work, we present computational results which give detailed insight into the atomic-scale structure and physical properties of the AlTiVCr HEA. Specifically, we describe the alloy's electronic structure and phase stability using a combination of first-principles DFT calculations, a perturbative concentration wave analysis~\cite{khan_statistical_2016, woodgate_compositional_2022, woodgate_modelling_2024}, and atomistic Monte Carlo simulations using a recovered pairwise form of the alloy's internal energy. We also examine the electronic transport properties of the material, inspecting how chemical ordering affects this key physical property. In alignment with the available experimental data, as well as with previous computational studies, we predict a B2 chemical ordering emerging in the alloy at high temperatures, in which Al and Ti move to separate sublattices while V and Cr express much weaker sublattice preferences. Counter-intuitively, this (partial) chemical ordering is found to result in an increased predicted residual resistivity for the alloy, a result attributed to a reduction in the electronic density of states at the Fermi level induced by the chemical ordering. Our atomistic Monte Carlo simulations then predict eventual ordering (at comparatively low temperatures) into a fully ordered single-phase ground state accompanied, naturally, by a vanishing residual resistivity. {Explicitly, the novelty of the present work as compared to prior studies (see, \textit{e.g.}, Refs.~\citenum{huang_order-disorder_2022} and \citenum{widom_first-principles_2024}) has two key aspects. First, the alloy's phase stability and transport properties are simultaneously assessed within a unified computational modelling framework, with particular emphasis on how the former affects the latter. Second, being based on first-principles electronic structure calculations, we are able to discuss the underlying physical mechanisms driving the experimentally observed B2 chemical ordering, as well as influencing the electrical transport properties of the alloy. Moreover, by studying the phase stability of the alloy using our concentration wave formalism, we are able to cross-validate existing simulation studies using alternative techniques.}

The remainder of this paper is structured as follows. 
{In Sec.~\ref{sec:results_and_discussion}, we present and discuss our results for the AlTiVCr HEA on which this work focuses, where our computational modelling predicts the aforementioned B2 chemical ordering emerging at high temperatures. We also elucidate the underlying physical mechanisms driving this chemical ordering and governing the material's transport properties. This is achieved by describing the details of the alloy's electronic structure in both disordered and ordered phases as modelled within the tenets of DFT. In Sec.~\ref{sec:summary_and_conclusions}, we summarise our results, draw conclusions, and give an outlook on potential future avenues of study. In Sec.~\ref{sec:methods}, we discuss our methods, including giving a detailed outline of our computationally efficient, multi-stage modelling approach for simulating the electronic structure, phase stability, and transport properties of HEAs.}

\section{Results and Discussion}
\label{sec:results_and_discussion}

\subsection{Electronic structure of the A2 phase}

We begin by describing the electronic structure of the AlTiVCr HEA in its disordered bcc (A2) phase. In the context of this particular alloy, it is important to treat magnetism appropriately; not only has AlTiVCr previously been proposed as a candidate `spin filter' material~\cite{stephen_synthesis_2016, venkateswara_competing_2018, stephen_electrical_2019, stephen_structural_2019}, but---more generally---magnetism can often play a key role in determining the free energy landscape of alloys~\cite{staunton_interaction_1987, woodgate_interplay_2023, ghosh_chemical_2024}. In this work, where all calculations are performed at the experimental lattice constant, we consistently find that elemental magnetic moments are quenched, \textit{i.e.}, they collapse to zero during the self-consistency cycle. This is in agreement with the supercell calculations of Ref.~\citenum{widom_first-principles_2024}, where (for a range of DFT exchange-correlation functionals) elemental magnetic moments were only reliably obtained when an expanded lattice parameter was employed. We therefore conclude that the appropriate magnetic state for our purposes is non-magnetic.

\begin{figure}[t]
    \centering
    \includegraphics[width=\linewidth]{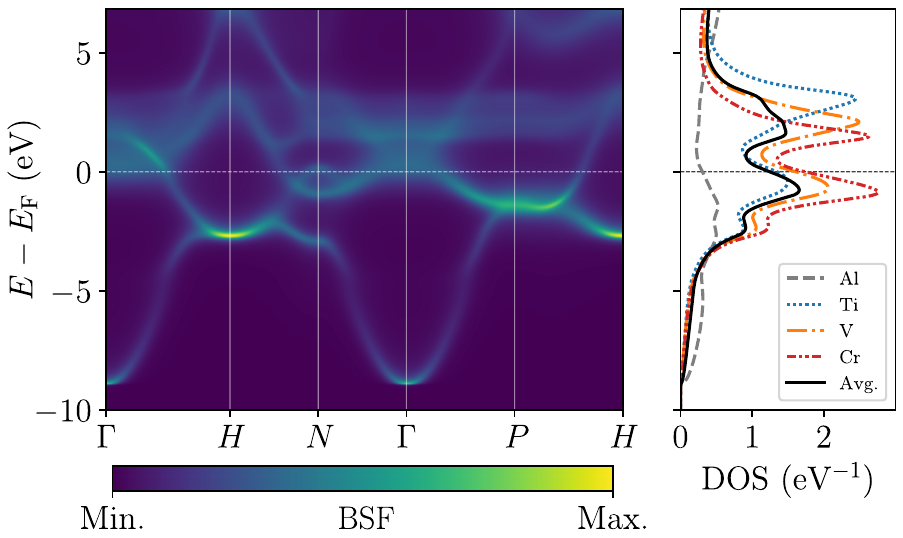}
    \caption{Bloch spectral function (a `band structure' for the disordered alloy) and species-resolved electronic DOS around the Fermi level, $E_\mathrm{F}$, for the AlTiVCr high-entropy alloy as described within the KKR-CPA for the A2 (disordered bcc) phase. A degree of hybridisation between the $sp$-like states of Al and the $3d$ states of the transition metals is indicated by the localised peaks/troughs in the species-resolved DOS for Al at energies where there are peaks/troughs in the DOS associated with the narrow $3d$ bands of the transition metals.}
    \label{fig:a2_electronic_dos}
\end{figure}

Shown in Fig.~\ref{fig:a2_electronic_dos} are plots of the Bloch spectral function (BSF) along high-symmetry lines of the irreducible Brillouin zone (IBZ) of the bcc lattice, as well as plots of the averaged and species-resolved electronic density of states (DOS) around the Fermi level for disordered bcc (A2) AlTiVCr. {These self-consistent calculations, further discussed in Sec.~\ref{sec:methods}, utilise the Korringa--Kohn--Rostoker (KKR) formulation of DFT, with the coherent potential approximation (CPA) employed to average over chemical disorder.} It can be seen that the band structure of the material is heavily smeared by the chemical disorder---a consequence of the lack of translational symmetry in a disordered alloy. We associate the dispersive, near-parabolic bands seen in the BSF plot with the $sp$-states of both Al and the transition metals, while the comparatively narrow bands around the Fermi level are associated with the $3d$ bands of Ti, V, and Cr. Related to the increasing valence numbers of the elements, it can be seen that---integrated up to the Fermi level---the area under the electronic DOS curve is smallest for Al, then increases for Ti, then V, through to Cr, for which it is largest. Finally, we note that a degree of hybridisation between the $sp$-like states of Al and the $3d$ states of the transition metals can be seen. The hallmark of this is the presence of localised peaks/troughs in the species-resolved DOS for Al which are aligned with localised peaks/troughs in the same curves for the transition metals.

\subsection{Concentration wave analysis}
\label{sec:concentration_wave_results}

\begin{figure}[t]
    \centering
    \includegraphics[width=\linewidth]{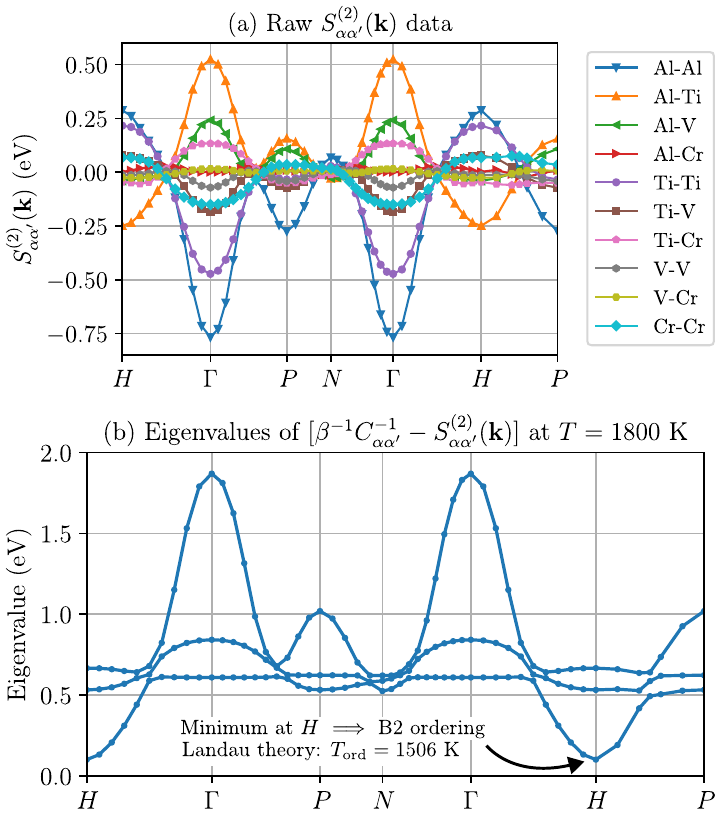}
    \caption{Plots of data associated with the concentration wave analysis performed using the alloy $S^{(2)}$ theory~\cite{khan_statistical_2016, woodgate_compositional_2022, woodgate_modelling_2024}. Panel (a) shows the raw $S^{(2)}_{\alpha \alpha'}(\mathbf{k})$ data plotted along selected high-symmetry lines of the IBZ of the bcc lattice, while panel (b) shows the eigenvalues of the chemical stability matrix along those same high-symmetry directions. That the minimum eigenvalue lies at $H$ in panel (b) indicates that a B2 chemical ordering is favoured.}
    \label{fig:concentration_wave_analysis}
\end{figure}

Proceeding, we perform a concentration wave analysis to garner information about the dominant atom-atom correlations and infer which chemical ordering(s) are likely to emerge as the alloy is cooled. Shown in Fig.~\ref{fig:concentration_wave_analysis} are plots of (a) $S^{(2)}_{\alpha \alpha'}(\mathbf{k})$ and (b) the eigenvalues of the chemical stability matrix{, the latter of which is evaluated at a temperature of $T=1800$~K. Both of these quantities are formally defined---and their meaning discussed---in the Methods, Sec.~\ref{sec:methods-concentration_waves}, Eqs.~\eqref{eq:s2_definition} and \eqref{eq:chemical_stability_reciprocal}. The data} are plotted along selected high-symmetry lines of the IBZ.

We consider first the raw $S^{(2)}_{\alpha \alpha'}(\mathbf{k})$ data, shown in panel (a) of Fig.~\ref{fig:concentration_wave_analysis}. When the quantity $S^{(2)}_{\alpha \alpha'}(\mathbf{k})$ takes large values and varies strongly as a function of $\mathbf{k}$, it indicates that a pair of elements interacts strongly, while when this quantity takes small values and varies weakly it indicates that a pair of elements interacts weakly. It can be seen that, in this case, the strongest atomic interactions are between Al-Al, Al-Ti, and Ti-Ti pairs, with other elements interacting more weakly. This is consistent with the strong propensity for Al-Ti alloys to form ordered intermetallic compounds~\cite{ohnuma_phase_2000}.

Next, we consider the eigenvalues of the chemical stability matrix, shown in panel (b) of Fig.~\ref{fig:concentration_wave_analysis}. Here, the lowest-lying eigenvalue is most relevant, as this indicates the dominant chemical order~\cite{woodgate_compositional_2022}. A minimum at the $H$ point, corresponding to $\mathbf{k} = (0,0,2\pi/a)$, describes a B2 chemical ordering~\cite{khachaturyan_ordering_1978, woodgate_short-range_2023}. The associated (normalised) eigenvector is $\Delta c_\alpha = (0.734, -0.656, -0.156, 0.078)$, while the predicted ordering temperature (the temperature at which this eigenvalue passes through zero) is $T_\mathrm{ord} = 1506$~K. The eigenvector indicates that Al moves strongly to one sublattice and Ti strongly to the other. The smaller $\Delta c_\alpha$ values associated with V and Cr indicate a weak preference for V moving to the Ti sublattice and Cr to the Al sublattice. If the amplitude of this predicted concentration wave is allowed to `grow' until one sublattice contains (at least) one atomic species whose concentration reaches zero, we obtain a partially ordered B2 structure whose (partial) site occupancies are provided in Table~\ref{table:b2_occupancies}. A self-consistent KKR-CPA calculation performed on this B2 (partially) ordered structure reveals that it is 53~meV/atom lower in energy than the disordered A2 structure.

\begin{table}[t]
\caption{Sublattice partial occupancies for the B2 structure predicted by the concentration wave analysis. The strongest site preferences are for Al and Ti, while V and Cr remain spread more evenly across both sublattices.}\label{table:b2_occupancies}
\begin{ruledtabular}
\begin{tabular}{lllll}
Lattice site & $c_\mathrm{Al}$ & $c_\mathrm{Ti}$ & $c_\mathrm{V}$ & $c_\mathrm{Cr}$ \\ \hline
1a           & 0.500           & 0.027           & 0.197          & 0.276           \\
1b           & 0.000           & 0.473           & 0.303          & 0.224          
\end{tabular}
\end{ruledtabular}
\end{table}

These data are broadly in alignment with several previous studies. For example, Huang \textit{et al.}~\cite{huang_order-disorder_2022} used CALPHAD to predict a B2 ordering temperature of 1239~K for the equiatomic composition. A separate study by Widom~\cite{widom_first-principles_2024} used first-principles calculations (PBE exchange-correlation functional) to predict a B2 ordering temperature of 1200~K, with the ordering characterised by Al moving to one sublattice, Ti to the other, and Cr and V expressing marginally weaker sublattice preferences, in excellent alignment with our own data. Experimentally, it is widely reported that this alloy forms a B2 phase across a wide temperature range~\cite{qiu_lightweight_2017}, which is in agreement with our own results. However, it should be noted that the nature of our predicted B2 chemical ordering differs from that proposed by Stephen \textit{et al.}~\cite{stephen_structural_2019}, who suggested that Al and Ti would occupy one sublattice while Cr and V occupy the other. We stress, though, that no assessment of the thermodynamic stability of this proposed structure is provided by these authors.

\begin{figure}[b]
    \centering
    \includegraphics[width=\linewidth]{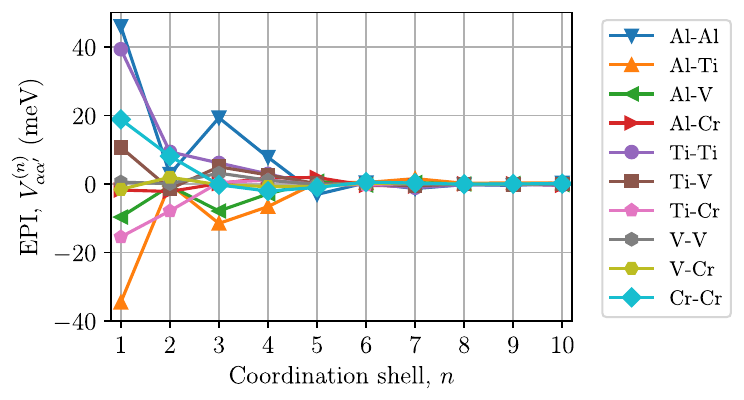}
    \caption{Calculated real-space atom-atom effective pair interactions (EPIs) for the AlTiVCr high-entropy alloy recovered from the \textit{ab initio} concentration wave analysis. A negative value of $V^{(n)}_{\alpha,\alpha'}$ indicates that it is energetically favourable to find the pair $\alpha$-$\alpha'$ on coordination shell $n$, while a positive value of $V^{(n)}_{\alpha,\alpha'}$ indicates the opposite. It can be seen that the strongest interactions are between Al-Al, Al-Ti, and Ti-Ti pairs, and also that interactions tail off quickly with increasing distance.}
    \label{fig:epis}
\end{figure}

From the reciprocal-space $S^{(2)}_{\alpha \alpha'}(\mathbf{k})$ data, we subsequently recover atom-atom EPIs, $V^{(n)}_{\alpha,\alpha'}$, suitable for use in atomistic simulations. EPIs for the AlTiVCr system for the first ten coordination shells of the bcc lattice are shown in Fig.~\ref{fig:epis}. Consistent with the discussion above, here it is seen that the strongest EPIs are for Al-Al, Al-Ti, and Ti-Ti pairs, which is most evident at nearest neighbour distance, \textit{i.e.}, for $n=1$. It can also be seen that the interactions are relatively short-ranged, which justifies our truncation of interactions beyond sixth-nearest neighbour distance in the fixed-lattice Monte Carlo simulations which are to follow.

\subsection{Atomistic, fixed-lattice Monte Carlo simulations}

\begin{figure}[t]
    \centering
    \includegraphics[width=\linewidth]{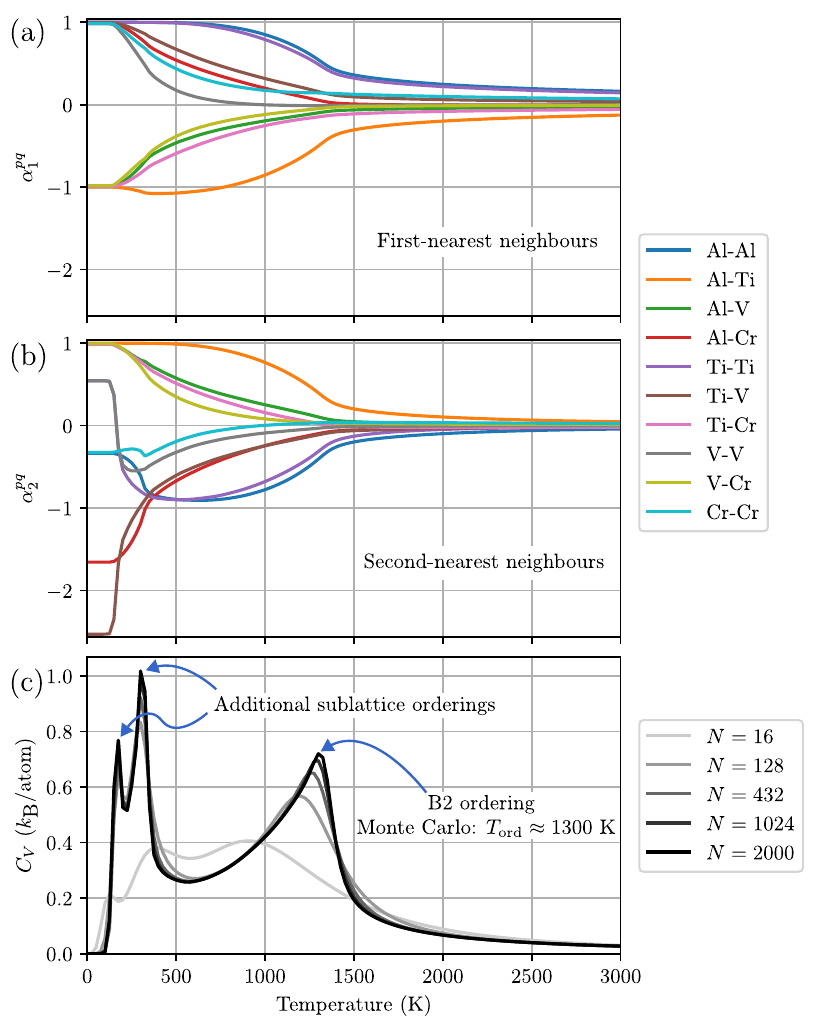}
    \caption{Plots of the Warren--Cowley ASRO parameters on the first (a) and second (b) coordination shells of the bcc lattice, together with (c) the specific heat as a function of temperature obtained using fixed-lattice Monte Carlo simulations. The data in panels (a) and (b) were obtained for a Wang--Landau sampling simulation on a system containing $N=2000$ atoms, while panel (c) evidences the convergence of results with respect to system size. The high-temperature peak in the specific heat data is associated with the initial B2 chemical ordering, while the low-temperature peaks are associated with additional sublattice ordering into an ordered single-phase ground state, shown in panel (a) of Fig.~\ref{fig:monte_carlo_configurations}.}
    \label{fig:monte_carlo_data}
\end{figure}

\begin{figure}
    \centering
    \includegraphics[width=\linewidth]{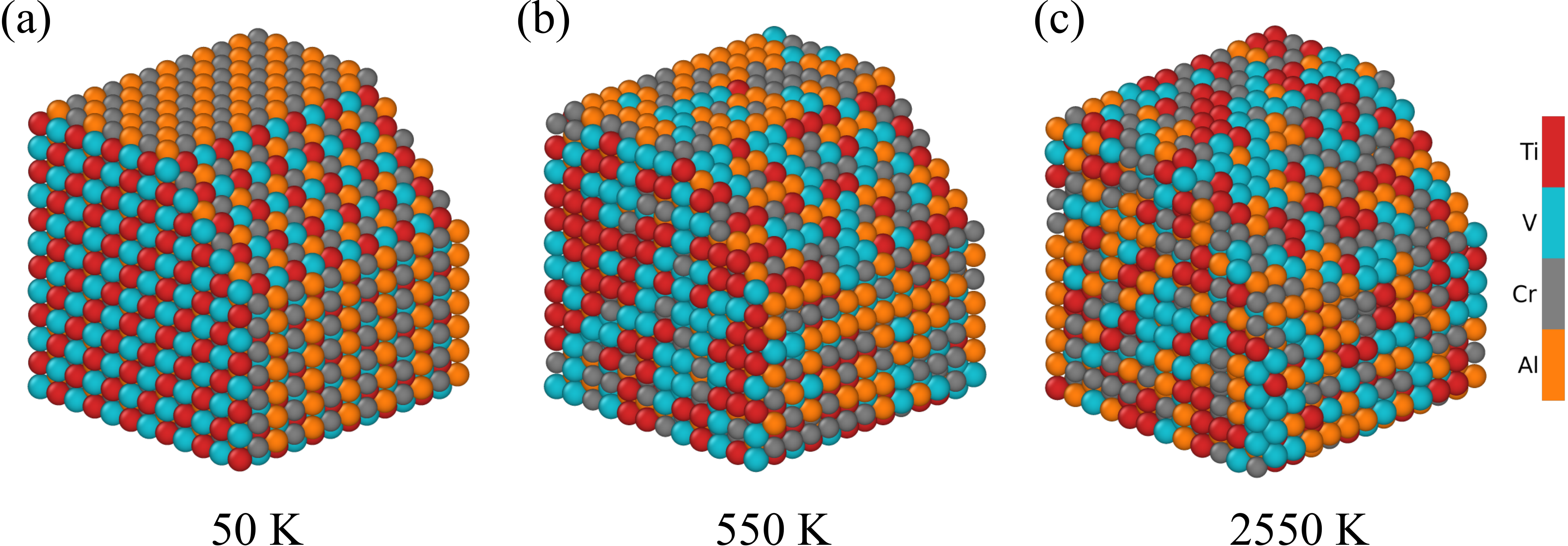}
    \caption{Representative equilibrium configurations for the AlTiVCr HEA at (a) 50~K, (b) 550~K, and (c) 2550~K. Al, Ti, V, and Cr are coloured orange, red, turquoise, and grey, respectively. The progression from fully disordered bcc to partially ordered B2 through to a fully ordered, single-phase ground state with decreasing simulation temperature can clearly be seen. Images generated using \textsc{Ovito}~\cite{stukowski_visualization_2010}.}
    \label{fig:monte_carlo_configurations}
\end{figure}

Using the recovered atom-atom EPIs graphed in Fig.~\ref{fig:epis}, we subsequently perform fixed-lattice Monte Carlo simulations to study the phase behaviour of the alloy in detail. This serves both to validate the mean-field Landau theory and to allow us to examine any additional phase transitions occurring at temperatures below the initial ordering temperature. Figure~\ref{fig:monte_carlo_data} shows plots of the temperature-dependent Warren--Cowley atomic short-range order (ASRO) parameters and simulation specific heat as obtained using Wang--Landau sampling, while Fig.~\ref{fig:monte_carlo_configurations} shows representative equilibrium atomic configurations at selected temperatures obtained using Metropolis--Hastings Monte Carlo. Convergence of simulation quantities with respect to system size was tested by performing simulations on cells ranging from $N=16$ atoms ($2\times2\times2$ cubic bcc unit cells) to $N=2000$ atoms ($10\times10\times10$ cubic bcc unit cells). Results for convergence of the specific heat with respect to system size are shown in panel (c) of Fig.~\ref{fig:monte_carlo_data}. The minimal changes in specific heat data between the cases $N=1024$ and $N=2000$ confirm that our simulation results are well converged with respect to this key simulation variable.

Inspection of the ASRO and specific heat data shown in Fig.~\ref{fig:monte_carlo_data} reveals three phase transitions. At high temperature, AlTiVCr is in a fully disordered bcc (A2) structure, a sample configuration of which is shown in panel (c) of Fig.~\ref{fig:monte_carlo_configurations}. As the alloy is cooled from this high-temperature state, a phase transition occurs at $T_\mathrm{ord}\approx1300$~K---indicated by a peak in the specific heat---and AlTiVCr takes on the partially ordered B2 (CsCl) structure, a sample configuration of which is shown in panel (b) of Fig.~\ref{fig:monte_carlo_configurations}. (That the ordering temperature predicted by the Monte Carlo simulations is lower than that predicted by the earlier concentration wave analysis is consistent with the fact that the former rigorously includes local thermal fluctuations, while the latter represents a Landau-type mean-field theory.) In this B2 structure, Al and Ti atoms move to separate sublattices. This is evidenced by the negative (positive) value of $\alpha^{pq}_1$ ($\alpha^{pq}_2$) for Al-Ti pairs---indicating these pairs are favoured as nearest neighbours and disfavoured as second-nearest neighbours---and positive (negative) values of $\alpha^{pq}_1$ ($\alpha^{pq}_2$) for Al-Al and Ti-Ti pairs---indicating these pairings are disfavoured as nearest neighbours and favoured as second-nearest neighbours. Weaker sublattice preferences for Cr and V can be seen, with Ti-Cr and Al-V pairs favoured as nearest neighbours and disfavoured as second-nearest neighbours, while Al-Cr and Ti-V pairs are disfavoured as nearest neighbours and favoured as second-nearest neighbours. This indicates that Cr atoms preferentially sit on the Al sublattice and V atoms on the Ti sublattice. This transition is therefore consistent with the mean-field analysis presented above in Sec.~\ref{sec:concentration_wave_results}, albeit with a reduction in the predicted ordering temperature, as is to be expected.

Upon further cooling, two additional phase transitions can be seen at $T \approx 300$~K and $T \approx 175$~K, indicating further sublattice orderings. The ordering at higher temperature corresponds to a layering of Al and Cr atoms on their sublattice, while the ordering at lower temperature corresponds to a layering of Ti and V atoms on their sublattice. The final, single-phase ground state is shown in panel (a) of Fig.~\ref{fig:monte_carlo_configurations}. The two sublattice orderings predicted by this model are not identical; on the Al/Cr sublattice some Al-Cr nearest neighbours remain, while on the V/Ti sublattice Ti-V nearest neighbours are expelled. This ground-state structure is thus similar in pair preferences, though not identical, to the Heusler-like structure shown in Fig.~1(a) of Ref.~\citenum{kalliney_x-ray_2026}, which those authors label the Y-I structure. We stress, however, that our atom-atom EPIs of Eq.~\eqref{eq:bragg-williams} are most reliable in the limit of high temperature, and that results for predicted ground-state structures must be interpreted cautiously. Additionally, the low predicted temperatures of these sublattice orderings in our calculations suggest that they may be challenging to observe experimentally.

\subsection{Impact of the B2 chemical ordering on the alloy's electronic structure and transport properties}

\begin{figure*}[t]
    \centering
    \includegraphics[width=\linewidth]{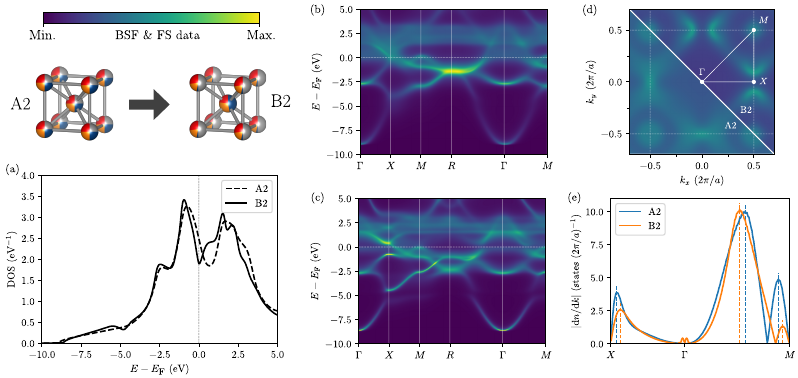}
    \caption{Changes to the electronic structure of the AlTiVCr HEA induced by the transition from the disordered bcc (A2) structure to the predicted B2-ordered structure with partial site occupancies as given in Table~\ref{table:b2_occupancies}. Panel (a) shows changes to the electronic density of states (DOS) around the Fermi level, $E_\mathrm{F}$. Panels (b) and (c) show plots of the Bloch spectral function (BSF) for the alloy in A2 and B2-ordered structures, respectively. The calculations for the A2 structure utilise a simple cubic representation of the bcc lattice with a 2-atom basis, for ease of comparison with the predicted B2-ordered structure. Panel (d) then shows $k_z=0$ cuts through the alloy's smeared-out Fermi surface (FS) compared for the two structures, demonstrating how it evolves as a result of the ordering. Finally, panel (e) shows plots of the quantity $|\mathrm{d}n/\mathrm{d}k|$ (defined in the main text) along two selected high-symmetry paths, which help to resolve separate FS sheets and their broadening. All BSF and FS data are plotted to the same colour scale. Partial B2 order results in sharper features in the DOS and BSF plots. Note that, as the predicted B2-ordered structure still has some sublattice disorder, a pristine band structure is not recovered.}
    \label{fig:ordering_edos_impact}
\end{figure*}

The study of the electronic transport properties of HEAs remains an active area of research~\cite{raghuraman_investigation_2021}, because the extreme compositional disorder in these complex systems results in heavily smeared Fermi surfaces~\cite{robarts_extreme_2020} and short electronic mean free paths~\cite{mu_uncovering_2019}. Chemical orderings, such as the high-temperature B2 ordering predicted in this work for AlTiVCr, can substantially impact these transport properties due to the associated changes to the symmetries and electronic structure of the material~\cite{woodgate_emergent_2025}.

Here, for the B2 ordering predicted by the \textit{ab initio} concentration wave analysis, we assess the impact of this chemical ordering on the {electronic structure and consequent} residual resistivity of the alloy{, where the latter is described within a Kubo--Greenwood linear response framework}. We define a B2 chemical order parameter, $\eta$, to quantify the degree of (partial) chemical ordering, where $\eta=0$ corresponds to the disordered solid solution, and $\eta=1$ corresponds to the largest permitted chemical fluctuation consistent with the predicted chemical polarisation of the concentration wave $\Delta c_\alpha (\mathbf{k} = (0,0,2\pi/a))$. This results in linear interpolation between the site occupancies of the fully disordered state and those of the B2 structure given in Table~\ref{table:b2_occupancies}. {We stress that this path is a model path in configuration space chosen for convenience in these calculations, and may not be the exact path followed experimentally by the alloy during the B2 chemical ordering.}

{Shown in Fig.~\ref{fig:ordering_edos_impact} are plots detailing how the electronic structure of the alloy evolves during this predicted chemical ordering, including the electronic DOS and BSF, $A_\mathrm{B}(\mathbf{k},E)$. The BSF can be thought of as a generalisation of a band structure to a disordered system~\cite{ebert_calculating_2011}. We show the BSF for a range of energies around $E_\mathrm{F}$ along a high-symmetry path through the irreducible Brillouin zone, as well as a constant-$E$ slice of this quantity in reciprocal space, \textit{i.e.}, the alloy's smeared-out Fermi surface (FS), together with the momentum-resolved occupation number derived from the BSF (discussed below). Data for the A2 structure are shown using the 2-atom simple cubic representation of this structure, with bands back-folded into the smaller Brillouin zone, to enable ease of comparison between it and the predicted B2-ordered structure.} Figure~\ref{fig:residual_resistivity_data} is {then} a plot of the calculated residual resistivity, $\rho_0$, as a function of $\eta$ for this system{, showing how this key intrinsic property evolves during chemical ordering}.

{We first discuss the observed changes in the electronic structure of the alloy as a result of the chemical ordering, shown in Fig.~\ref{fig:ordering_edos_impact}. Considering panel (a) first, it can be seen that---compared to the A2 structure---the DOS for the predicted B2-ordered structure has some sharper features at low energies, indicative of the formation of bonding states, and a reduced value at the Fermi level. This shift of $E_\mathrm{F}$ into a DOS minimum can be thought of as a pseudogap-like depression~\cite{widom_first-principles_2024}, as widely discussed for Al-rich transition-metal intermetallics and attributed to electron diffraction at prominent Bragg planes of the reciprocal lattice together with strong Al($sp$)--transition-metal($d$) hybridisation~\cite{fu_influence_2007}. Further insight is gained by considering details of the alloy's smeared-out band structures, panels (b) and (c). Compared to the A2 structure, where both sites in the simple cubic unit cell are crystallographically equivalent, in our predicted B2-ordered structure the degeneracy of some bands is lifted on account of the altered symmetry. Notably, at the $R$ point, the smeared-out band of high spectral weight approximately 1~eV below $E_\mathrm{F}$ is split into two smeared-out bands, while at the $X$ point an avoided crossing shifts spectral weight to above and below the Fermi level. Considering next details of the alloy's smeared-out FS, shown as a slice in the $k_z=0$ plane, panel (d), we note that for both the A2 and predicted B2-ordered structures there is significant smearing and non-zero spectral weight across the entire considered slice, indicative of short electronic mean free paths~\cite{robarts_extreme_2020}. The overall spectral weight in the FS cut also appears reduced upon entering the predicted B2-ordered structure. At the $M$ point, the A2 ring evolves into a spot in B2 (pocket closure), implying a smaller FS, while at the $X$ point symmetry breaking induces an avoided band crossing such that the connected contour seen for the A2 structure transforms into two crescents for the B2-ordered structure. These Fermi-surface changes are consistent with the reduction in the DOS at $E_\mathrm{F}$ discussed above.

\begin{figure}[t]
    \centering
    \includegraphics[width=\linewidth]{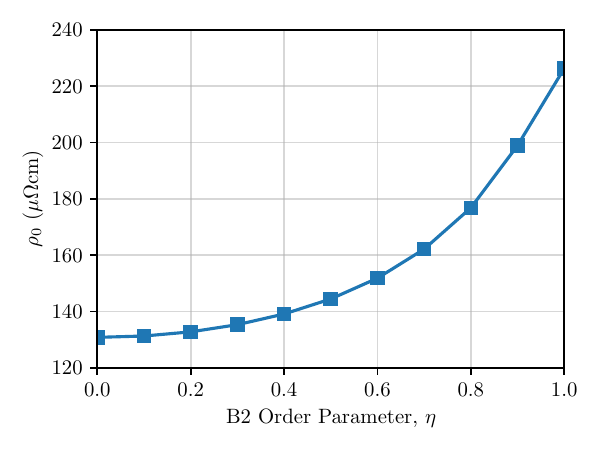}
    \caption{Calculated residual resistivity of the AlTiVCr HEA as a function of B2 chemical order parameter, $\eta$. Counter-intuitively, the (partial) chemical ordering results in an increased predicted residual resistivity as compared to the fully disordered (A2) alloy. (The A2 phase corresponds to $\eta=0$.)}
    \label{fig:residual_resistivity_data}
\end{figure}

We now consider how the alloy's residual resistivity evolves during this chemical ordering, \textit{i.e.}, the data shown in Fig.~\ref{fig:residual_resistivity_data}. Two key observations should be made here. The first is that, even compared to other HEAs~\cite{mu_uncovering_2019, raghuraman_investigation_2021, woodgate_emergent_2025}, the calculated residual resistivity values are large for AlTiVCr for all values of $\eta$---this material is a `bad' metal. Phenomenologically, we attribute this result to the differing nature of the electronic valence states associated with Al (a simple metal with no $d$ electrons) as compared to the transition metals (all of which have $3d$ states near the Fermi level) present in the composition, which results in strong electronic scattering. The strong electronic scattering caused by $3d$ solutes in Al-based binary alloys is discussed further by Samolyuk, Shyam, and Yoon~\cite{samolyuk_first_2026}, and we assert that similar scattering mechanisms are at play in this complex HEA. The second key observation is that the calculated residual resistivity increases as the system transitions from the fully disordered A2 structure to the predicted B2-ordered structure (with partial B2 order). This result is perhaps unexpected---chemical ordering typically represents the restoration of (at least some level of) translational symmetry and thus is expected to result in a reduced resistivity. However, it is reminiscent of the so-called \textit{Komplex} or $K$-state phenomenon, first observed in Ni-Cr alloys~\cite{thomas_uber_1951}, where a degree of ASRO can increase the residual resistivity of an alloy~\cite{lowitzer_ab_2010}. Increased residual resistivity values as a result of chemical ordering have also been previously predicted in other Al-containing HEAs~\cite{woodgate_emergent_2025}.

Though the factors determining alloys' residual resistivity values are myriad and complex~\cite{raghuraman_investigation_2021}, and within the Kubo--Greenwood formalism ultimately determined by evaluation of quantities integrated over the entire Brillouin zone, we interpret this result most naturally as originating from the reduced value of the DOS at the Fermi level. This is because the DOS at the Fermi level dictates the number of available charge carriers at low temperatures, and thus is key in determining electronic transport properties. As the electronic structure remains heavily smeared by the residual chemical disorder in the predicted B2-ordered structure, we therefore deduce that the increased residual resistivity originates from the reduction of available states at $E_\mathrm{F}$ induced by the ordering, similar to the mechanism discussed in Ref.~\citenum{woodgate_emergent_2025}.

To scrutinise this interpretation further, we consider the momentum-resolved occupation number, 
\begin{equation}
n(\mathbf{k}) = \int_{-\infty}^{E_\mathrm{F}} A_\mathrm{B}(\mathbf{k},E)\,\mathrm{d}E,
\end{equation}
obtained by integrating the BSF over the occupied states, following the approach of Robarts \textit{et al.}~\cite{robarts_extreme_2020}. For each Fermi-surface sheet, $n(\mathbf{k})$ takes the form of a disorder-smeared step, which the derivative $|\mathrm{d}n/\mathrm{d}k|$ resolves as a peak. These results are given in panel (e) of Fig.~\ref{fig:ordering_edos_impact}, where the peak position marks the Fermi wave vector, $k_\mathrm{F}$, while the peak width reflects the disorder-induced smearing of the state and is inversely proportional to its coherence length, $\xi = v_\mathrm{F}\tau$, with $v_\mathrm{F}$ the Fermi velocity and $\tau$ the state lifetime~\cite{robarts_extreme_2020}. Comparing the A2 and predicted B2-ordered structures, the peak widths are qualitatively unchanged upon ordering, indicating that the electronic mean free path is essentially unaffected by the partial B2 order. In contrast, the dominant peak positions shift towards $\Gamma$, indicating a contraction of $k_\mathrm{F}$ and hence a shrinking of the FS upon ordering, consistent with the pocket closure and contour changes discussed above. In a Drude-like picture, the conductivity, $\sigma$, may be expressed as $\sigma \propto \lambda\,S_\mathrm{F}$, with $\lambda$ an average mean free path and $S_\mathrm{F}$ the FS area~\cite{robarts_extreme_2020}. Thus, these observations support the interpretation that the increase in residual resistivity upon ordering is driven primarily by the shrinking of the FS---the momentum-space representation of the reduced DOS at $E_\mathrm{F}$---rather than by strengthened disorder scattering.
}

\section{Summary and Conclusions}
\label{sec:summary_and_conclusions}

In summary, results have been presented analysing the electronic structure, phase stability, and {transport} properties of the lightweight AlTiVCr HEA. The methods used include self-consistent DFT calculations within the KKR-CPA, an \textit{ab initio} concentration wave analysis, an electronic transport analysis using Kubo--Greenwood linear response theory, and fixed-lattice Monte Carlo simulations using both the Metropolis--Hastings and Wang--Landau sampling algorithms.

In alignment with existing experimental and computational data, a B2 chemical ordering was predicted to occur in the alloy at high temperatures. ($T_\mathrm{ord} \approx 1300$~K, as determined from fixed-lattice Monte Carlo simulations.) This ordering is driven primarily by Al and Ti moving to separate sublattices, with V and Cr expressing weaker site preferences. The influence of this chemical ordering on the electronic transport properties of the alloy was subsequently examined, and it was found that the chemical ordering resulted in an \textit{increased} residual resistivity as compared to the disordered alloy, which is understood to originate primarily from a reduction in the electronic DOS at the Fermi level. Finally, in addition to the high-temperature B2 ordering, atomistic Monte Carlo simulations revealed the predicted low-temperature ground state of the alloy within this modelling approach to be a fully ordered, single-phase structure.

The main implications of this work are threefold. First, these results provide detailed insight into the {electronic structure, phase stability, and transport properties} of this well-studied HEA, aligning with existing experimental data---as well as computational studies performed using alternative modelling approaches. This therefore serves as an important validation study. Second, our residual resistivity calculations indicate that measurement of electronic transport properties could be a useful tool to detect chemical order in this and related alloy compositions. Third, the work serves as further evidence of the reliability of the KKR-CPA and associated analyses for studying disorder--order transitions in high-entropy alloys. {Though non-magnetic calculations are well justified in the context of this study, where we focus primarily on the alloy's phase stability at elevated temperatures and where local magnetic moments collapsed during the electronic self-consistency cycle, f}uture work on the AlTiVCr HEA could seek to examine beyond-DFT methods---such as dynamical mean-field theory (DMFT)~\cite{minar_multiple-scattering_2005}---to improve the description of electronic correlations and examine the apparently subtle magnetism of this complex material. Future work using the outlined concentration-wave-based modelling approach for assessing HEA phase stability could seek to perform high-throughput studies searching for new potential HEA compositions with desirable physical properties.

\section{Methods}
\label{sec:methods}

\subsection{Alloy thermodynamics and phase stability}

\subsubsection{Electronic structure of the solid solution: The coherent potential approximation (CPA)}

In this work, all electronic structure calculations are performed using the KKR~\cite{korringa_calculation_1947, kohn_solution_1954, ebert_calculating_2011} formulation of DFT~\cite{hohenberg_inhomogeneous_1964, kohn_self-consistent_1965, martin_electronic_2004}, which uses multiple-scattering theory~\cite{faulkner_multiple_2018} to construct the Green's function for the Kohn--Sham equations. By virtue of the KKR method's formulation in terms of scattering of electronic waves from an array of atomic lattice sites (scattering centres), in the context of disordered materials, such as alloys, it is possible to develop effective medium theories to average over the disorder. These theories aim to construct configurationally averaged Green's functions representing the averaged electronic scattering properties of a given disordered medium, such as a substitutionally disordered alloy~\cite{elliott_theory_1974}. The most widely used effective medium theory within the KKR method is the CPA~\cite{ebert_calculating_2011, faulkner_multiple_2018}. First introduced by Soven~\cite{soven_coherent-potential_1967} in the context of scattering from Dirac $\delta$-function potentials in a 1D lattice, the CPA was subsequently generalised to three dimensions and potentials of finite spatial extent~\cite{soven_application_1970, gyorffy_coherent-potential_1972, stocks_complete_1978}, as is required for use in the context of DFT calculations. The CPA has previously been utilised---with success---to model a wide range of HEA compositions and to study a variety of physical phenomena. Previous studies have shown that the CPA accurately captures details of HEAs' electronic structure~\cite{robarts_extreme_2020, redka_interplay_2024} and magnetism~\cite{billington_bulk_2020, bista_fast_2025}, as well as a range of transport~\cite{samolyuk_temperature_2018, mu_uncovering_2019, raghuraman_investigation_2021} and structural~\cite{tian_structural_2013, tian_impact_2017, huang_elasticity_2018} properties. We are therefore confident of its suitability for studying the AlTiVCr HEA in the present work.

\subsubsection{Energetic cost of atomic-scale chemical fluctuations: A concentration wave analysis}
\label{sec:methods-concentration_waves}

To study the phase stability of a particular alloy composition, it is necessary to consider how atoms of different chemical species are arranged on the underlying crystal lattice, which is bcc in the context of the present work. In brief, an arrangement of atoms (a `configuration') on a crystal lattice with lattice positions $\{\mathbf{R}_i\}$ is specified by a set of \textit{site-occupation numbers}, $\{\xi_{i\alpha}\}$, where $\xi_{i\alpha}=1$ if site $i$ is occupied by an atom of chemical species $\alpha$, and $\xi_{i\alpha}=0$ otherwise. It is then natural to consider the average value of these site-occupation numbers, leading to the definition of the site-wise concentrations,
\begin{equation}
    c_{i\alpha} := \langle \xi_{i\alpha} \rangle,
    \label{eq:site-wise_concentrations}
\end{equation}
where $\langle \cdot \rangle$ denotes an average taken with respect to the relevant thermodynamic ensemble. These site-wise concentrations represent atomic long-range order (ALRO) parameters classifying potential chemically ordered phases.

\begin{figure}[b]
    \centering
    \includegraphics[width=\linewidth]{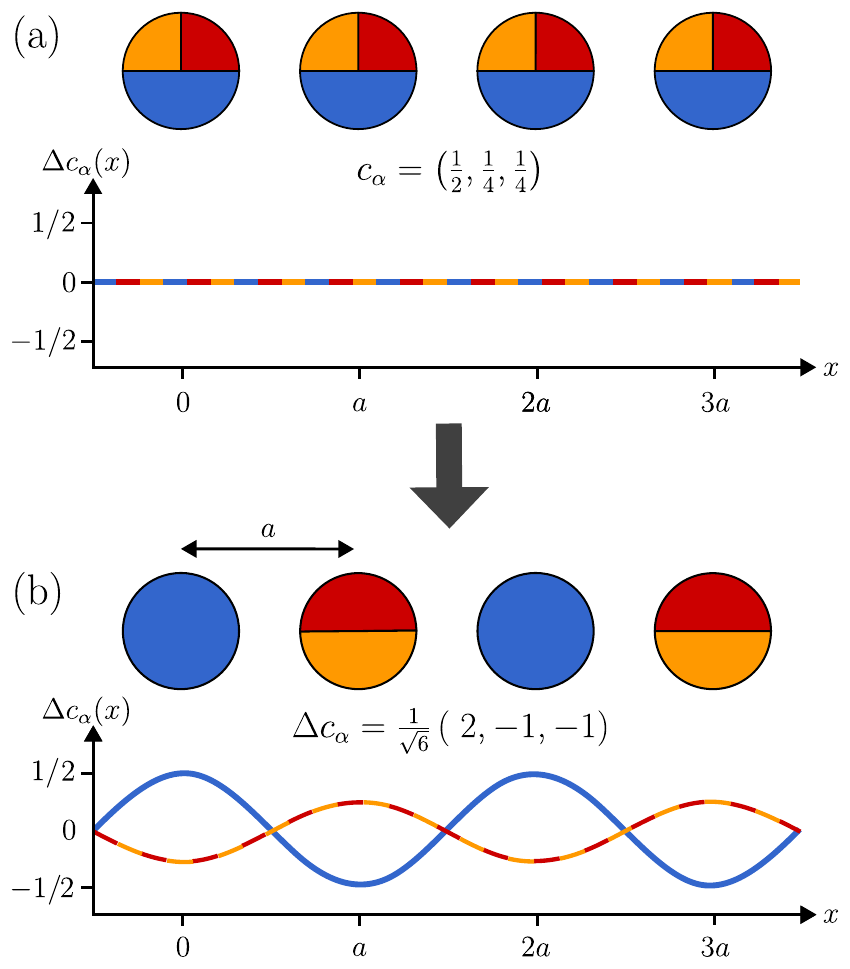}
    \caption{Schematic illustration of a particular \textit{concentration wave} modulating partial atomic lattice site occupancies in a toy, one-dimensional $A_2BC$ alloy. In order, the chemical species are coloured blue, orange, and red. The first state of the alloy, shown in panel (a), is homogeneous, with all lattice site occupancies equal. The second state of the alloy, shown in panel (b), is obtained by modulating the occupancies of panel (a) with a concentration wave of wave vector $k=\frac{\pi}{a}$ and (normalised) chemical polarisation $\Delta c_\alpha = \frac{1}{\sqrt{6}}(2, -1, -1)$.}
    \label{fig:concentration_wave_illustration}
\end{figure}

In the limit of high temperature, where the free energy landscape is dominated by entropy, the alloy is fully disordered and the site-wise concentrations are spatially homogeneous, \textit{i.e.}, any atom can occupy any lattice site with probability equal to its average concentration in the alloy, denoted $c_\alpha$. This disordered state is typically referred to as a \textit{solid solution}. With decreasing temperature, however, chemical order can emerge. The (now inhomogeneous) site-wise concentrations associated with a given chemical ordering can be written as a fluctuation about the homogeneous high-temperature reference state in a real-space representation as $c_{i\alpha} = c_\alpha + \Delta c_{i \alpha}$. Alternatively, the inhomogeneous state can be described in reciprocal space using the language of \textit{concentration waves}~\cite{khachaturyan_ordering_1978, gyorffy_concentration_1983} via
\begin{equation}
c_{i\alpha} = c_\alpha + \sum_{\mathbf{k}} e^{i \mathbf{k} \cdot \mathbf{R}_i} \Delta c_\alpha(\mathbf{k}),
\label{eq:concentration_wave}
\end{equation}
where $\Delta c_\alpha (\mathbf{k})$ represents a static concentration wave with wave vector $\mathbf{k}$ and chemical polarisation $\Delta c_\alpha$. A conceptual illustration of a concentration wave modulating partial lattice site occupancies in a one-dimensional `alloy' is given in Fig.~\ref{fig:concentration_wave_illustration}. In three dimensions, for most chemical orderings, the sum over $\mathbf{k}$ in Eq.~\eqref{eq:concentration_wave} typically runs over a handful of high-symmetry $\mathbf{k}$-vectors. For example, the B2 structure shown in Fig.~\ref{fig:a2_b2} is described by a concentration wave with wave vector $\mathbf{k} = (0,0,2\pi/a)$ and equivalent~\cite{khachaturyan_ordering_1978}.

The internal energy of the homogeneous, disordered alloy (\textit{i.e.}, $E_\mathrm{CPA}(\{c_{i\alpha}\})$ for $c_{i\alpha} \equiv c_\alpha$) can be evaluated within the CPA~\cite{johnson_density-functional_1986, johnson_total-energy_1990}. However, to assess the (free) energetic cost of atomic-scale chemical fluctuations represented by a given concentration wave it is necessary to consider how this quantity changes as the site-wise concentrations vary. To this end, concentration derivatives of $E_\mathrm{CPA}(\{c_{i\alpha}\})$ need to be considered. To lowest order, the relevant quantity is the second concentration derivative of the internal energy~\cite{khan_statistical_2016}, denoted $S^{(2)}_{i\alpha;j\alpha'}$ and defined as
\begin{equation}
S^{(2)}_{i\alpha;j\alpha'} \equiv \frac{\partial^2 E_\mathrm{CPA}}{\partial c_{i \alpha} \partial c_{j \alpha'}}\Big|_{\{c_\alpha\}}.
\label{eq:s2_definition}
\end{equation}
Due to the translational symmetry of the underlying crystal, it is most convenient to evaluate the lattice Fourier transform of this quantity, $S^{(2)}_{\alpha\alpha'}(\mathbf{k})$. The coupled set of equations which must be evaluated to obtain this quantity, including fully the rearrangement of charge due to chemical fluctuations, is given in Ref.~\citenum{khan_statistical_2016}, and we omit discussion of the details here for brevity. We emphasise here, though, that the alloy $S^{(2)}$ theory~\cite{khan_statistical_2016, woodgate_compositional_2022, woodgate_modelling_2024}---which is a multicomponent generalisation of an earlier theory developed solely for binary alloys~\cite{staunton_compositional_1994, johnson_first-principles_1994, clark_van_1995}---has previously been applied, with success, to the thermodynamics and phase stability of various Al-containing~\cite{woodgate_structure_2024, woodgate_emergent_2025, woodgate_formation_nodate}, Ti-containing~\cite{woodgate_competition_2024, woodgate_emergent_2025}, and refractory HEAs~\cite{woodgate_short-range_2023, woodgate_competition_2024}. We are therefore confident of its reliability in the context of the present work.

Proceeding, the change in free energy, $\Delta \Omega$, induced by a given (set of) concentration wave(s) modulating the site occupancies of the solid solution is then given by
\begin{equation}
    \Delta \Omega = \frac{1}{2} \sum_{\bf k} \sum_{\alpha, \alpha'} \Delta c_\alpha({\bf k}) [\beta^{-1} C^{-1}_{\alpha \alpha'} -S^{(2)}_{\alpha \alpha'}({\bf k})] \Delta c_{\alpha'}({\bf k}),
\label{eq:chemical_stability_reciprocal}
\end{equation}
where $\beta = \frac{1}{k_\mathrm{B} T}$, and $C_{\alpha \alpha'}^{-1} = \frac{\delta_{\alpha \alpha'}}{c_\alpha}$ is associated with entropic contributions to the free energy. The matrix in square brackets, $[\beta^{-1} C^{-1}_{\alpha \alpha'} -S^{(2)}_{\alpha \alpha'}({\bf k})]$, we refer to as the \textit{chemical stability matrix}. At high temperatures its eigenvalues are strictly positive for all wave vectors, indicating that the solid solution is thermodynamically stable. However, with decreasing temperature some eigenvalue is anticipated to pass through zero at a given temperature, $T_\mathrm{ord}$, for a given wave vector, $\mathbf{k}_\mathrm{ord}$, and an ordering is inferred~\cite{khan_statistical_2016}. The eigenvector associated with the eigenvalue passing through zero then describes which chemical species partition themselves onto which sublattice(s)~\cite{khan_statistical_2016, woodgate_compositional_2022, woodgate_modelling_2024}. This analysis constitutes a mean-field, Landau-type theory for alloy phase stability, and is naturally anticipated to marginally overestimate the true chemical ordering temperature.

\subsubsection{Recovery of atom-atom effective pair interactions}

Going beyond the above Landau-type mean-field theory, the quantity $S^{(2)}_{i\alpha;j\alpha'}$ can be related to a pairwise form of the alloy internal energy~\cite{khan_statistical_2016, woodgate_compositional_2022, woodgate_modelling_2024} which we refer to as the Bragg--Williams model~\cite{bragg_effect_1934, bragg_effect_1935}. The Hamiltonian for this model takes the form
\begin{equation}
    H(\{\xi_{i\alpha}\}) = \frac{1}{2}\sum_{i \alpha;j\alpha'} V_{i\alpha;j\alpha'} \xi_{i \alpha} \xi_{j \alpha'},
    \label{eq:bragg-williams}
\end{equation}
where $V_{i\alpha;j\alpha'} = -S^{(2)}_{i\alpha;j\alpha'}$  denotes the atom-atom \textit{effective pair interaction} (EPI) between an atom of chemical species $\alpha$ on lattice site $i$ and an atom of chemical species $\alpha'$ on lattice site $j$. As the underlying CPA reference medium is---on the average---spatially homogeneous and isotropic, so too are the atom-atom EPIs. We therefore write $V^{(n)}_{\alpha, \alpha'}$ to denote the interaction between chemical species $\alpha$ and $\alpha'$ on coordination shell $n$, \textit{i.e.}, at $n^\mathrm{th}$-nearest neighbour distance. Additionally, the EPIs are typically short-ranged and can be set to zero beyond some finite `cutoff' distance, which further simplifies evaluation of Eq.~\eqref{eq:bragg-williams} in practical calculations.

{It should be stressed that the EPIs, $V^{(n)}_{\alpha, \alpha'}$, are most reliable in the limit of high temperature, and predictions of stable low-temperature phases obtained using the model Hamiltonian of Eq.~\eqref{eq:bragg-williams} should be treated cautiously. Moreover, being a fixed-lattice model, the effects of local lattice distortions~\cite{kostiuchenko_impact_2019} and differences in vibrational entropy between phases~\cite{van_de_walle_effect_2002} are neglected. Typically, it is expected that such effects would reduce predicted chemical ordering temperatures when transitioning from a disordered alloy to a (partially) ordered intermetallic. Therefore, the simulations presented in this work may overestimate the true B2 transition temperature of the alloy. Finally, we note that all calculations presented here are non-magnetic, so any magnetic entropy differences between phases---or magnetic orderings at low temperatures---are not captured. Nonetheless, we believe that the alignment of our results with existing experimental and computational data suggests that the key mechanisms are well captured by our approach.}

\subsubsection{Atomistic Monte Carlo simulations}

Using the Hamiltonian of Eq.~\eqref{eq:bragg-williams}, lattice-based Monte Carlo simulations can be used to investigate the phase behaviour of an alloy system in detail. In this work, two Monte Carlo schemes are used: the Metropolis--Hastings method~\cite{metropolis_equation_1953, landau_guide_2014} and Wang--Landau sampling~\cite{wang_efficient_2001, landau_guide_2014}. Metropolis--Kawasaki dynamics~\cite{kawasaki_diffusion_1966}, meaning that Monte Carlo moves consist of pairs of atoms being exchanged within the simulation cell, are used throughout. This ensures that the overall alloy composition remains fixed during a simulation run.

In the Metropolis--Hastings algorithm~\cite{metropolis_equation_1953, landau_guide_2014}, trial moves from an initial configuration, labelled $n$, to a proposed candidate configuration, labelled $m$, are accepted with probability
\begin{equation}
    P_{n\rightarrow m} =
    \begin{cases}
        \exp(-\Delta E / k_\mathrm{B} T), & \Delta E > 0,\\[4pt]
        1, & \Delta E \le 0,
    \end{cases}
    \label{eq:metropolis_transition_probability}
\end{equation}
where $\Delta E = E_m - E_n$ is the corresponding change in internal energy, $k_\mathrm{B}$ is the usual Boltzmann constant, and $T$ is the simulation temperature. In this work, the Metropolis--Hastings algorithm is primarily employed to generate equilibrated configurations at chosen temperatures for visualisation. Simulations begin from a randomly generated supercell matching the target species concentrations, after which Monte Carlo sampling is carried out at the desired temperature. After a specified number of Monte Carlo steps, the system reaches an approximately equilibrated configuration suitable for visualisation and further analysis.

In contrast to the Metropolis--Hastings method, the Wang--Landau sampling algorithm~\cite{wang_efficient_2001, landau_guide_2014} is a flat-histogram scheme that enables direct determination of the configurational DOS, $g(E)$, of any model with a Hamiltonian that defines the internal energy as a function of the model configuration. Because $g(E)$ does not depend on temperature, once obtained it enables the computation of thermodynamic properties at arbitrary temperatures. Determination of $g(E)$ is achieved by grouping the spectrum of energies available to the system into distinct macrostates and constructing a histogram of energies representative of the true $g(E)$. An initial guess of $g(E)$ (\textit{e.g.}, $g(E)\equiv 1$) is iteratively refined over the course of a simulation run until convergence is achieved. A detailed discussion of this procedure in the context of simulation of alloys can be found in other works~\cite{woodgate_emergent_2025, naguszewski_optimal_2026}, so we omit those details here.

With $g(E)$ available following convergence of a Wang--Landau sampling run, the probability of observing a given energy $E_j$ at temperature $T$ is determined via
\begin{equation}
    P(E_j, T) = \frac{g(E_j)\, e^{-E_j/k_\mathrm{B} T}}{Z},
    \label{eq:prob_dist}
\end{equation}
where $E_j$ labels an energy (macrostate) of the Wang--Landau histogram, and $Z$ is the usual partition function. Knowledge of this probability distribution provides access to a range of thermodynamic observables. An example is the temperature-dependent specific heat, $C_V$, which is computed via~\cite{landau_guide_2014}
\begin{equation}
    C_V(T) = \frac{\langle E^2 \rangle - \langle E \rangle^2}{k_\mathrm{B} T^2},
\end{equation}
where the required ensemble averages, $\langle \cdot \rangle$, are evaluated using the probability distribution of Eq.~\eqref{eq:prob_dist}.

To quantify atomic short-range order in a simulation, the multicomponent Warren--Cowley atomic short-range order (ASRO) parameters~\cite{cowley_approximate_1950, cowley_short-range_1965}, $\alpha^{pq}_n$, are used. These are defined as
\begin{equation}
    \alpha^{pq}_n = 1 - \frac{P^{pq}_n}{c_q},
\end{equation}
where $P^{pq}_n$ is the conditional probability of finding a pair of atoms of chemical species $p$ and chemical species $q$ on coordination shell $n$, and $c_q$ is the overall concentration of element $q$. Positive $\alpha^{pq}_n$ values indicate that $p$-$q$ pairs occur less frequently than in a fully random alloy; negative values instead signal a preference for such pairings; and $\alpha^{pq}_n = 0$ corresponds to complete disorder.

These ASRO parameters are first evaluated as functions of energy across the Wang--Landau-sampled range (via averages over the configurations in each energy bin) and are then converted into temperature-dependent quantities using a weighted sum over the energies~\cite{woodgate_emergent_2025}, where the weights are simply the probabilities of Eq.~\eqref{eq:prob_dist}. Though these ASRO parameters are incomplete~\cite{sheriff_quantifying_2024}, \textit{i.e.}, they do not capture all details of all possible local chemical environments, we consider them sufficient for the purposes of the present study.

\subsection{Electronic transport properties}

Given its construction as a scattering theory for electronic waves, it is perhaps unsurprising that the KKR-CPA is well suited to the study of the electronic transport properties of disordered systems~\cite{butler_theory_1985}. In this work, we focus solely on the residual resistivity, an intrinsic physical quantity which is finite for a disordered (or partially ordered) alloy in which translational symmetry is broken and Bloch states are not eigenstates of the Hamiltonian.

From the configurationally averaged Green's function, $G$, obtained using the KKR-CPA, the Kubo--Greenwood formula~\cite{kubo_statistical-mechanical_1957, greenwood_boltzmann_1958} can be applied and the residual resistivity calculated according to~\cite{butler_theory_1985, mu_uncovering_2019, woodgate_emergent_2025}
\begin{equation}
    \sigma_{\mu\nu} = \frac{\hbar}{\pi V} \operatorname{Tr} \langle j_\mu \operatorname{Im} G^+ (E_\mathrm{F}) j_\nu \operatorname{Im} G^+ (E_\mathrm{F}) \rangle_\mathrm{CPA},
\label{eq:kubo_greenwood}
\end{equation}
where $V$ is the simulation cell volume; $G^+(E_\mathrm{F})$ is the retarded Green's function at the Fermi level, $E_\mathrm{F}$; and $j_\mu$ is the current density operator with Cartesian coordinate indices $\mu$ and $\nu$. Angled brackets, $\langle \cdot \rangle_\mathrm{CPA}$, are used to indicate the CPA average over substitutional disorder. From the conductivity tensor, the residual resistivity of a material is recovered in a simple post-processing step. {We note here that our presented residual resistivity calculations do not incorporate the effects of local lattice distortions, which are typically expected to increase calculated residual resistivity values on account of the increased smearing of electronic bands around the Fermi surface~\cite{mu_influence_2019}. Such effects can be incorporated, in an approximate way, within the alloy analogy model~\cite{ebert_calculating_2015}, but this is beyond the scope of the present work.}

\subsection{Computational details}

All self-consistent KKR-CPA calculations were performed within the atomic sphere approximation (ASA)~\cite{andersen_linear_1975} using the all-electron SPR-KKR package~\cite{ebert_calculating_2011, sprkkr_website}. Throughout, the lattice parameter of the system was fixed to the experimental value of 3.075~\AA~\cite{qiu_lightweight_2017}. The local density approximation (LDA) was used, specifically the parametrisation of Vosko, Wilk, and Nusair~\cite{vosko_accurate_1980}. For self-consistent calculations, Brillouin zone integrations were performed using an SPR-KKR parameter setting of \texttt{NKTAB=5000}, which results in a $62 \times 62 \times 62$ $\mathbf{k}$-point mesh over the first Brillouin zone in the case of the disordered bcc (A2) structure. A 32-point semi-circular contour in the complex plane was used for energy integrations during self-consistent calculations. Transport calculations were performed using the Kubo--Greenwood formalism as implemented in SPR-KKR, with Brillouin-zone integrations performed using a parameter of \texttt{NKTAB=150000} (full Brillouin zone), resulting in a dense $53 \times 53 \times 53$ $\mathbf{k}$-point mesh over the first Brillouin zone of the B2 crystal structure. In the case of initial self-consistent calculations and the concentration wave analysis, scalar-relativistic calculations were used, while for the transport analysis fully relativistic calculations were utilised. {The use of the ASA and LDA in this study can be justified on the basis that we are studying a densely packed, metallic system, where these approximations are robust. Moreover, in the context of studying phase transitions, we are concerned primarily with energy \textit{differences} between phases, and the ASA provides a satisfactory description of these. Holding the lattice parameter fixed at the experimental value, which is somewhat larger than the LDA-optimised value, naturally helps to incorporate the effects of thermal expansion in our modelling.}

The concentration wave analysis was performed using an in-house implementation of the $S^{(2)}$ theory for multicomponent alloys, the computational details of which have been discussed in earlier works~\cite{khan_statistical_2016, woodgate_compositional_2022, woodgate_modelling_2024}. (This code includes an adaptive meshing scheme for efficient Brillouin zone integrations~\cite{bruno_algorithms_1997}.) The quantity $S^{(2)}_{\alpha\alpha'}(\mathbf{k})$ was sampled at a total of 56 $\mathbf{k}$-points in the irreducible Brillouin zone, including high-symmetry points. From an initial analysis of atom-atom EPIs, $V_{\alpha \alpha'}^{(n)}$, fitted to the first 10 coordination shells of the bcc lattice, it was determined that a fit to the first 6 coordination shells of the bcc lattice, corresponding to a real-space `cutoff' of 6.15~\AA, was sufficient to capture the reciprocal-space $S^{(2)}_{\alpha \alpha'}(\mathbf{k})$ data with good accuracy.

All fixed-lattice Monte Carlo simulations (Metropolis--Hastings and Wang--Landau) were performed using the \textsc{Brawl} package~\cite{naguszewski_brawl_2025}, with details of the employed parallelisation and parallel load-balancing schemes for the Wang--Landau sampling algorithm as discussed in Ref.~\citenum{naguszewski_optimal_2026}. The final Wang--Landau sampling run used a simulation supercell consisting of $10 \times 10 \times 10$ bcc cubic unit cells for a total of 2000 atoms. The total energy range for the 2000-atom simulation cell, from $-65$~meV/atom to 10~meV/atom, was divided into 1024 energy bins. (For tests of convergence of key quantities such as the specific heat with respect to the size of the simulation cell, the energy range and number of energy bins were scaled appropriately.) The applied Wang--Landau flatness tolerance, used to determine convergence of the energy histogram at a given iteration, was set to be $F>0.8$. The Wang--Landau update factor tolerance, $\log f$, was set to be $1\times10^{-12}$ for all simulation runs.

\section*{Acknowledgements}
C.D.W.\ thanks Prof.\ Julie B.\ Staunton (Department of Physics, University of Warwick, United Kingdom) for provision of the code implementing the $S^{(2)}$ theory for multicomponent alloys as used in this work, and for helpful discussions.
C.D.W.\ acknowledges support from a UK Engineering and Physical Sciences Research Council (EPSRC) Doctoral Prize Fellowship at the University of Bristol, Grant EP/W524414/1. 
H.J.N.\ is supported by a studentship within the EPSRC Centre for Doctoral Training in Modelling of Heterogeneous Systems, Grant EP/S022848/1. 
D.R.\ acknowledges financial support from the Deutsche Forschungsgemeinschaft (DFG) under Project No. 528706678.

The authors also acknowledge the use of compute resources provided by the Isambard 3 high-performance computing (HPC) facility. Isambard 3 is hosted by the University of Bristol and operated by the GW4 Alliance (\href{https://gw4.ac.uk}{https://gw4.ac.uk}) and is funded by UK Research and Innovation; and the EPSRC, Grant EP/X039137/1. Additional compute resources were provided by the Advanced Computing Research Centre (ACRC) of the University of Bristol, and by the Scientific Computing Research Technology Platform (SCRTP) of the University of Warwick.

\section*{Author contributions}

C.D.W.\ conceived of the approach, with input from all authors. N.F.P.\ performed self-consistent SPR-KKR calculations for the disordered A2 phase under the supervision of D.R\@. C.D.W.\ performed the concentration wave analysis and recovered the atom-atom effective pair interactions. H.J.N.\ implemented the Wang--Landau sampling algorithm in the \textsc{Brawl} package and performed the fixed-lattice Monte Carlo simulations. N.F.P.\ performed self-consistent SPR-KKR calculations and electronic transport analysis for the predicted B2-ordered structures under the supervision of D.R\@. C.D.W.\ prepared the first draft of the manuscript with input from H.J.N\@. Subsequently, all authors revised the manuscript and approved its final version.\\

\section*{Data Availability}

All data supporting the findings of this study are openly available at the following DOI/URL: \href{https://doi.org/10.5281/zenodo.18095932}{https://doi.org/10.5281/zenodo.18095932}. The all-electron SPR-KKR package, used for performing self-consistent KKR-CPA calculations and the electronic transport analysis, is available at \href{https://sprkkr.org}{https://sprkkr.org}. The \textsc{Brawl} package, used for performing the fixed-lattice Monte Carlo simulations, is available at \href{https://github.com/ChrisWoodgate/BraWl}{https://github.com/ChrisWoodgate/BraWl}. The code implementing the $S^{(2)}$ theory for multicomponent alloys used in the concentration wave analysis is available from C.D.W.\ upon reasonable request.

\section*{Declarations}

\textbf{Competing interests.} The authors declare that they have no known competing financial interests or personal relationships that could have appeared to influence the work reported in this paper.


\begin{thebibliography}{116}%
\makeatletter
\providecommand \@ifxundefined [1]{%
 \@ifx{#1\undefined}
}%
\providecommand \@ifnum [1]{%
 \ifnum #1\expandafter \@firstoftwo
 \else \expandafter \@secondoftwo
 \fi
}%
\providecommand \@ifx [1]{%
 \ifx #1\expandafter \@firstoftwo
 \else \expandafter \@secondoftwo
 \fi
}%
\providecommand \natexlab [1]{#1}%
\providecommand \enquote  [1]{``#1''}%
\providecommand \bibnamefont  [1]{#1}%
\providecommand \bibfnamefont [1]{#1}%
\providecommand \citenamefont [1]{#1}%
\providecommand \href@noop [0]{\@secondoftwo}%
\providecommand \href [0]{\begingroup \@sanitize@url \@href}%
\providecommand \@href[1]{\@@startlink{#1}\@@href}%
\providecommand \@@href[1]{\endgroup#1\@@endlink}%
\providecommand \@sanitize@url [0]{\catcode `\\12\catcode `\$12\catcode `\&12\catcode `\#12\catcode `\^12\catcode `\_12\catcode `\%12\relax}%
\providecommand \@@startlink[1]{}%
\providecommand \@@endlink[0]{}%
\providecommand \url  [0]{\begingroup\@sanitize@url \@url }%
\providecommand \@url [1]{\endgroup\@href {#1}{\urlprefix }}%
\providecommand \urlprefix  [0]{URL }%
\providecommand \Eprint [0]{\href }%
\providecommand \doibase [0]{https://doi.org/}%
\providecommand \selectlanguage [0]{\@gobble}%
\providecommand \bibinfo  [0]{\@secondoftwo}%
\providecommand \bibfield  [0]{\@secondoftwo}%
\providecommand \translation [1]{[#1]}%
\providecommand \BibitemOpen [0]{}%
\providecommand \bibitemStop [0]{}%
\providecommand \bibitemNoStop [0]{.\EOS\space}%
\providecommand \EOS [0]{\spacefactor3000\relax}%
\providecommand \BibitemShut  [1]{\csname bibitem#1\endcsname}%
\let\auto@bib@innerbib\@empty
\bibitem [{\citenamefont {Yeh}\ \emph {et~al.}(2004)\citenamefont {Yeh}, \citenamefont {Chen}, \citenamefont {Lin}, \citenamefont {Gan}, \citenamefont {Chin}, \citenamefont {Shun}, \citenamefont {Tsau},\ and\ \citenamefont {Chang}}]{yeh_nanostructured_2004}%
  \BibitemOpen
  \bibfield  {author} {\bibinfo {author} {\bibfnamefont {J.-W.}\ \bibnamefont {Yeh}}, \bibinfo {author} {\bibfnamefont {S.-K.}\ \bibnamefont {Chen}}, \bibinfo {author} {\bibfnamefont {S.-J.}\ \bibnamefont {Lin}}, \bibinfo {author} {\bibfnamefont {J.-Y.}\ \bibnamefont {Gan}}, \bibinfo {author} {\bibfnamefont {T.-S.}\ \bibnamefont {Chin}}, \bibinfo {author} {\bibfnamefont {T.-T.}\ \bibnamefont {Shun}}, \bibinfo {author} {\bibfnamefont {C.-H.}\ \bibnamefont {Tsau}},\ and\ \bibinfo {author} {\bibfnamefont {S.-Y.}\ \bibnamefont {Chang}},\ }\bibfield  {title} {\bibinfo {title} {Nanostructured {High}-{Entropy} {Alloys} with {Multiple} {Principal} {Elements}: {Novel} {Alloy} {Design} {Concepts} and {Outcomes}},\ }\href {https://doi.org/10.1002/adem.200300567} {\bibfield  {journal} {\bibinfo  {journal} {Advanced Engineering Materials}\ }\textbf {\bibinfo {volume} {6}},\ \bibinfo {pages} {299} (\bibinfo {year} {2004})}\BibitemShut {NoStop}%
\bibitem [{\citenamefont {Cantor}\ \emph {et~al.}(2004)\citenamefont {Cantor}, \citenamefont {Chang}, \citenamefont {Knight},\ and\ \citenamefont {Vincent}}]{cantor_microstructural_2004}%
  \BibitemOpen
  \bibfield  {author} {\bibinfo {author} {\bibfnamefont {B.}~\bibnamefont {Cantor}}, \bibinfo {author} {\bibfnamefont {I.~T.~H.}\ \bibnamefont {Chang}}, \bibinfo {author} {\bibfnamefont {P.}~\bibnamefont {Knight}},\ and\ \bibinfo {author} {\bibfnamefont {A.~J.~B.}\ \bibnamefont {Vincent}},\ }\bibfield  {title} {\bibinfo {title} {Microstructural development in equiatomic multicomponent alloys},\ }\href {https://doi.org/10.1016/j.msea.2003.10.257} {\bibfield  {journal} {\bibinfo  {journal} {Materials Science and Engineering: A}\ }\textbf {\bibinfo {volume} {375-377}},\ \bibinfo {pages} {213} (\bibinfo {year} {2004})}\BibitemShut {NoStop}%
\bibitem [{\citenamefont {Miracle}\ and\ \citenamefont {Senkov}(2017)}]{miracle_critical_2017}%
  \BibitemOpen
  \bibfield  {author} {\bibinfo {author} {\bibfnamefont {D.~B.}\ \bibnamefont {Miracle}}\ and\ \bibinfo {author} {\bibfnamefont {O.~N.}\ \bibnamefont {Senkov}},\ }\bibfield  {title} {\bibinfo {title} {A critical review of high entropy alloys and related concepts},\ }\href {https://doi.org/10.1016/j.actamat.2016.08.081} {\bibfield  {journal} {\bibinfo  {journal} {Acta Materialia}\ }\textbf {\bibinfo {volume} {122}},\ \bibinfo {pages} {448} (\bibinfo {year} {2017})}\BibitemShut {NoStop}%
\bibitem [{\citenamefont {George}\ \emph {et~al.}(2019)\citenamefont {George}, \citenamefont {Raabe},\ and\ \citenamefont {Ritchie}}]{george_high-entropy_2019}%
  \BibitemOpen
  \bibfield  {author} {\bibinfo {author} {\bibfnamefont {E.~P.}\ \bibnamefont {George}}, \bibinfo {author} {\bibfnamefont {D.}~\bibnamefont {Raabe}},\ and\ \bibinfo {author} {\bibfnamefont {R.~O.}\ \bibnamefont {Ritchie}},\ }\bibfield  {title} {\bibinfo {title} {High-entropy alloys},\ }\href {https://doi.org/10.1038/s41578-019-0121-4} {\bibfield  {journal} {\bibinfo  {journal} {Nature Reviews Materials}\ }\textbf {\bibinfo {volume} {4}},\ \bibinfo {pages} {515} (\bibinfo {year} {2019})}\BibitemShut {NoStop}%
\bibitem [{\citenamefont {Gludovatz}\ \emph {et~al.}(2014)\citenamefont {Gludovatz}, \citenamefont {Hohenwarter}, \citenamefont {Catoor}, \citenamefont {Chang}, \citenamefont {George},\ and\ \citenamefont {Ritchie}}]{gludovatz_fracture-resistant_2014}%
  \BibitemOpen
  \bibfield  {author} {\bibinfo {author} {\bibfnamefont {B.}~\bibnamefont {Gludovatz}}, \bibinfo {author} {\bibfnamefont {A.}~\bibnamefont {Hohenwarter}}, \bibinfo {author} {\bibfnamefont {D.}~\bibnamefont {Catoor}}, \bibinfo {author} {\bibfnamefont {E.~H.}\ \bibnamefont {Chang}}, \bibinfo {author} {\bibfnamefont {E.~P.}\ \bibnamefont {George}},\ and\ \bibinfo {author} {\bibfnamefont {R.~O.}\ \bibnamefont {Ritchie}},\ }\bibfield  {title} {\bibinfo {title} {A fracture-resistant high-entropy alloy for cryogenic applications},\ }\href {https://doi.org/10.1126/science.1254581} {\bibfield  {journal} {\bibinfo  {journal} {Science}\ }\textbf {\bibinfo {volume} {345}},\ \bibinfo {pages} {1153} (\bibinfo {year} {2014})}\BibitemShut {NoStop}%
\bibitem [{\citenamefont {Gludovatz}\ \emph {et~al.}(2016)\citenamefont {Gludovatz}, \citenamefont {Hohenwarter}, \citenamefont {Thurston}, \citenamefont {Bei}, \citenamefont {Wu}, \citenamefont {George},\ and\ \citenamefont {Ritchie}}]{gludovatz_exceptional_2016}%
  \BibitemOpen
  \bibfield  {author} {\bibinfo {author} {\bibfnamefont {B.}~\bibnamefont {Gludovatz}}, \bibinfo {author} {\bibfnamefont {A.}~\bibnamefont {Hohenwarter}}, \bibinfo {author} {\bibfnamefont {K.~V.~S.}\ \bibnamefont {Thurston}}, \bibinfo {author} {\bibfnamefont {H.}~\bibnamefont {Bei}}, \bibinfo {author} {\bibfnamefont {Z.}~\bibnamefont {Wu}}, \bibinfo {author} {\bibfnamefont {E.~P.}\ \bibnamefont {George}},\ and\ \bibinfo {author} {\bibfnamefont {R.~O.}\ \bibnamefont {Ritchie}},\ }\bibfield  {title} {\bibinfo {title} {Exceptional damage-tolerance of a medium-entropy alloy {CrCoNi} at cryogenic temperatures},\ }\href {https://doi.org/10.1038/ncomms10602} {\bibfield  {journal} {\bibinfo  {journal} {Nature Communications}\ }\textbf {\bibinfo {volume} {7}},\ \bibinfo {pages} {10602} (\bibinfo {year} {2016})}\BibitemShut {NoStop}%
\bibitem [{\citenamefont {Liu}\ \emph {et~al.}(2022)\citenamefont {Liu}, \citenamefont {Yu}, \citenamefont {Kabra}, \citenamefont {Jiang}, \citenamefont {Forna-Kreutzer}, \citenamefont {Zhang}, \citenamefont {Payne}, \citenamefont {Walsh}, \citenamefont {Gludovatz}, \citenamefont {Asta}, \citenamefont {Minor}, \citenamefont {George},\ and\ \citenamefont {Ritchie}}]{liu_exceptional_2022}%
  \BibitemOpen
  \bibfield  {author} {\bibinfo {author} {\bibfnamefont {D.}~\bibnamefont {Liu}}, \bibinfo {author} {\bibfnamefont {Q.}~\bibnamefont {Yu}}, \bibinfo {author} {\bibfnamefont {S.}~\bibnamefont {Kabra}}, \bibinfo {author} {\bibfnamefont {M.}~\bibnamefont {Jiang}}, \bibinfo {author} {\bibfnamefont {P.}~\bibnamefont {Forna-Kreutzer}}, \bibinfo {author} {\bibfnamefont {R.}~\bibnamefont {Zhang}}, \bibinfo {author} {\bibfnamefont {M.}~\bibnamefont {Payne}}, \bibinfo {author} {\bibfnamefont {F.}~\bibnamefont {Walsh}}, \bibinfo {author} {\bibfnamefont {B.}~\bibnamefont {Gludovatz}}, \bibinfo {author} {\bibfnamefont {M.}~\bibnamefont {Asta}}, \bibinfo {author} {\bibfnamefont {A.~M.}\ \bibnamefont {Minor}}, \bibinfo {author} {\bibfnamefont {E.~P.}\ \bibnamefont {George}},\ and\ \bibinfo {author} {\bibfnamefont {R.~O.}\ \bibnamefont {Ritchie}},\ }\bibfield  {title} {\bibinfo {title} {Exceptional fracture toughness of {CrCoNi}-based medium- and high-entropy alloys at 20 kelvin},\ }\href
  {https://doi.org/10.1126/science.abp8070} {\bibfield  {journal} {\bibinfo  {journal} {Science}\ }\textbf {\bibinfo {volume} {378}},\ \bibinfo {pages} {978} (\bibinfo {year} {2022})}\BibitemShut {NoStop}%
\bibitem [{\citenamefont {Qiu}\ \emph {et~al.}(2017{\natexlab{a}})\citenamefont {Qiu}, \citenamefont {Thomas}, \citenamefont {Gibson}, \citenamefont {Fraser},\ and\ \citenamefont {Birbilis}}]{qiu_corrosion_2017}%
  \BibitemOpen
  \bibfield  {author} {\bibinfo {author} {\bibfnamefont {Y.}~\bibnamefont {Qiu}}, \bibinfo {author} {\bibfnamefont {S.}~\bibnamefont {Thomas}}, \bibinfo {author} {\bibfnamefont {M.~A.}\ \bibnamefont {Gibson}}, \bibinfo {author} {\bibfnamefont {H.~L.}\ \bibnamefont {Fraser}},\ and\ \bibinfo {author} {\bibfnamefont {N.}~\bibnamefont {Birbilis}},\ }\bibfield  {title} {\bibinfo {title} {Corrosion of high entropy alloys},\ }\href {https://doi.org/10.1038/s41529-017-0009-y} {\bibfield  {journal} {\bibinfo  {journal} {npj Materials Degradation}\ }\textbf {\bibinfo {volume} {1}},\ \bibinfo {pages} {15} (\bibinfo {year} {2017}{\natexlab{a}})}\BibitemShut {NoStop}%
\bibitem [{\citenamefont {El-Atwani}\ \emph {et~al.}(2019)\citenamefont {El-Atwani}, \citenamefont {Li}, \citenamefont {Li}, \citenamefont {Devaraj}, \citenamefont {Baldwin}, \citenamefont {Schneider}, \citenamefont {Sobieraj}, \citenamefont {Wróbel}, \citenamefont {Nguyen-Manh}, \citenamefont {Maloy},\ and\ \citenamefont {Martinez}}]{el-atwani_outstanding_2019}%
  \BibitemOpen
  \bibfield  {author} {\bibinfo {author} {\bibfnamefont {O.}~\bibnamefont {El-Atwani}}, \bibinfo {author} {\bibfnamefont {N.}~\bibnamefont {Li}}, \bibinfo {author} {\bibfnamefont {M.}~\bibnamefont {Li}}, \bibinfo {author} {\bibfnamefont {A.}~\bibnamefont {Devaraj}}, \bibinfo {author} {\bibfnamefont {J.~K.~S.}\ \bibnamefont {Baldwin}}, \bibinfo {author} {\bibfnamefont {M.~M.}\ \bibnamefont {Schneider}}, \bibinfo {author} {\bibfnamefont {D.}~\bibnamefont {Sobieraj}}, \bibinfo {author} {\bibfnamefont {J.~S.}\ \bibnamefont {Wróbel}}, \bibinfo {author} {\bibfnamefont {D.}~\bibnamefont {Nguyen-Manh}}, \bibinfo {author} {\bibfnamefont {S.~A.}\ \bibnamefont {Maloy}},\ and\ \bibinfo {author} {\bibfnamefont {E.}~\bibnamefont {Martinez}},\ }\bibfield  {title} {\bibinfo {title} {Outstanding radiation resistance of tungsten-based high-entropy alloys},\ }\href {https://doi.org/10.1126/sciadv.aav2002} {\bibfield  {journal} {\bibinfo  {journal} {Science Advances}\ }\textbf {\bibinfo {volume} {5}},\ \bibinfo {pages}
  {eaav2002} (\bibinfo {year} {2019})}\BibitemShut {NoStop}%
\bibitem [{\citenamefont {El~Atwani}\ \emph {et~al.}(2023)\citenamefont {El~Atwani}, \citenamefont {Vo}, \citenamefont {Tunes}, \citenamefont {Lee}, \citenamefont {Alvarado}, \citenamefont {Krienke}, \citenamefont {Poplawsky}, \citenamefont {Kohnert}, \citenamefont {Gigax}, \citenamefont {Chen}, \citenamefont {Li}, \citenamefont {Wang}, \citenamefont {Wróbel}, \citenamefont {Nguyen-Manh}, \citenamefont {Baldwin}, \citenamefont {Tukac}, \citenamefont {Aydogan}, \citenamefont {Fensin},\ and\ \citenamefont {Martinez}}]{el_atwani_quinary_2023}%
  \BibitemOpen
  \bibfield  {author} {\bibinfo {author} {\bibfnamefont {O.}~\bibnamefont {El~Atwani}}, \bibinfo {author} {\bibfnamefont {H.~T.}\ \bibnamefont {Vo}}, \bibinfo {author} {\bibfnamefont {M.~A.}\ \bibnamefont {Tunes}}, \bibinfo {author} {\bibfnamefont {C.}~\bibnamefont {Lee}}, \bibinfo {author} {\bibfnamefont {A.}~\bibnamefont {Alvarado}}, \bibinfo {author} {\bibfnamefont {N.}~\bibnamefont {Krienke}}, \bibinfo {author} {\bibfnamefont {J.~D.}\ \bibnamefont {Poplawsky}}, \bibinfo {author} {\bibfnamefont {A.~A.}\ \bibnamefont {Kohnert}}, \bibinfo {author} {\bibfnamefont {J.}~\bibnamefont {Gigax}}, \bibinfo {author} {\bibfnamefont {W.-Y.}\ \bibnamefont {Chen}}, \bibinfo {author} {\bibfnamefont {M.}~\bibnamefont {Li}}, \bibinfo {author} {\bibfnamefont {Y.~Q.}\ \bibnamefont {Wang}}, \bibinfo {author} {\bibfnamefont {J.~S.}\ \bibnamefont {Wróbel}}, \bibinfo {author} {\bibfnamefont {D.}~\bibnamefont {Nguyen-Manh}}, \bibinfo {author} {\bibfnamefont {J.~K.~S.}\ \bibnamefont {Baldwin}}, \bibinfo {author} {\bibfnamefont
  {O.~U.}\ \bibnamefont {Tukac}}, \bibinfo {author} {\bibfnamefont {E.}~\bibnamefont {Aydogan}}, \bibinfo {author} {\bibfnamefont {S.}~\bibnamefont {Fensin}},\ and\ \bibinfo {author} {\bibfnamefont {E.}~\bibnamefont {Martinez}},\ }\bibfield  {title} {\bibinfo {title} {A quinary {WTaCrVHf} nanocrystalline refractory high-entropy alloy withholding extreme irradiation environments},\ }\href {https://doi.org/10.1038/s41467-023-38000-y} {\bibfield  {journal} {\bibinfo  {journal} {Nature Communications}\ }\textbf {\bibinfo {volume} {14}},\ \bibinfo {pages} {2516} (\bibinfo {year} {2023})}\BibitemShut {NoStop}%
\bibitem [{\citenamefont {Praveen}\ and\ \citenamefont {Kim}(2018)}]{praveen_highentropy_2018}%
  \BibitemOpen
  \bibfield  {author} {\bibinfo {author} {\bibfnamefont {S.}~\bibnamefont {Praveen}}\ and\ \bibinfo {author} {\bibfnamefont {H.~S.}\ \bibnamefont {Kim}},\ }\bibfield  {title} {\bibinfo {title} {High‐{Entropy} {Alloys}: {Potential} {Candidates} for {High}‐{Temperature} {Applications} – {An} {Overview}},\ }\href {https://doi.org/10.1002/adem.201700645} {\bibfield  {journal} {\bibinfo  {journal} {Advanced Engineering Materials}\ }\textbf {\bibinfo {volume} {20}},\ \bibinfo {pages} {1700645} (\bibinfo {year} {2018})}\BibitemShut {NoStop}%
\bibitem [{\citenamefont {Chen}\ \emph {et~al.}(2018)\citenamefont {Chen}, \citenamefont {Zhou}, \citenamefont {Wang}, \citenamefont {Liu}, \citenamefont {Lv}, \citenamefont {Yang}, \citenamefont {Xu},\ and\ \citenamefont {Liu}}]{chen_review_2018}%
  \BibitemOpen
  \bibfield  {author} {\bibinfo {author} {\bibfnamefont {J.}~\bibnamefont {Chen}}, \bibinfo {author} {\bibfnamefont {X.}~\bibnamefont {Zhou}}, \bibinfo {author} {\bibfnamefont {W.}~\bibnamefont {Wang}}, \bibinfo {author} {\bibfnamefont {B.}~\bibnamefont {Liu}}, \bibinfo {author} {\bibfnamefont {Y.}~\bibnamefont {Lv}}, \bibinfo {author} {\bibfnamefont {W.}~\bibnamefont {Yang}}, \bibinfo {author} {\bibfnamefont {D.}~\bibnamefont {Xu}},\ and\ \bibinfo {author} {\bibfnamefont {Y.}~\bibnamefont {Liu}},\ }\bibfield  {title} {\bibinfo {title} {A review on fundamental of high entropy alloys with promising high–temperature properties},\ }\href {https://doi.org/10.1016/j.jallcom.2018.05.067} {\bibfield  {journal} {\bibinfo  {journal} {Journal of Alloys and Compounds}\ }\textbf {\bibinfo {volume} {760}},\ \bibinfo {pages} {15} (\bibinfo {year} {2018})}\BibitemShut {NoStop}%
\bibitem [{\citenamefont {Koželj}\ \emph {et~al.}(2014)\citenamefont {Koželj}, \citenamefont {Vrtnik}, \citenamefont {Jelen}, \citenamefont {Jazbec}, \citenamefont {Jagličić}, \citenamefont {Maiti}, \citenamefont {Feuerbacher}, \citenamefont {Steurer},\ and\ \citenamefont {Dolinšek}}]{kozelj_discovery_2014}%
  \BibitemOpen
  \bibfield  {author} {\bibinfo {author} {\bibfnamefont {P.}~\bibnamefont {Koželj}}, \bibinfo {author} {\bibfnamefont {S.}~\bibnamefont {Vrtnik}}, \bibinfo {author} {\bibfnamefont {A.}~\bibnamefont {Jelen}}, \bibinfo {author} {\bibfnamefont {S.}~\bibnamefont {Jazbec}}, \bibinfo {author} {\bibfnamefont {Z.}~\bibnamefont {Jagličić}}, \bibinfo {author} {\bibfnamefont {S.}~\bibnamefont {Maiti}}, \bibinfo {author} {\bibfnamefont {M.}~\bibnamefont {Feuerbacher}}, \bibinfo {author} {\bibfnamefont {W.}~\bibnamefont {Steurer}},\ and\ \bibinfo {author} {\bibfnamefont {J.}~\bibnamefont {Dolinšek}},\ }\bibfield  {title} {\bibinfo {title} {Discovery of a {Superconducting} {High}-{Entropy} {Alloy}},\ }\href {https://doi.org/10.1103/PhysRevLett.113.107001} {\bibfield  {journal} {\bibinfo  {journal} {Physical Review Letters}\ }\textbf {\bibinfo {volume} {113}},\ \bibinfo {pages} {107001} (\bibinfo {year} {2014})}\BibitemShut {NoStop}%
\bibitem [{\citenamefont {Billington}\ \emph {et~al.}(2020)\citenamefont {Billington}, \citenamefont {James}, \citenamefont {Harris-Lee}, \citenamefont {Lagos}, \citenamefont {O'Neill}, \citenamefont {Tsuda}, \citenamefont {Toyoki}, \citenamefont {Kotani}, \citenamefont {Nakamura}, \citenamefont {Bei}, \citenamefont {Mu}, \citenamefont {Samolyuk}, \citenamefont {Stocks}, \citenamefont {Duffy}, \citenamefont {Taylor}, \citenamefont {Giblin},\ and\ \citenamefont {Dugdale}}]{billington_bulk_2020}%
  \BibitemOpen
  \bibfield  {author} {\bibinfo {author} {\bibfnamefont {D.}~\bibnamefont {Billington}}, \bibinfo {author} {\bibfnamefont {A.~D.~N.}\ \bibnamefont {James}}, \bibinfo {author} {\bibfnamefont {E.~I.}\ \bibnamefont {Harris-Lee}}, \bibinfo {author} {\bibfnamefont {D.~A.}\ \bibnamefont {Lagos}}, \bibinfo {author} {\bibfnamefont {D.}~\bibnamefont {O'Neill}}, \bibinfo {author} {\bibfnamefont {N.}~\bibnamefont {Tsuda}}, \bibinfo {author} {\bibfnamefont {K.}~\bibnamefont {Toyoki}}, \bibinfo {author} {\bibfnamefont {Y.}~\bibnamefont {Kotani}}, \bibinfo {author} {\bibfnamefont {T.}~\bibnamefont {Nakamura}}, \bibinfo {author} {\bibfnamefont {H.}~\bibnamefont {Bei}}, \bibinfo {author} {\bibfnamefont {S.}~\bibnamefont {Mu}}, \bibinfo {author} {\bibfnamefont {G.~D.}\ \bibnamefont {Samolyuk}}, \bibinfo {author} {\bibfnamefont {G.~M.}\ \bibnamefont {Stocks}}, \bibinfo {author} {\bibfnamefont {J.~A.}\ \bibnamefont {Duffy}}, \bibinfo {author} {\bibfnamefont {J.~W.}\ \bibnamefont {Taylor}}, \bibinfo {author} {\bibfnamefont
  {S.~R.}\ \bibnamefont {Giblin}},\ and\ \bibinfo {author} {\bibfnamefont {S.~B.}\ \bibnamefont {Dugdale}},\ }\bibfield  {title} {\bibinfo {title} {Bulk and element-specific magnetism of medium-entropy and high-entropy {Cantor}-{Wu} alloys},\ }\href {https://doi.org/10.1103/PhysRevB.102.174405} {\bibfield  {journal} {\bibinfo  {journal} {Physical Review B}\ }\textbf {\bibinfo {volume} {102}},\ \bibinfo {pages} {174405} (\bibinfo {year} {2020})}\BibitemShut {NoStop}%
\bibitem [{\citenamefont {Sales}\ \emph {et~al.}(2016)\citenamefont {Sales}, \citenamefont {Jin}, \citenamefont {Bei}, \citenamefont {Stocks}, \citenamefont {Samolyuk}, \citenamefont {May},\ and\ \citenamefont {McGuire}}]{sales_quantum_2016}%
  \BibitemOpen
  \bibfield  {author} {\bibinfo {author} {\bibfnamefont {B.~C.}\ \bibnamefont {Sales}}, \bibinfo {author} {\bibfnamefont {K.}~\bibnamefont {Jin}}, \bibinfo {author} {\bibfnamefont {H.}~\bibnamefont {Bei}}, \bibinfo {author} {\bibfnamefont {G.~M.}\ \bibnamefont {Stocks}}, \bibinfo {author} {\bibfnamefont {G.~D.}\ \bibnamefont {Samolyuk}}, \bibinfo {author} {\bibfnamefont {A.~F.}\ \bibnamefont {May}},\ and\ \bibinfo {author} {\bibfnamefont {M.~A.}\ \bibnamefont {McGuire}},\ }\bibfield  {title} {\bibinfo {title} {Quantum {Critical} {Behavior} in a {Concentrated} {Ternary} {Solid} {Solution}},\ }\href {https://doi.org/10.1038/srep26179} {\bibfield  {journal} {\bibinfo  {journal} {Scientific Reports}\ }\textbf {\bibinfo {volume} {6}},\ \bibinfo {pages} {26179} (\bibinfo {year} {2016})}\BibitemShut {NoStop}%
\bibitem [{\citenamefont {Robarts}\ \emph {et~al.}(2020)\citenamefont {Robarts}, \citenamefont {Millichamp}, \citenamefont {Lagos}, \citenamefont {Laverock}, \citenamefont {Billington}, \citenamefont {Duffy}, \citenamefont {O’Neill}, \citenamefont {Giblin}, \citenamefont {Taylor}, \citenamefont {Kontrym-Sznajd}, \citenamefont {Samsel-Czekała}, \citenamefont {Bei}, \citenamefont {Mu}, \citenamefont {Samolyuk}, \citenamefont {Stocks},\ and\ \citenamefont {Dugdale}}]{robarts_extreme_2020}%
  \BibitemOpen
  \bibfield  {author} {\bibinfo {author} {\bibfnamefont {H.~C.}\ \bibnamefont {Robarts}}, \bibinfo {author} {\bibfnamefont {T.~E.}\ \bibnamefont {Millichamp}}, \bibinfo {author} {\bibfnamefont {D.~A.}\ \bibnamefont {Lagos}}, \bibinfo {author} {\bibfnamefont {J.}~\bibnamefont {Laverock}}, \bibinfo {author} {\bibfnamefont {D.}~\bibnamefont {Billington}}, \bibinfo {author} {\bibfnamefont {J.~A.}\ \bibnamefont {Duffy}}, \bibinfo {author} {\bibfnamefont {D.}~\bibnamefont {O’Neill}}, \bibinfo {author} {\bibfnamefont {S.~R.}\ \bibnamefont {Giblin}}, \bibinfo {author} {\bibfnamefont {J.~W.}\ \bibnamefont {Taylor}}, \bibinfo {author} {\bibfnamefont {G.}~\bibnamefont {Kontrym-Sznajd}}, \bibinfo {author} {\bibfnamefont {M.}~\bibnamefont {Samsel-Czekała}}, \bibinfo {author} {\bibfnamefont {H.}~\bibnamefont {Bei}}, \bibinfo {author} {\bibfnamefont {S.}~\bibnamefont {Mu}}, \bibinfo {author} {\bibfnamefont {G.~D.}\ \bibnamefont {Samolyuk}}, \bibinfo {author} {\bibfnamefont {G.~M.}\ \bibnamefont {Stocks}},\ and\ \bibinfo
  {author} {\bibfnamefont {S.~B.}\ \bibnamefont {Dugdale}},\ }\bibfield  {title} {\bibinfo {title} {Extreme {Fermi} {Surface} {Smearing} in a {Maximally} {Disordered} {Concentrated} {Solid} {Solution}},\ }\href {https://doi.org/10.1103/PhysRevLett.124.046402} {\bibfield  {journal} {\bibinfo  {journal} {Physical Review Letters}\ }\textbf {\bibinfo {volume} {124}},\ \bibinfo {pages} {046402} (\bibinfo {year} {2020})}\BibitemShut {NoStop}%
\bibitem [{\citenamefont {Qiu}\ \emph {et~al.}(2017{\natexlab{b}})\citenamefont {Qiu}, \citenamefont {Hu}, \citenamefont {Taylor}, \citenamefont {Styles}, \citenamefont {Marceau}, \citenamefont {Ceguerra}, \citenamefont {Gibson}, \citenamefont {Liu}, \citenamefont {Fraser},\ and\ \citenamefont {Birbilis}}]{qiu_lightweight_2017}%
  \BibitemOpen
  \bibfield  {author} {\bibinfo {author} {\bibfnamefont {Y.}~\bibnamefont {Qiu}}, \bibinfo {author} {\bibfnamefont {Y.}~\bibnamefont {Hu}}, \bibinfo {author} {\bibfnamefont {A.}~\bibnamefont {Taylor}}, \bibinfo {author} {\bibfnamefont {M.}~\bibnamefont {Styles}}, \bibinfo {author} {\bibfnamefont {R.}~\bibnamefont {Marceau}}, \bibinfo {author} {\bibfnamefont {A.}~\bibnamefont {Ceguerra}}, \bibinfo {author} {\bibfnamefont {M.}~\bibnamefont {Gibson}}, \bibinfo {author} {\bibfnamefont {Z.}~\bibnamefont {Liu}}, \bibinfo {author} {\bibfnamefont {H.}~\bibnamefont {Fraser}},\ and\ \bibinfo {author} {\bibfnamefont {N.}~\bibnamefont {Birbilis}},\ }\bibfield  {title} {\bibinfo {title} {A lightweight single-phase {AlTiVCr} compositionally complex alloy},\ }\href {https://doi.org/10.1016/j.actamat.2016.10.037} {\bibfield  {journal} {\bibinfo  {journal} {Acta Materialia}\ }\textbf {\bibinfo {volume} {123}},\ \bibinfo {pages} {115} (\bibinfo {year} {2017}{\natexlab{b}})}\BibitemShut {NoStop}%
\bibitem [{\citenamefont {Qiu}\ \emph {et~al.}(2018)\citenamefont {Qiu}, \citenamefont {Thomas}, \citenamefont {Gibson}, \citenamefont {Fraser}, \citenamefont {Pohl},\ and\ \citenamefont {Birbilis}}]{qiu_microstructure_2018}%
  \BibitemOpen
  \bibfield  {author} {\bibinfo {author} {\bibfnamefont {Y.}~\bibnamefont {Qiu}}, \bibinfo {author} {\bibfnamefont {S.}~\bibnamefont {Thomas}}, \bibinfo {author} {\bibfnamefont {M.}~\bibnamefont {Gibson}}, \bibinfo {author} {\bibfnamefont {H.}~\bibnamefont {Fraser}}, \bibinfo {author} {\bibfnamefont {K.}~\bibnamefont {Pohl}},\ and\ \bibinfo {author} {\bibfnamefont {N.}~\bibnamefont {Birbilis}},\ }\bibfield  {title} {\bibinfo {title} {Microstructure and corrosion properties of the low-density single-phase compositionally complex alloy {AlTiVCr}},\ }\href {https://doi.org/10.1016/j.corsci.2018.01.035} {\bibfield  {journal} {\bibinfo  {journal} {Corrosion Science}\ }\textbf {\bibinfo {volume} {133}},\ \bibinfo {pages} {386} (\bibinfo {year} {2018})}\BibitemShut {NoStop}%
\bibitem [{\citenamefont {Huang}\ \emph {et~al.}(2019)\citenamefont {Huang}, \citenamefont {Miao},\ and\ \citenamefont {Luo}}]{huang_lightweight_2019}%
  \BibitemOpen
  \bibfield  {author} {\bibinfo {author} {\bibfnamefont {X.}~\bibnamefont {Huang}}, \bibinfo {author} {\bibfnamefont {J.}~\bibnamefont {Miao}},\ and\ \bibinfo {author} {\bibfnamefont {A.~A.}\ \bibnamefont {Luo}},\ }\bibfield  {title} {\bibinfo {title} {Lightweight {AlCrTiV} high-entropy alloys with dual-phase microstructure via microalloying},\ }\href {https://doi.org/10.1007/s10853-018-2970-4} {\bibfield  {journal} {\bibinfo  {journal} {Journal of Materials Science}\ }\textbf {\bibinfo {volume} {54}},\ \bibinfo {pages} {2271} (\bibinfo {year} {2019})}\BibitemShut {NoStop}%
\bibitem [{\citenamefont {Huang}\ \emph {et~al.}(2022)\citenamefont {Huang}, \citenamefont {Miao},\ and\ \citenamefont {Luo}}]{huang_order-disorder_2022}%
  \BibitemOpen
  \bibfield  {author} {\bibinfo {author} {\bibfnamefont {X.}~\bibnamefont {Huang}}, \bibinfo {author} {\bibfnamefont {J.}~\bibnamefont {Miao}},\ and\ \bibinfo {author} {\bibfnamefont {A.~A.}\ \bibnamefont {Luo}},\ }\bibfield  {title} {\bibinfo {title} {Order-disorder transition and its mechanical effects in lightweight {AlCrTiV} high entropy alloys},\ }\href {https://doi.org/10.1016/j.scriptamat.2021.114462} {\bibfield  {journal} {\bibinfo  {journal} {Scripta Materialia}\ }\textbf {\bibinfo {volume} {210}},\ \bibinfo {pages} {114462} (\bibinfo {year} {2022})}\BibitemShut {NoStop}%
\bibitem [{\citenamefont {Senkov}\ \emph {et~al.}(2014{\natexlab{a}})\citenamefont {Senkov}, \citenamefont {Senkova},\ and\ \citenamefont {Woodward}}]{senkov_effect_2014}%
  \BibitemOpen
  \bibfield  {author} {\bibinfo {author} {\bibfnamefont {O.}~\bibnamefont {Senkov}}, \bibinfo {author} {\bibfnamefont {S.}~\bibnamefont {Senkova}},\ and\ \bibinfo {author} {\bibfnamefont {C.}~\bibnamefont {Woodward}},\ }\bibfield  {title} {\bibinfo {title} {Effect of aluminum on the microstructure and properties of two refractory high-entropy alloys},\ }\href {https://doi.org/10.1016/j.actamat.2014.01.029} {\bibfield  {journal} {\bibinfo  {journal} {Acta Materialia}\ }\textbf {\bibinfo {volume} {68}},\ \bibinfo {pages} {214} (\bibinfo {year} {2014}{\natexlab{a}})}\BibitemShut {NoStop}%
\bibitem [{\citenamefont {Senkov}\ \emph {et~al.}(2014{\natexlab{b}})\citenamefont {Senkov}, \citenamefont {Woodward},\ and\ \citenamefont {Miracle}}]{senkov_microstructure_2014}%
  \BibitemOpen
  \bibfield  {author} {\bibinfo {author} {\bibfnamefont {O.~N.}\ \bibnamefont {Senkov}}, \bibinfo {author} {\bibfnamefont {C.}~\bibnamefont {Woodward}},\ and\ \bibinfo {author} {\bibfnamefont {D.~B.}\ \bibnamefont {Miracle}},\ }\bibfield  {title} {\bibinfo {title} {Microstructure and {Properties} of {Aluminum}-{Containing} {Refractory} {High}-{Entropy} {Alloys}},\ }\href {https://doi.org/10.1007/s11837-014-1066-0} {\bibfield  {journal} {\bibinfo  {journal} {JOM}\ }\textbf {\bibinfo {volume} {66}},\ \bibinfo {pages} {2030} (\bibinfo {year} {2014}{\natexlab{b}})}\BibitemShut {NoStop}%
\bibitem [{\citenamefont {Miracle}\ \emph {et~al.}(2020)\citenamefont {Miracle}, \citenamefont {Tsai}, \citenamefont {Senkov}, \citenamefont {Soni},\ and\ \citenamefont {Banerjee}}]{miracle_refractory_2020}%
  \BibitemOpen
  \bibfield  {author} {\bibinfo {author} {\bibfnamefont {D.~B.}\ \bibnamefont {Miracle}}, \bibinfo {author} {\bibfnamefont {M.-H.}\ \bibnamefont {Tsai}}, \bibinfo {author} {\bibfnamefont {O.~N.}\ \bibnamefont {Senkov}}, \bibinfo {author} {\bibfnamefont {V.}~\bibnamefont {Soni}},\ and\ \bibinfo {author} {\bibfnamefont {R.}~\bibnamefont {Banerjee}},\ }\bibfield  {title} {\bibinfo {title} {Refractory high entropy superalloys ({RSAs})},\ }\href {https://doi.org/10.1016/j.scriptamat.2020.06.048} {\bibfield  {journal} {\bibinfo  {journal} {Scripta Materialia}\ }\textbf {\bibinfo {volume} {187}},\ \bibinfo {pages} {445} (\bibinfo {year} {2020})}\BibitemShut {NoStop}%
\bibitem [{\citenamefont {Widom}(2018)}]{widom_modeling_2018}%
  \BibitemOpen
  \bibfield  {author} {\bibinfo {author} {\bibfnamefont {M.}~\bibnamefont {Widom}},\ }\bibfield  {title} {\bibinfo {title} {Modeling the structure and thermodynamics of high-entropy alloys},\ }\href {https://doi.org/10.1557/jmr.2018.222} {\bibfield  {journal} {\bibinfo  {journal} {Journal of Materials Research}\ }\textbf {\bibinfo {volume} {33}},\ \bibinfo {pages} {2881} (\bibinfo {year} {2018})}\BibitemShut {NoStop}%
\bibitem [{\citenamefont {Eisenbach}\ \emph {et~al.}(2019)\citenamefont {Eisenbach}, \citenamefont {Pei},\ and\ \citenamefont {Liu}}]{eisenbach_first-principles_2019}%
  \BibitemOpen
  \bibfield  {author} {\bibinfo {author} {\bibfnamefont {M.}~\bibnamefont {Eisenbach}}, \bibinfo {author} {\bibfnamefont {Z.}~\bibnamefont {Pei}},\ and\ \bibinfo {author} {\bibfnamefont {X.}~\bibnamefont {Liu}},\ }\bibfield  {title} {\bibinfo {title} {First-principles study of order-disorder transitions in multicomponent solid-solution alloys},\ }\href {https://doi.org/10.1088/1361-648X/ab13d8} {\bibfield  {journal} {\bibinfo  {journal} {Journal of Physics: Condensed Matter}\ }\textbf {\bibinfo {volume} {31}},\ \bibinfo {pages} {273002} (\bibinfo {year} {2019})}\BibitemShut {NoStop}%
\bibitem [{\citenamefont {Ferrari}\ \emph {et~al.}(2020)\citenamefont {Ferrari}, \citenamefont {Dutta}, \citenamefont {Gubaev}, \citenamefont {Ikeda}, \citenamefont {Srinivasan}, \citenamefont {Grabowski},\ and\ \citenamefont {Körmann}}]{ferrari_frontiers_2020}%
  \BibitemOpen
  \bibfield  {author} {\bibinfo {author} {\bibfnamefont {A.}~\bibnamefont {Ferrari}}, \bibinfo {author} {\bibfnamefont {B.}~\bibnamefont {Dutta}}, \bibinfo {author} {\bibfnamefont {K.}~\bibnamefont {Gubaev}}, \bibinfo {author} {\bibfnamefont {Y.}~\bibnamefont {Ikeda}}, \bibinfo {author} {\bibfnamefont {P.}~\bibnamefont {Srinivasan}}, \bibinfo {author} {\bibfnamefont {B.}~\bibnamefont {Grabowski}},\ and\ \bibinfo {author} {\bibfnamefont {F.}~\bibnamefont {Körmann}},\ }\bibfield  {title} {\bibinfo {title} {Frontiers in atomistic simulations of high entropy alloys},\ }\href {https://doi.org/10.1063/5.0025310} {\bibfield  {journal} {\bibinfo  {journal} {Journal of Applied Physics}\ }\textbf {\bibinfo {volume} {128}},\ \bibinfo {pages} {150901} (\bibinfo {year} {2020})}\BibitemShut {NoStop}%
\bibitem [{\citenamefont {Ferrari}\ \emph {et~al.}(2023)\citenamefont {Ferrari}, \citenamefont {Körmann}, \citenamefont {Asta},\ and\ \citenamefont {Neugebauer}}]{ferrari_simulating_2023}%
  \BibitemOpen
  \bibfield  {author} {\bibinfo {author} {\bibfnamefont {A.}~\bibnamefont {Ferrari}}, \bibinfo {author} {\bibfnamefont {F.}~\bibnamefont {Körmann}}, \bibinfo {author} {\bibfnamefont {M.}~\bibnamefont {Asta}},\ and\ \bibinfo {author} {\bibfnamefont {J.}~\bibnamefont {Neugebauer}},\ }\bibfield  {title} {\bibinfo {title} {Simulating short-range order in compositionally complex materials},\ }\href {https://doi.org/10.1038/s43588-023-00407-4} {\bibfield  {journal} {\bibinfo  {journal} {Nature Computational Science}\ }\textbf {\bibinfo {volume} {3}},\ \bibinfo {pages} {221} (\bibinfo {year} {2023})}\BibitemShut {NoStop}%
\bibitem [{\citenamefont {Widom}\ \emph {et~al.}(2014)\citenamefont {Widom}, \citenamefont {Huhn}, \citenamefont {Maiti},\ and\ \citenamefont {Steurer}}]{widom_hybrid_2014}%
  \BibitemOpen
  \bibfield  {author} {\bibinfo {author} {\bibfnamefont {M.}~\bibnamefont {Widom}}, \bibinfo {author} {\bibfnamefont {W.~P.}\ \bibnamefont {Huhn}}, \bibinfo {author} {\bibfnamefont {S.}~\bibnamefont {Maiti}},\ and\ \bibinfo {author} {\bibfnamefont {W.}~\bibnamefont {Steurer}},\ }\bibfield  {title} {\bibinfo {title} {Hybrid {Monte} {Carlo}/{Molecular} {Dynamics} {Simulation} of a {Refractory} {Metal} {High} {Entropy} {Alloy}},\ }\href {https://doi.org/10.1007/s11661-013-2000-8} {\bibfield  {journal} {\bibinfo  {journal} {Metallurgical and Materials Transactions A}\ }\textbf {\bibinfo {volume} {45}},\ \bibinfo {pages} {196} (\bibinfo {year} {2014})}\BibitemShut {NoStop}%
\bibitem [{\citenamefont {Tamm}\ \emph {et~al.}(2015)\citenamefont {Tamm}, \citenamefont {Aabloo}, \citenamefont {Klintenberg}, \citenamefont {Stocks},\ and\ \citenamefont {Caro}}]{tamm_atomic-scale_2015}%
  \BibitemOpen
  \bibfield  {author} {\bibinfo {author} {\bibfnamefont {A.}~\bibnamefont {Tamm}}, \bibinfo {author} {\bibfnamefont {A.}~\bibnamefont {Aabloo}}, \bibinfo {author} {\bibfnamefont {M.}~\bibnamefont {Klintenberg}}, \bibinfo {author} {\bibfnamefont {M.}~\bibnamefont {Stocks}},\ and\ \bibinfo {author} {\bibfnamefont {A.}~\bibnamefont {Caro}},\ }\bibfield  {title} {\bibinfo {title} {Atomic-scale properties of {Ni}-based {FCC} ternary, and quaternary alloys},\ }\href {https://doi.org/10.1016/j.actamat.2015.08.015} {\bibfield  {journal} {\bibinfo  {journal} {Acta Materialia}\ }\textbf {\bibinfo {volume} {99}},\ \bibinfo {pages} {307} (\bibinfo {year} {2015})}\BibitemShut {NoStop}%
\bibitem [{\citenamefont {Widom}(2024)}]{widom_first-principles_2024}%
  \BibitemOpen
  \bibfield  {author} {\bibinfo {author} {\bibfnamefont {M.}~\bibnamefont {Widom}},\ }\bibfield  {title} {\bibinfo {title} {First-principles study of the order-disorder transition in the {AlCrTiV} high entropy alloy},\ }\href {https://doi.org/10.1103/PhysRevMaterials.8.093603} {\bibfield  {journal} {\bibinfo  {journal} {Physical Review Materials}\ }\textbf {\bibinfo {volume} {8}},\ \bibinfo {pages} {093603} (\bibinfo {year} {2024})}\BibitemShut {NoStop}%
\bibitem [{\citenamefont {Kostiuchenko}\ \emph {et~al.}(2019)\citenamefont {Kostiuchenko}, \citenamefont {Körmann}, \citenamefont {Neugebauer},\ and\ \citenamefont {Shapeev}}]{kostiuchenko_impact_2019}%
  \BibitemOpen
  \bibfield  {author} {\bibinfo {author} {\bibfnamefont {T.}~\bibnamefont {Kostiuchenko}}, \bibinfo {author} {\bibfnamefont {F.}~\bibnamefont {Körmann}}, \bibinfo {author} {\bibfnamefont {J.}~\bibnamefont {Neugebauer}},\ and\ \bibinfo {author} {\bibfnamefont {A.}~\bibnamefont {Shapeev}},\ }\bibfield  {title} {\bibinfo {title} {Impact of lattice relaxations on phase transitions in a high-entropy alloy studied by machine-learning potentials},\ }\href {https://doi.org/10.1038/s41524-019-0195-y} {\bibfield  {journal} {\bibinfo  {journal} {npj Computational Materials}\ }\textbf {\bibinfo {volume} {5}},\ \bibinfo {pages} {55} (\bibinfo {year} {2019})}\BibitemShut {NoStop}%
\bibitem [{\citenamefont {Rosenbrock}\ \emph {et~al.}(2021)\citenamefont {Rosenbrock}, \citenamefont {Gubaev}, \citenamefont {Shapeev}, \citenamefont {Pártay}, \citenamefont {Bernstein}, \citenamefont {Csányi},\ and\ \citenamefont {Hart}}]{rosenbrock_machine-learned_2021}%
  \BibitemOpen
  \bibfield  {author} {\bibinfo {author} {\bibfnamefont {C.~W.}\ \bibnamefont {Rosenbrock}}, \bibinfo {author} {\bibfnamefont {K.}~\bibnamefont {Gubaev}}, \bibinfo {author} {\bibfnamefont {A.~V.}\ \bibnamefont {Shapeev}}, \bibinfo {author} {\bibfnamefont {L.~B.}\ \bibnamefont {Pártay}}, \bibinfo {author} {\bibfnamefont {N.}~\bibnamefont {Bernstein}}, \bibinfo {author} {\bibfnamefont {G.}~\bibnamefont {Csányi}},\ and\ \bibinfo {author} {\bibfnamefont {G.~L.~W.}\ \bibnamefont {Hart}},\ }\bibfield  {title} {\bibinfo {title} {Machine-learned interatomic potentials for alloys and alloy phase diagrams},\ }\href {https://doi.org/10.1038/s41524-020-00477-2} {\bibfield  {journal} {\bibinfo  {journal} {npj Computational Materials}\ }\textbf {\bibinfo {volume} {7}},\ \bibinfo {pages} {24} (\bibinfo {year} {2021})}\BibitemShut {NoStop}%
\bibitem [{\citenamefont {Körmann}\ \emph {et~al.}(2021)\citenamefont {Körmann}, \citenamefont {Kostiuchenko}, \citenamefont {Shapeev},\ and\ \citenamefont {Neugebauer}}]{kormann_b2_2021}%
  \BibitemOpen
  \bibfield  {author} {\bibinfo {author} {\bibfnamefont {F.}~\bibnamefont {Körmann}}, \bibinfo {author} {\bibfnamefont {T.}~\bibnamefont {Kostiuchenko}}, \bibinfo {author} {\bibfnamefont {A.}~\bibnamefont {Shapeev}},\ and\ \bibinfo {author} {\bibfnamefont {J.}~\bibnamefont {Neugebauer}},\ }\bibfield  {title} {\bibinfo {title} {B2 ordering in body-centered-cubic {AlNbTiV} refractory high-entropy alloys},\ }\href {https://doi.org/10.1103/PhysRevMaterials.5.053803} {\bibfield  {journal} {\bibinfo  {journal} {Physical Review Materials}\ }\textbf {\bibinfo {volume} {5}},\ \bibinfo {pages} {053803} (\bibinfo {year} {2021})}\BibitemShut {NoStop}%
\bibitem [{\citenamefont {Zhang}\ \emph {et~al.}(2025)\citenamefont {Zhang}, \citenamefont {Sorkin}, \citenamefont {Aitken}, \citenamefont {Politano}, \citenamefont {Behler}, \citenamefont {P~Thompson}, \citenamefont {Ko}, \citenamefont {Ong}, \citenamefont {Chalykh}, \citenamefont {Korogod}, \citenamefont {Podryabinkin}, \citenamefont {Shapeev}, \citenamefont {Li}, \citenamefont {Mishin}, \citenamefont {Pei}, \citenamefont {Liu}, \citenamefont {Kim}, \citenamefont {Park}, \citenamefont {Hwang}, \citenamefont {Han}, \citenamefont {Sheriff}, \citenamefont {Cao},\ and\ \citenamefont {Freitas}}]{zhang_roadmap_2025}%
  \BibitemOpen
  \bibfield  {author} {\bibinfo {author} {\bibfnamefont {Y.-W.}\ \bibnamefont {Zhang}}, \bibinfo {author} {\bibfnamefont {V.}~\bibnamefont {Sorkin}}, \bibinfo {author} {\bibfnamefont {Z.~H.}\ \bibnamefont {Aitken}}, \bibinfo {author} {\bibfnamefont {A.}~\bibnamefont {Politano}}, \bibinfo {author} {\bibfnamefont {J.}~\bibnamefont {Behler}}, \bibinfo {author} {\bibfnamefont {A.}~\bibnamefont {P~Thompson}}, \bibinfo {author} {\bibfnamefont {T.~W.}\ \bibnamefont {Ko}}, \bibinfo {author} {\bibfnamefont {S.~P.}\ \bibnamefont {Ong}}, \bibinfo {author} {\bibfnamefont {O.}~\bibnamefont {Chalykh}}, \bibinfo {author} {\bibfnamefont {D.}~\bibnamefont {Korogod}}, \bibinfo {author} {\bibfnamefont {E.}~\bibnamefont {Podryabinkin}}, \bibinfo {author} {\bibfnamefont {A.}~\bibnamefont {Shapeev}}, \bibinfo {author} {\bibfnamefont {J.}~\bibnamefont {Li}}, \bibinfo {author} {\bibfnamefont {Y.}~\bibnamefont {Mishin}}, \bibinfo {author} {\bibfnamefont {Z.}~\bibnamefont {Pei}}, \bibinfo {author} {\bibfnamefont {X.}~\bibnamefont
  {Liu}}, \bibinfo {author} {\bibfnamefont {J.}~\bibnamefont {Kim}}, \bibinfo {author} {\bibfnamefont {Y.}~\bibnamefont {Park}}, \bibinfo {author} {\bibfnamefont {S.}~\bibnamefont {Hwang}}, \bibinfo {author} {\bibfnamefont {S.}~\bibnamefont {Han}}, \bibinfo {author} {\bibfnamefont {K.}~\bibnamefont {Sheriff}}, \bibinfo {author} {\bibfnamefont {Y.}~\bibnamefont {Cao}},\ and\ \bibinfo {author} {\bibfnamefont {R.}~\bibnamefont {Freitas}},\ }\bibfield  {title} {\bibinfo {title} {Roadmap for the development of machine learning-based interatomic potentials},\ }\href {https://doi.org/10.1088/1361-651X/ad9d63} {\bibfield  {journal} {\bibinfo  {journal} {Modelling and Simulation in Materials Science and Engineering}\ }\textbf {\bibinfo {volume} {33}},\ \bibinfo {pages} {023301} (\bibinfo {year} {2025})}\BibitemShut {NoStop}%
\bibitem [{\citenamefont {Cao}\ \emph {et~al.}(2025)\citenamefont {Cao}, \citenamefont {Sheriff},\ and\ \citenamefont {Freitas}}]{cao_capturing_2025}%
  \BibitemOpen
  \bibfield  {author} {\bibinfo {author} {\bibfnamefont {Y.}~\bibnamefont {Cao}}, \bibinfo {author} {\bibfnamefont {K.}~\bibnamefont {Sheriff}},\ and\ \bibinfo {author} {\bibfnamefont {R.}~\bibnamefont {Freitas}},\ }\bibfield  {title} {\bibinfo {title} {Capturing short-range order in high-entropy alloys with machine learning potentials},\ }\href {https://doi.org/10.1038/s41524-025-01722-2} {\bibfield  {journal} {\bibinfo  {journal} {npj Computational Materials}\ }\textbf {\bibinfo {volume} {11}},\ \bibinfo {pages} {268} (\bibinfo {year} {2025})}\BibitemShut {NoStop}%
\bibitem [{\citenamefont {Fernández-Caballero}\ \emph {et~al.}(2017)\citenamefont {Fernández-Caballero}, \citenamefont {Wróbel}, \citenamefont {Mummery},\ and\ \citenamefont {Nguyen-Manh}}]{fernandez-caballero_short-range_2017}%
  \BibitemOpen
  \bibfield  {author} {\bibinfo {author} {\bibfnamefont {A.}~\bibnamefont {Fernández-Caballero}}, \bibinfo {author} {\bibfnamefont {J.~S.}\ \bibnamefont {Wróbel}}, \bibinfo {author} {\bibfnamefont {P.~M.}\ \bibnamefont {Mummery}},\ and\ \bibinfo {author} {\bibfnamefont {D.}~\bibnamefont {Nguyen-Manh}},\ }\bibfield  {title} {\bibinfo {title} {Short-{Range} {Order} in {High} {Entropy} {Alloys}: {Theoretical} {Formulation} and {Application} to {Mo}-{Nb}-{Ta}-{V}-{W} {System}},\ }\href {https://doi.org/10.1007/s11669-017-0582-3} {\bibfield  {journal} {\bibinfo  {journal} {Journal of Phase Equilibria and Diffusion}\ }\textbf {\bibinfo {volume} {38}},\ \bibinfo {pages} {391} (\bibinfo {year} {2017})}\BibitemShut {NoStop}%
\bibitem [{\citenamefont {Sobieraj}\ \emph {et~al.}(2020)\citenamefont {Sobieraj}, \citenamefont {Wróbel}, \citenamefont {Rygier}, \citenamefont {Kurzydłowski}, \citenamefont {El~Atwani}, \citenamefont {Devaraj}, \citenamefont {Martinez~Saez},\ and\ \citenamefont {Nguyen-Manh}}]{sobieraj_chemical_2020}%
  \BibitemOpen
  \bibfield  {author} {\bibinfo {author} {\bibfnamefont {D.}~\bibnamefont {Sobieraj}}, \bibinfo {author} {\bibfnamefont {J.~S.}\ \bibnamefont {Wróbel}}, \bibinfo {author} {\bibfnamefont {T.}~\bibnamefont {Rygier}}, \bibinfo {author} {\bibfnamefont {K.~J.}\ \bibnamefont {Kurzydłowski}}, \bibinfo {author} {\bibfnamefont {O.}~\bibnamefont {El~Atwani}}, \bibinfo {author} {\bibfnamefont {A.}~\bibnamefont {Devaraj}}, \bibinfo {author} {\bibfnamefont {E.}~\bibnamefont {Martinez~Saez}},\ and\ \bibinfo {author} {\bibfnamefont {D.}~\bibnamefont {Nguyen-Manh}},\ }\bibfield  {title} {\bibinfo {title} {Chemical short-range order in derivative {Cr}–{Ta}–{Ti}–{V}–{W} high entropy alloys from the first-principles thermodynamic study},\ }\href {https://doi.org/10.1039/D0CP03764H} {\bibfield  {journal} {\bibinfo  {journal} {Physical Chemistry Chemical Physics}\ }\textbf {\bibinfo {volume} {22}},\ \bibinfo {pages} {23929} (\bibinfo {year} {2020})}\BibitemShut {NoStop}%
\bibitem [{\citenamefont {Kim}\ and\ \citenamefont {Widom}(2023)}]{kim_interaction_2023}%
  \BibitemOpen
  \bibfield  {author} {\bibinfo {author} {\bibfnamefont {A.~D.}\ \bibnamefont {Kim}}\ and\ \bibinfo {author} {\bibfnamefont {M.}~\bibnamefont {Widom}},\ }\bibfield  {title} {\bibinfo {title} {Interaction models and configurational entropies of binary {MoTa} and the {MoNbTaW} high entropy alloy},\ }\href {https://doi.org/10.1103/PhysRevMaterials.7.063803} {\bibfield  {journal} {\bibinfo  {journal} {Physical Review Materials}\ }\textbf {\bibinfo {volume} {7}},\ \bibinfo {pages} {063803} (\bibinfo {year} {2023})}\BibitemShut {NoStop}%
\bibitem [{\citenamefont {Vazquez}\ \emph {et~al.}(2024)\citenamefont {Vazquez}, \citenamefont {Sauceda},\ and\ \citenamefont {Arróyave}}]{vazquez_deciphering_2024}%
  \BibitemOpen
  \bibfield  {author} {\bibinfo {author} {\bibfnamefont {G.}~\bibnamefont {Vazquez}}, \bibinfo {author} {\bibfnamefont {D.}~\bibnamefont {Sauceda}},\ and\ \bibinfo {author} {\bibfnamefont {R.}~\bibnamefont {Arróyave}},\ }\bibfield  {title} {\bibinfo {title} {Deciphering chemical ordering in {High} {Entropy} {Materials}: {A} machine learning-accelerated high-throughput cluster expansion approach},\ }\href {https://doi.org/10.1016/j.actamat.2024.120137} {\bibfield  {journal} {\bibinfo  {journal} {Acta Materialia}\ }\textbf {\bibinfo {volume} {276}},\ \bibinfo {pages} {120137} (\bibinfo {year} {2024})}\BibitemShut {NoStop}%
\bibitem [{\citenamefont {Zhang}\ and\ \citenamefont {Yang}(2022)}]{zhang_calphad_2022}%
  \BibitemOpen
  \bibfield  {author} {\bibinfo {author} {\bibfnamefont {C.}~\bibnamefont {Zhang}}\ and\ \bibinfo {author} {\bibfnamefont {Y.}~\bibnamefont {Yang}},\ }\bibfield  {title} {\bibinfo {title} {The {CALPHAD} approach for {HEAs}: {Challenges} and opportunities},\ }\href {https://doi.org/10.1557/s43577-022-00284-8} {\bibfield  {journal} {\bibinfo  {journal} {MRS Bulletin}\ }\textbf {\bibinfo {volume} {47}},\ \bibinfo {pages} {158} (\bibinfo {year} {2022})}\BibitemShut {NoStop}%
\bibitem [{\citenamefont {Li}\ \emph {et~al.}(2023)\citenamefont {Li}, \citenamefont {Wang}, \citenamefont {Fan}, \citenamefont {Lu}, \citenamefont {Wang}, \citenamefont {Li},\ and\ \citenamefont {Liaw}}]{li_calphad-aided_2023}%
  \BibitemOpen
  \bibfield  {author} {\bibinfo {author} {\bibfnamefont {T.}~\bibnamefont {Li}}, \bibinfo {author} {\bibfnamefont {S.}~\bibnamefont {Wang}}, \bibinfo {author} {\bibfnamefont {W.}~\bibnamefont {Fan}}, \bibinfo {author} {\bibfnamefont {Y.}~\bibnamefont {Lu}}, \bibinfo {author} {\bibfnamefont {T.}~\bibnamefont {Wang}}, \bibinfo {author} {\bibfnamefont {T.}~\bibnamefont {Li}},\ and\ \bibinfo {author} {\bibfnamefont {P.~K.}\ \bibnamefont {Liaw}},\ }\bibfield  {title} {\bibinfo {title} {{CALPHAD}-aided design for superior thermal stability and mechanical behavior in a {TiZrHfNb} refractory high-entropy alloy},\ }\href {https://doi.org/10.1016/j.actamat.2023.118728} {\bibfield  {journal} {\bibinfo  {journal} {Acta Materialia}\ }\textbf {\bibinfo {volume} {246}},\ \bibinfo {pages} {118728} (\bibinfo {year} {2023})}\BibitemShut {NoStop}%
\bibitem [{\citenamefont {Singh}\ \emph {et~al.}(2015)\citenamefont {Singh}, \citenamefont {Smirnov},\ and\ \citenamefont {Johnson}}]{singh_atomic_2015}%
  \BibitemOpen
  \bibfield  {author} {\bibinfo {author} {\bibfnamefont {P.}~\bibnamefont {Singh}}, \bibinfo {author} {\bibfnamefont {A.~V.}\ \bibnamefont {Smirnov}},\ and\ \bibinfo {author} {\bibfnamefont {D.~D.}\ \bibnamefont {Johnson}},\ }\bibfield  {title} {\bibinfo {title} {Atomic short-range order and incipient long-range order in high-entropy alloys},\ }\href {https://doi.org/10.1103/PhysRevB.91.224204} {\bibfield  {journal} {\bibinfo  {journal} {Physical Review B}\ }\textbf {\bibinfo {volume} {91}},\ \bibinfo {pages} {224204} (\bibinfo {year} {2015})}\BibitemShut {NoStop}%
\bibitem [{\citenamefont {Körmann}\ \emph {et~al.}(2017)\citenamefont {Körmann}, \citenamefont {Ruban},\ and\ \citenamefont {Sluiter}}]{kormann_long-ranged_2017}%
  \BibitemOpen
  \bibfield  {author} {\bibinfo {author} {\bibfnamefont {F.}~\bibnamefont {Körmann}}, \bibinfo {author} {\bibfnamefont {A.~V.}\ \bibnamefont {Ruban}},\ and\ \bibinfo {author} {\bibfnamefont {M.~H.}\ \bibnamefont {Sluiter}},\ }\bibfield  {title} {\bibinfo {title} {Long-ranged interactions in bcc {NbMoTaW} high-entropy alloys},\ }\href {https://doi.org/10.1080/21663831.2016.1198837} {\bibfield  {journal} {\bibinfo  {journal} {Materials Research Letters}\ }\textbf {\bibinfo {volume} {5}},\ \bibinfo {pages} {35} (\bibinfo {year} {2017})}\BibitemShut {NoStop}%
\bibitem [{\citenamefont {Singh}\ \emph {et~al.}(2018)\citenamefont {Singh}, \citenamefont {Smirnov},\ and\ \citenamefont {Johnson}}]{singh_ta-nb-mo-w_2018}%
  \BibitemOpen
  \bibfield  {author} {\bibinfo {author} {\bibfnamefont {P.}~\bibnamefont {Singh}}, \bibinfo {author} {\bibfnamefont {A.~V.}\ \bibnamefont {Smirnov}},\ and\ \bibinfo {author} {\bibfnamefont {D.~D.}\ \bibnamefont {Johnson}},\ }\bibfield  {title} {\bibinfo {title} {Ta-{Nb}-{Mo}-{W} refractory high-entropy alloys: {Anomalous} ordering behavior and its intriguing electronic origin},\ }\href {https://doi.org/10.1103/PhysRevMaterials.2.055004} {\bibfield  {journal} {\bibinfo  {journal} {Physical Review Materials}\ }\textbf {\bibinfo {volume} {2}},\ \bibinfo {pages} {055004} (\bibinfo {year} {2018})}\BibitemShut {NoStop}%
\bibitem [{\citenamefont {Momma}\ and\ \citenamefont {Izumi}(2011)}]{momma_vesta_2011}%
  \BibitemOpen
  \bibfield  {author} {\bibinfo {author} {\bibfnamefont {K.}~\bibnamefont {Momma}}\ and\ \bibinfo {author} {\bibfnamefont {F.}~\bibnamefont {Izumi}},\ }\bibfield  {title} {\bibinfo {title} {\textit{{VESTA} 3} for three-dimensional visualization of crystal, volumetric and morphology data},\ }\href {https://doi.org/10.1107/S0021889811038970} {\bibfield  {journal} {\bibinfo  {journal} {Journal of Applied Crystallography}\ }\textbf {\bibinfo {volume} {44}},\ \bibinfo {pages} {1272} (\bibinfo {year} {2011})}\BibitemShut {NoStop}%
\bibitem [{\citenamefont {Khan}\ \emph {et~al.}(2016)\citenamefont {Khan}, \citenamefont {Staunton},\ and\ \citenamefont {Stocks}}]{khan_statistical_2016}%
  \BibitemOpen
  \bibfield  {author} {\bibinfo {author} {\bibfnamefont {S.~N.}\ \bibnamefont {Khan}}, \bibinfo {author} {\bibfnamefont {J.~B.}\ \bibnamefont {Staunton}},\ and\ \bibinfo {author} {\bibfnamefont {G.~M.}\ \bibnamefont {Stocks}},\ }\bibfield  {title} {\bibinfo {title} {Statistical physics of multicomponent alloys using {KKR}-{CPA}},\ }\href {https://doi.org/10.1103/PhysRevB.93.054206} {\bibfield  {journal} {\bibinfo  {journal} {Physical Review B}\ }\textbf {\bibinfo {volume} {93}},\ \bibinfo {pages} {054206} (\bibinfo {year} {2016})}\BibitemShut {NoStop}%
\bibitem [{\citenamefont {Woodgate}\ and\ \citenamefont {Staunton}(2022)}]{woodgate_compositional_2022}%
  \BibitemOpen
  \bibfield  {author} {\bibinfo {author} {\bibfnamefont {C.~D.}\ \bibnamefont {Woodgate}}\ and\ \bibinfo {author} {\bibfnamefont {J.~B.}\ \bibnamefont {Staunton}},\ }\bibfield  {title} {\bibinfo {title} {Compositional phase stability in medium-entropy and high-entropy {Cantor}-{Wu} alloys from an \textit{ab initio} all-electron {Landau}-type theory and atomistic modeling},\ }\href {https://doi.org/10.1103/PhysRevB.105.115124} {\bibfield  {journal} {\bibinfo  {journal} {Physical Review B}\ }\textbf {\bibinfo {volume} {105}},\ \bibinfo {pages} {115124} (\bibinfo {year} {2022})}\BibitemShut {NoStop}%
\bibitem [{\citenamefont {Woodgate}(2024)}]{woodgate_modelling_2024}%
  \BibitemOpen
  \bibfield  {author} {\bibinfo {author} {\bibfnamefont {C.~D.}\ \bibnamefont {Woodgate}},\ }\href {https://doi.org/10.1007/978-3-031-62021-8} {\emph {\bibinfo {title} {Modelling {Atomic} {Arrangements} in {Multicomponent} {Alloys}: {A} {Perturbative}, {First}-{Principles}-{Based} {Approach}}}},\ \bibinfo {series} {Springer {Series} in {Materials} {Science}}, Vol.\ \bibinfo {volume} {346}\ (\bibinfo  {publisher} {Springer Nature Switzerland},\ \bibinfo {address} {Cham},\ \bibinfo {year} {2024})\BibitemShut {NoStop}%
\bibitem [{\citenamefont {Stephen}\ \emph {et~al.}(2016)\citenamefont {Stephen}, \citenamefont {McDonald}, \citenamefont {Lejeune}, \citenamefont {Lewis},\ and\ \citenamefont {Heiman}}]{stephen_synthesis_2016}%
  \BibitemOpen
  \bibfield  {author} {\bibinfo {author} {\bibfnamefont {G.~M.}\ \bibnamefont {Stephen}}, \bibinfo {author} {\bibfnamefont {I.}~\bibnamefont {McDonald}}, \bibinfo {author} {\bibfnamefont {B.}~\bibnamefont {Lejeune}}, \bibinfo {author} {\bibfnamefont {L.~H.}\ \bibnamefont {Lewis}},\ and\ \bibinfo {author} {\bibfnamefont {D.}~\bibnamefont {Heiman}},\ }\bibfield  {title} {\bibinfo {title} {Synthesis of low-moment {CrVTiAl}: {A} potential room temperature spin filter},\ }\href {https://doi.org/10.1063/1.4971826} {\bibfield  {journal} {\bibinfo  {journal} {Applied Physics Letters}\ }\textbf {\bibinfo {volume} {109}},\ \bibinfo {pages} {242401} (\bibinfo {year} {2016})}\BibitemShut {NoStop}%
\bibitem [{\citenamefont {Venkateswara}\ \emph {et~al.}(2018)\citenamefont {Venkateswara}, \citenamefont {Gupta}, \citenamefont {Samatham}, \citenamefont {Varma}, \citenamefont {{Enamullah}}, \citenamefont {Suresh},\ and\ \citenamefont {Alam}}]{venkateswara_competing_2018}%
  \BibitemOpen
  \bibfield  {author} {\bibinfo {author} {\bibfnamefont {Y.}~\bibnamefont {Venkateswara}}, \bibinfo {author} {\bibfnamefont {S.}~\bibnamefont {Gupta}}, \bibinfo {author} {\bibfnamefont {S.~S.}\ \bibnamefont {Samatham}}, \bibinfo {author} {\bibfnamefont {M.~R.}\ \bibnamefont {Varma}}, \bibinfo {author} {\bibnamefont {{Enamullah}}}, \bibinfo {author} {\bibfnamefont {K.~G.}\ \bibnamefont {Suresh}},\ and\ \bibinfo {author} {\bibfnamefont {A.}~\bibnamefont {Alam}},\ }\bibfield  {title} {\bibinfo {title} {Competing magnetic and spin-gapless semiconducting behavior in fully compensated ferrimagnetic {CrVTiAl}: {Theory} and experiment},\ }\href {https://doi.org/10.1103/PhysRevB.97.054407} {\bibfield  {journal} {\bibinfo  {journal} {Physical Review B}\ }\textbf {\bibinfo {volume} {97}},\ \bibinfo {pages} {054407} (\bibinfo {year} {2018})}\BibitemShut {NoStop}%
\bibitem [{\citenamefont {Stephen}\ \emph {et~al.}(2019{\natexlab{a}})\citenamefont {Stephen}, \citenamefont {Lane}, \citenamefont {Buda}, \citenamefont {Graf}, \citenamefont {Kaprzyk}, \citenamefont {Barbiellini}, \citenamefont {Bansil},\ and\ \citenamefont {Heiman}}]{stephen_electrical_2019}%
  \BibitemOpen
  \bibfield  {author} {\bibinfo {author} {\bibfnamefont {G.~M.}\ \bibnamefont {Stephen}}, \bibinfo {author} {\bibfnamefont {C.}~\bibnamefont {Lane}}, \bibinfo {author} {\bibfnamefont {G.}~\bibnamefont {Buda}}, \bibinfo {author} {\bibfnamefont {D.}~\bibnamefont {Graf}}, \bibinfo {author} {\bibfnamefont {S.}~\bibnamefont {Kaprzyk}}, \bibinfo {author} {\bibfnamefont {B.}~\bibnamefont {Barbiellini}}, \bibinfo {author} {\bibfnamefont {A.}~\bibnamefont {Bansil}},\ and\ \bibinfo {author} {\bibfnamefont {D.}~\bibnamefont {Heiman}},\ }\bibfield  {title} {\bibinfo {title} {Electrical and magnetic properties of thin films of the spin-filter material {CrVTiAl}},\ }\href {https://doi.org/10.1103/PhysRevB.99.224207} {\bibfield  {journal} {\bibinfo  {journal} {Physical Review B}\ }\textbf {\bibinfo {volume} {99}},\ \bibinfo {pages} {224207} (\bibinfo {year} {2019}{\natexlab{a}})}\BibitemShut {NoStop}%
\bibitem [{\citenamefont {Stephen}\ \emph {et~al.}(2019{\natexlab{b}})\citenamefont {Stephen}, \citenamefont {Buda}, \citenamefont {Jamer}, \citenamefont {Lane}, \citenamefont {Kaprzyk}, \citenamefont {Barbiellini}, \citenamefont {Graf}, \citenamefont {Lewis}, \citenamefont {Bansil},\ and\ \citenamefont {Heiman}}]{stephen_structural_2019}%
  \BibitemOpen
  \bibfield  {author} {\bibinfo {author} {\bibfnamefont {G.~M.}\ \bibnamefont {Stephen}}, \bibinfo {author} {\bibfnamefont {G.}~\bibnamefont {Buda}}, \bibinfo {author} {\bibfnamefont {M.~E.}\ \bibnamefont {Jamer}}, \bibinfo {author} {\bibfnamefont {C.}~\bibnamefont {Lane}}, \bibinfo {author} {\bibfnamefont {S.}~\bibnamefont {Kaprzyk}}, \bibinfo {author} {\bibfnamefont {B.}~\bibnamefont {Barbiellini}}, \bibinfo {author} {\bibfnamefont {D.}~\bibnamefont {Graf}}, \bibinfo {author} {\bibfnamefont {L.~H.}\ \bibnamefont {Lewis}}, \bibinfo {author} {\bibfnamefont {A.}~\bibnamefont {Bansil}},\ and\ \bibinfo {author} {\bibfnamefont {D.}~\bibnamefont {Heiman}},\ }\bibfield  {title} {\bibinfo {title} {Structural and electronic properties of the spin-filter material {CrVTiAl} with disorder},\ }\href {https://doi.org/10.1063/1.5079749} {\bibfield  {journal} {\bibinfo  {journal} {Journal of Applied Physics}\ }\textbf {\bibinfo {volume} {125}},\ \bibinfo {pages} {123903} (\bibinfo {year} {2019}{\natexlab{b}})}\BibitemShut
  {NoStop}%
\bibitem [{\citenamefont {Staunton}\ \emph {et~al.}(1987)\citenamefont {Staunton}, \citenamefont {Johnson},\ and\ \citenamefont {Gy\H{o}rffy}}]{staunton_interaction_1987}%
  \BibitemOpen
  \bibfield  {author} {\bibinfo {author} {\bibfnamefont {J.~B.}\ \bibnamefont {Staunton}}, \bibinfo {author} {\bibfnamefont {D.~D.}\ \bibnamefont {Johnson}},\ and\ \bibinfo {author} {\bibfnamefont {B.~L.}\ \bibnamefont {Gy\H{o}rffy}},\ }\bibfield  {title} {\bibinfo {title} {Interaction between magnetic and compositional order in {Ni}-rich {Ni}$_{c}${Fe}$_{1-c}$ alloys (invited)},\ }\href {https://doi.org/10.1063/1.338664} {\bibfield  {journal} {\bibinfo  {journal} {Journal of Applied Physics}\ }\textbf {\bibinfo {volume} {61}},\ \bibinfo {pages} {3693} (\bibinfo {year} {1987})}\BibitemShut {NoStop}%
\bibitem [{\citenamefont {Woodgate}\ \emph {et~al.}(2023)\citenamefont {Woodgate}, \citenamefont {Hedlund}, \citenamefont {Lewis},\ and\ \citenamefont {Staunton}}]{woodgate_interplay_2023}%
  \BibitemOpen
  \bibfield  {author} {\bibinfo {author} {\bibfnamefont {C.~D.}\ \bibnamefont {Woodgate}}, \bibinfo {author} {\bibfnamefont {D.}~\bibnamefont {Hedlund}}, \bibinfo {author} {\bibfnamefont {L.~H.}\ \bibnamefont {Lewis}},\ and\ \bibinfo {author} {\bibfnamefont {J.~B.}\ \bibnamefont {Staunton}},\ }\bibfield  {title} {\bibinfo {title} {Interplay between magnetism and short-range order in medium- and high-entropy alloys: {CrCoNi}, {CrFeCoNi}, and {CrMnFeCoNi}},\ }\href {https://doi.org/10.1103/PhysRevMaterials.7.053801} {\bibfield  {journal} {\bibinfo  {journal} {Physical Review Materials}\ }\textbf {\bibinfo {volume} {7}},\ \bibinfo {pages} {053801} (\bibinfo {year} {2023})}\BibitemShut {NoStop}%
\bibitem [{\citenamefont {Ghosh}\ \emph {et~al.}(2024)\citenamefont {Ghosh}, \citenamefont {Ueltzen}, \citenamefont {George}, \citenamefont {Neugebauer},\ and\ \citenamefont {Körmann}}]{ghosh_chemical_2024}%
  \BibitemOpen
  \bibfield  {author} {\bibinfo {author} {\bibfnamefont {S.}~\bibnamefont {Ghosh}}, \bibinfo {author} {\bibfnamefont {K.}~\bibnamefont {Ueltzen}}, \bibinfo {author} {\bibfnamefont {J.}~\bibnamefont {George}}, \bibinfo {author} {\bibfnamefont {J.}~\bibnamefont {Neugebauer}},\ and\ \bibinfo {author} {\bibfnamefont {F.}~\bibnamefont {Körmann}},\ }\bibfield  {title} {\bibinfo {title} {Chemical ordering and magnetism in face-centered cubic {CrCoNi} alloy},\ }\href {https://doi.org/10.1038/s41524-024-01439-8} {\bibfield  {journal} {\bibinfo  {journal} {npj Computational Materials}\ }\textbf {\bibinfo {volume} {10}},\ \bibinfo {pages} {284} (\bibinfo {year} {2024})}\BibitemShut {NoStop}%
\bibitem [{\citenamefont {Ohnuma}\ \emph {et~al.}(2000)\citenamefont {Ohnuma}, \citenamefont {Fujita}, \citenamefont {Mitsui}, \citenamefont {Ishikawa}, \citenamefont {Kainuma},\ and\ \citenamefont {Ishida}}]{ohnuma_phase_2000}%
  \BibitemOpen
  \bibfield  {author} {\bibinfo {author} {\bibfnamefont {I.}~\bibnamefont {Ohnuma}}, \bibinfo {author} {\bibfnamefont {Y.}~\bibnamefont {Fujita}}, \bibinfo {author} {\bibfnamefont {H.}~\bibnamefont {Mitsui}}, \bibinfo {author} {\bibfnamefont {K.}~\bibnamefont {Ishikawa}}, \bibinfo {author} {\bibfnamefont {R.}~\bibnamefont {Kainuma}},\ and\ \bibinfo {author} {\bibfnamefont {K.}~\bibnamefont {Ishida}},\ }\bibfield  {title} {\bibinfo {title} {Phase equilibria in the {Ti}–{Al} binary system},\ }\href {https://doi.org/10.1016/S1359-6454(00)00118-X} {\bibfield  {journal} {\bibinfo  {journal} {Acta Materialia}\ }\textbf {\bibinfo {volume} {48}},\ \bibinfo {pages} {3113} (\bibinfo {year} {2000})}\BibitemShut {NoStop}%
\bibitem [{\citenamefont {Khachaturyan}(1978)}]{khachaturyan_ordering_1978}%
  \BibitemOpen
  \bibfield  {author} {\bibinfo {author} {\bibfnamefont {A.~G.}\ \bibnamefont {Khachaturyan}},\ }\bibfield  {title} {\bibinfo {title} {Ordering in substitutional and interstitial solid solutions},\ }\href {https://doi.org/10.1016/0079-6425(78)90003-8} {\bibfield  {journal} {\bibinfo  {journal} {Progress in Materials Science}\ }\textbf {\bibinfo {volume} {22}},\ \bibinfo {pages} {1} (\bibinfo {year} {1978})}\BibitemShut {NoStop}%
\bibitem [{\citenamefont {Woodgate}\ and\ \citenamefont {Staunton}(2023)}]{woodgate_short-range_2023}%
  \BibitemOpen
  \bibfield  {author} {\bibinfo {author} {\bibfnamefont {C.~D.}\ \bibnamefont {Woodgate}}\ and\ \bibinfo {author} {\bibfnamefont {J.~B.}\ \bibnamefont {Staunton}},\ }\bibfield  {title} {\bibinfo {title} {Short-range order and compositional phase stability in refractory high-entropy alloys via first-principles theory and atomistic modeling: {NbMoTa}, {NbMoTaW}, and {VNbMoTaW}},\ }\href {https://doi.org/10.1103/PhysRevMaterials.7.013801} {\bibfield  {journal} {\bibinfo  {journal} {Physical Review Materials}\ }\textbf {\bibinfo {volume} {7}},\ \bibinfo {pages} {013801} (\bibinfo {year} {2023})}\BibitemShut {NoStop}%
\bibitem [{\citenamefont {Stukowski}(2010)}]{stukowski_visualization_2010}%
  \BibitemOpen
  \bibfield  {author} {\bibinfo {author} {\bibfnamefont {A.}~\bibnamefont {Stukowski}},\ }\bibfield  {title} {\bibinfo {title} {Visualization and analysis of atomistic simulation data with {OVITO}–the {Open} {Visualization} {Tool}},\ }\href {https://doi.org/10.1088/0965-0393/18/1/015012} {\bibfield  {journal} {\bibinfo  {journal} {Modelling and Simulation in Materials Science and Engineering}\ }\textbf {\bibinfo {volume} {18}},\ \bibinfo {pages} {015012} (\bibinfo {year} {2010})}\BibitemShut {NoStop}%
\bibitem [{\citenamefont {Kalliney}\ and\ \citenamefont {Widom}(2026)}]{kalliney_x-ray_2026}%
  \BibitemOpen
  \bibfield  {author} {\bibinfo {author} {\bibfnamefont {N.}~\bibnamefont {Kalliney}}\ and\ \bibinfo {author} {\bibfnamefont {M.}~\bibnamefont {Widom}},\ }\bibfield  {title} {\bibinfo {title} {X-ray and neutron diffraction patterns of the {AlCrTiV} high-entropy alloy and quaternary {Heusler} structures},\ }\href {https://doi.org/10.1039/D5FD00079C} {\bibfield  {journal} {\bibinfo  {journal} {Faraday Discussions}\ }\textbf {\bibinfo {volume} {\textnormal{Advance Article}}},\ \bibinfo {pages} {10.1039/D5FD00079C} (\bibinfo {year} {2026})}\BibitemShut {NoStop}%
\bibitem [{\citenamefont {Raghuraman}\ \emph {et~al.}(2021)\citenamefont {Raghuraman}, \citenamefont {Wang},\ and\ \citenamefont {Widom}}]{raghuraman_investigation_2021}%
  \BibitemOpen
  \bibfield  {author} {\bibinfo {author} {\bibfnamefont {V.}~\bibnamefont {Raghuraman}}, \bibinfo {author} {\bibfnamefont {Y.}~\bibnamefont {Wang}},\ and\ \bibinfo {author} {\bibfnamefont {M.}~\bibnamefont {Widom}},\ }\bibfield  {title} {\bibinfo {title} {An investigation of high entropy alloy conductivity using first-principles calculations},\ }\href {https://doi.org/10.1063/5.0065239} {\bibfield  {journal} {\bibinfo  {journal} {Applied Physics Letters}\ }\textbf {\bibinfo {volume} {119}},\ \bibinfo {pages} {121903} (\bibinfo {year} {2021})}\BibitemShut {NoStop}%
\bibitem [{\citenamefont {Mu}\ \emph {et~al.}(2019{\natexlab{a}})\citenamefont {Mu}, \citenamefont {Samolyuk}, \citenamefont {Wimmer}, \citenamefont {Troparevsky}, \citenamefont {Khan}, \citenamefont {Mankovsky}, \citenamefont {Ebert},\ and\ \citenamefont {Stocks}}]{mu_uncovering_2019}%
  \BibitemOpen
  \bibfield  {author} {\bibinfo {author} {\bibfnamefont {S.}~\bibnamefont {Mu}}, \bibinfo {author} {\bibfnamefont {G.~D.}\ \bibnamefont {Samolyuk}}, \bibinfo {author} {\bibfnamefont {S.}~\bibnamefont {Wimmer}}, \bibinfo {author} {\bibfnamefont {M.~C.}\ \bibnamefont {Troparevsky}}, \bibinfo {author} {\bibfnamefont {S.~N.}\ \bibnamefont {Khan}}, \bibinfo {author} {\bibfnamefont {S.}~\bibnamefont {Mankovsky}}, \bibinfo {author} {\bibfnamefont {H.}~\bibnamefont {Ebert}},\ and\ \bibinfo {author} {\bibfnamefont {G.~M.}\ \bibnamefont {Stocks}},\ }\bibfield  {title} {\bibinfo {title} {Uncovering electron scattering mechanisms in {NiFeCoCrMn} derived concentrated solid solution and high entropy alloys},\ }\href {https://doi.org/10.1038/s41524-018-0138-z} {\bibfield  {journal} {\bibinfo  {journal} {npj Computational Materials}\ }\textbf {\bibinfo {volume} {5}},\ \bibinfo {pages} {1} (\bibinfo {year} {2019}{\natexlab{a}})}\BibitemShut {NoStop}%
\bibitem [{\citenamefont {Woodgate}\ \emph {et~al.}(2025)\citenamefont {Woodgate}, \citenamefont {Naguszewski}, \citenamefont {Redka}, \citenamefont {Ján}, \citenamefont {David},\ and\ \citenamefont {Staunton}}]{woodgate_emergent_2025}%
  \BibitemOpen
  \bibfield  {author} {\bibinfo {author} {\bibfnamefont {C.~D.}\ \bibnamefont {Woodgate}}, \bibinfo {author} {\bibfnamefont {H.~J.}\ \bibnamefont {Naguszewski}}, \bibinfo {author} {\bibfnamefont {D.}~\bibnamefont {Redka}}, \bibinfo {author} {\bibfnamefont {M.}~\bibnamefont {Ján}}, \bibinfo {author} {\bibfnamefont {Q.}~\bibnamefont {David}},\ and\ \bibinfo {author} {\bibfnamefont {J.~B.}\ \bibnamefont {Staunton}},\ }\bibfield  {title} {\bibinfo {title} {Emergent {B2} chemical orderings in the {AlTiVNb} and {AlTiCrMo} refractory high-entropy superalloys studied via first-principles theory and atomistic modelling},\ }\href {https://doi.org/10.1088/2515-7639/adf468} {\bibfield  {journal} {\bibinfo  {journal} {Journal of Physics: Materials}\ }\textbf {\bibinfo {volume} {8}},\ \bibinfo {pages} {045002} (\bibinfo {year} {2025})}\BibitemShut {NoStop}%
\bibitem [{\citenamefont {Ebert}\ \emph {et~al.}(2011)\citenamefont {Ebert}, \citenamefont {Ködderitzsch},\ and\ \citenamefont {Minár}}]{ebert_calculating_2011}%
  \BibitemOpen
  \bibfield  {author} {\bibinfo {author} {\bibfnamefont {H.}~\bibnamefont {Ebert}}, \bibinfo {author} {\bibfnamefont {D.}~\bibnamefont {Ködderitzsch}},\ and\ \bibinfo {author} {\bibfnamefont {J.}~\bibnamefont {Minár}},\ }\bibfield  {title} {\bibinfo {title} {Calculating condensed matter properties using the {KKR}-{Green}'s function method—recent developments and applications},\ }\href {https://doi.org/10.1088/0034-4885/74/9/096501} {\bibfield  {journal} {\bibinfo  {journal} {Reports on Progress in Physics}\ }\textbf {\bibinfo {volume} {74}},\ \bibinfo {pages} {096501} (\bibinfo {year} {2011})}\BibitemShut {NoStop}%
\bibitem [{\citenamefont {Fu}\ \emph {et~al.}(2007)\citenamefont {Fu}, \citenamefont {Chen}, \citenamefont {Cheng}, \citenamefont {Gao},\ and\ \citenamefont {Yang}}]{fu_influence_2007}%
  \BibitemOpen
  \bibfield  {author} {\bibinfo {author} {\bibfnamefont {H.}~\bibnamefont {Fu}}, \bibinfo {author} {\bibfnamefont {D.}~\bibnamefont {Chen}}, \bibinfo {author} {\bibfnamefont {X.}~\bibnamefont {Cheng}}, \bibinfo {author} {\bibfnamefont {T.}~\bibnamefont {Gao}},\ and\ \bibinfo {author} {\bibfnamefont {X.}~\bibnamefont {Yang}},\ }\bibfield  {title} {\bibinfo {title} {The influence of the \textit{{X}} atoms and {Al} 3\textit{p} occupied states in {AlTi}\textit{{X}}$_{\textrm{2}}$ (\textit{{X}} = {Fe}, {Cu}, {Co}, {Ni}) intermetallics},\ }\href {https://doi.org/10.1016/j.physb.2006.06.152} {\bibfield  {journal} {\bibinfo  {journal} {Physica B: Condensed Matter}\ }\textbf {\bibinfo {volume} {388}},\ \bibinfo {pages} {303} (\bibinfo {year} {2007})}\BibitemShut {NoStop}%
\bibitem [{\citenamefont {Samolyuk}\ \emph {et~al.}(2026)\citenamefont {Samolyuk}, \citenamefont {Shyam},\ and\ \citenamefont {Yoon}}]{samolyuk_first_2026}%
  \BibitemOpen
  \bibfield  {author} {\bibinfo {author} {\bibfnamefont {G.}~\bibnamefont {Samolyuk}}, \bibinfo {author} {\bibfnamefont {A.}~\bibnamefont {Shyam}},\ and\ \bibinfo {author} {\bibfnamefont {M.}~\bibnamefont {Yoon}},\ }\bibfield  {title} {\bibinfo {title} {A first principles approach for determining solute effect on electrical resistivity of {Aluminum} solid solution},\ }\href {https://doi.org/10.1016/j.scriptamat.2025.117138} {\bibfield  {journal} {\bibinfo  {journal} {Scripta Materialia}\ }\textbf {\bibinfo {volume} {274}},\ \bibinfo {pages} {117138} (\bibinfo {year} {2026})}\BibitemShut {NoStop}%
\bibitem [{\citenamefont {Thomas}(1951)}]{thomas_uber_1951}%
  \BibitemOpen
  \bibfield  {author} {\bibinfo {author} {\bibfnamefont {H.}~\bibnamefont {Thomas}},\ }\bibfield  {title} {\bibinfo {title} {Über {Widerstandslegierungen}},\ }\href {https://doi.org/10.1007/BF01333398} {\bibfield  {journal} {\bibinfo  {journal} {Zeitschrift für Physik}\ }\textbf {\bibinfo {volume} {129}},\ \bibinfo {pages} {219} (\bibinfo {year} {1951})}\BibitemShut {NoStop}%
\bibitem [{\citenamefont {Lowitzer}\ \emph {et~al.}(2010)\citenamefont {Lowitzer}, \citenamefont {Ködderitzsch}, \citenamefont {Ebert}, \citenamefont {Tulip}, \citenamefont {Marmodoro},\ and\ \citenamefont {Staunton}}]{lowitzer_ab_2010}%
  \BibitemOpen
  \bibfield  {author} {\bibinfo {author} {\bibfnamefont {S.}~\bibnamefont {Lowitzer}}, \bibinfo {author} {\bibfnamefont {D.}~\bibnamefont {Ködderitzsch}}, \bibinfo {author} {\bibfnamefont {H.}~\bibnamefont {Ebert}}, \bibinfo {author} {\bibfnamefont {P.~R.}\ \bibnamefont {Tulip}}, \bibinfo {author} {\bibfnamefont {A.}~\bibnamefont {Marmodoro}},\ and\ \bibinfo {author} {\bibfnamefont {J.~B.}\ \bibnamefont {Staunton}},\ }\bibfield  {title} {\bibinfo {title} {An ab initio investigation of how residual resistivity can decrease when an alloy is deformed},\ }\href {https://doi.org/10.1209/0295-5075/92/37009} {\bibfield  {journal} {\bibinfo  {journal} {Europhysics Letters}\ }\textbf {\bibinfo {volume} {92}},\ \bibinfo {pages} {37009} (\bibinfo {year} {2010})}\BibitemShut {NoStop}%
\bibitem [{\citenamefont {Minár}\ \emph {et~al.}(2005)\citenamefont {Minár}, \citenamefont {Chioncel}, \citenamefont {Perlov}, \citenamefont {Ebert}, \citenamefont {Katsnelson},\ and\ \citenamefont {Lichtenstein}}]{minar_multiple-scattering_2005}%
  \BibitemOpen
  \bibfield  {author} {\bibinfo {author} {\bibfnamefont {J.}~\bibnamefont {Minár}}, \bibinfo {author} {\bibfnamefont {L.}~\bibnamefont {Chioncel}}, \bibinfo {author} {\bibfnamefont {A.}~\bibnamefont {Perlov}}, \bibinfo {author} {\bibfnamefont {H.}~\bibnamefont {Ebert}}, \bibinfo {author} {\bibfnamefont {M.~I.}\ \bibnamefont {Katsnelson}},\ and\ \bibinfo {author} {\bibfnamefont {A.~I.}\ \bibnamefont {Lichtenstein}},\ }\bibfield  {title} {\bibinfo {title} {Multiple-scattering formalism for correlated systems: {A} {KKR}-{DMFT} approach},\ }\href {https://doi.org/10.1103/physrevb.72.045125} {\bibfield  {journal} {\bibinfo  {journal} {Physical Review B}\ }\textbf {\bibinfo {volume} {72}},\ \bibinfo {pages} {045125} (\bibinfo {year} {2005})}\BibitemShut {NoStop}%
\bibitem [{\citenamefont {Korringa}(1947)}]{korringa_calculation_1947}%
  \BibitemOpen
  \bibfield  {author} {\bibinfo {author} {\bibfnamefont {J.}~\bibnamefont {Korringa}},\ }\bibfield  {title} {\bibinfo {title} {On the calculation of the energy of a {Bloch} wave in a metal},\ }\href {https://doi.org/10.1016/0031-8914(47)90013-X} {\bibfield  {journal} {\bibinfo  {journal} {Physica}\ }\textbf {\bibinfo {volume} {13}},\ \bibinfo {pages} {392} (\bibinfo {year} {1947})}\BibitemShut {NoStop}%
\bibitem [{\citenamefont {Kohn}\ and\ \citenamefont {Rostoker}(1954)}]{kohn_solution_1954}%
  \BibitemOpen
  \bibfield  {author} {\bibinfo {author} {\bibfnamefont {W.}~\bibnamefont {Kohn}}\ and\ \bibinfo {author} {\bibfnamefont {N.}~\bibnamefont {Rostoker}},\ }\bibfield  {title} {\bibinfo {title} {Solution of the {Schrödinger} {Equation} in {Periodic} {Lattices} with an {Application} to {Metallic} {Lithium}},\ }\href {https://doi.org/10.1103/PhysRev.94.1111} {\bibfield  {journal} {\bibinfo  {journal} {Physical Review}\ }\textbf {\bibinfo {volume} {94}},\ \bibinfo {pages} {1111} (\bibinfo {year} {1954})}\BibitemShut {NoStop}%
\bibitem [{\citenamefont {Hohenberg}\ and\ \citenamefont {Kohn}(1964)}]{hohenberg_inhomogeneous_1964}%
  \BibitemOpen
  \bibfield  {author} {\bibinfo {author} {\bibfnamefont {P.}~\bibnamefont {Hohenberg}}\ and\ \bibinfo {author} {\bibfnamefont {W.}~\bibnamefont {Kohn}},\ }\bibfield  {title} {\bibinfo {title} {Inhomogeneous {Electron} {Gas}},\ }\href {https://doi.org/10.1103/PhysRev.136.B864} {\bibfield  {journal} {\bibinfo  {journal} {Physical Review}\ }\textbf {\bibinfo {volume} {136}},\ \bibinfo {pages} {B864} (\bibinfo {year} {1964})}\BibitemShut {NoStop}%
\bibitem [{\citenamefont {Kohn}\ and\ \citenamefont {Sham}(1965)}]{kohn_self-consistent_1965}%
  \BibitemOpen
  \bibfield  {author} {\bibinfo {author} {\bibfnamefont {W.}~\bibnamefont {Kohn}}\ and\ \bibinfo {author} {\bibfnamefont {L.~J.}\ \bibnamefont {Sham}},\ }\bibfield  {title} {\bibinfo {title} {Self-{Consistent} {Equations} {Including} {Exchange} and {Correlation} {Effects}},\ }\href {https://doi.org/10.1103/PhysRev.140.A1133} {\bibfield  {journal} {\bibinfo  {journal} {Physical Review}\ }\textbf {\bibinfo {volume} {140}},\ \bibinfo {pages} {A1133} (\bibinfo {year} {1965})}\BibitemShut {NoStop}%
\bibitem [{\citenamefont {Martin}(2004)}]{martin_electronic_2004}%
  \BibitemOpen
  \bibfield  {author} {\bibinfo {author} {\bibfnamefont {R.~M.}\ \bibnamefont {Martin}},\ }\href@noop {} {\emph {\bibinfo {title} {Electronic {Structure}: {Basic} {Theory} and {Practical} {Methods}}}}\ (\bibinfo  {publisher} {Cambridge University Press},\ \bibinfo {address} {Cambridge, UK},\ \bibinfo {year} {2004})\BibitemShut {NoStop}%
\bibitem [{\citenamefont {Faulkner}\ \emph {et~al.}(2018)\citenamefont {Faulkner}, \citenamefont {Stocks},\ and\ \citenamefont {Wang}}]{faulkner_multiple_2018}%
  \BibitemOpen
  \bibfield  {author} {\bibinfo {author} {\bibfnamefont {J.~S.}\ \bibnamefont {Faulkner}}, \bibinfo {author} {\bibfnamefont {G.~M.}\ \bibnamefont {Stocks}},\ and\ \bibinfo {author} {\bibfnamefont {Y.}~\bibnamefont {Wang}},\ }\href {https://doi.org/10.1088/2053-2563/aae7d8} {\emph {\bibinfo {title} {Multiple {Scattering} {Theory}: {Electronic} {Structure} of {Solids}}}},\ \bibinfo {edition} {1st}\ ed.\ (\bibinfo  {publisher} {IOP Publishing},\ \bibinfo {address} {Bristol, UK},\ \bibinfo {year} {2018})\BibitemShut {NoStop}%
\bibitem [{\citenamefont {Elliott}\ \emph {et~al.}(1974)\citenamefont {Elliott}, \citenamefont {Krumhansl},\ and\ \citenamefont {Leath}}]{elliott_theory_1974}%
  \BibitemOpen
  \bibfield  {author} {\bibinfo {author} {\bibfnamefont {R.~J.}\ \bibnamefont {Elliott}}, \bibinfo {author} {\bibfnamefont {J.~A.}\ \bibnamefont {Krumhansl}},\ and\ \bibinfo {author} {\bibfnamefont {P.~L.}\ \bibnamefont {Leath}},\ }\bibfield  {title} {\bibinfo {title} {The theory and properties of randomly disordered crystals and related physical systems},\ }\href {https://doi.org/10.1103/RevModPhys.46.465} {\bibfield  {journal} {\bibinfo  {journal} {Reviews of Modern Physics}\ }\textbf {\bibinfo {volume} {46}},\ \bibinfo {pages} {465} (\bibinfo {year} {1974})}\BibitemShut {NoStop}%
\bibitem [{\citenamefont {Soven}(1967)}]{soven_coherent-potential_1967}%
  \BibitemOpen
  \bibfield  {author} {\bibinfo {author} {\bibfnamefont {P.}~\bibnamefont {Soven}},\ }\bibfield  {title} {\bibinfo {title} {Coherent-{Potential} {Model} of {Substitutional} {Disordered} {Alloys}},\ }\href {https://doi.org/10.1103/PhysRev.156.809} {\bibfield  {journal} {\bibinfo  {journal} {Physical Review}\ }\textbf {\bibinfo {volume} {156}},\ \bibinfo {pages} {809} (\bibinfo {year} {1967})}\BibitemShut {NoStop}%
\bibitem [{\citenamefont {Soven}(1970)}]{soven_application_1970}%
  \BibitemOpen
  \bibfield  {author} {\bibinfo {author} {\bibfnamefont {P.}~\bibnamefont {Soven}},\ }\bibfield  {title} {\bibinfo {title} {Application of the {Coherent} {Potential} {Approximation} to a {System} of {Muffin}-{Tin} {Potentials}},\ }\href {https://doi.org/10.1103/PhysRevB.2.4715} {\bibfield  {journal} {\bibinfo  {journal} {Physical Review B}\ }\textbf {\bibinfo {volume} {2}},\ \bibinfo {pages} {4715} (\bibinfo {year} {1970})}\BibitemShut {NoStop}%
\bibitem [{\citenamefont {Győrffy}(1972)}]{gyorffy_coherent-potential_1972}%
  \BibitemOpen
  \bibfield  {author} {\bibinfo {author} {\bibfnamefont {B.~L.}\ \bibnamefont {Győrffy}},\ }\bibfield  {title} {\bibinfo {title} {Coherent-{Potential} {Approximation} for a {Nonoverlapping}-{Muffin}-{Tin}-{Potential} {Model} of {Random} {Substitutional} {Alloys}},\ }\href {https://doi.org/10.1103/PhysRevB.5.2382} {\bibfield  {journal} {\bibinfo  {journal} {Physical Review B}\ }\textbf {\bibinfo {volume} {5}},\ \bibinfo {pages} {2382} (\bibinfo {year} {1972})}\BibitemShut {NoStop}%
\bibitem [{\citenamefont {Stocks}\ \emph {et~al.}(1978)\citenamefont {Stocks}, \citenamefont {Temmerman},\ and\ \citenamefont {Győrffy}}]{stocks_complete_1978}%
  \BibitemOpen
  \bibfield  {author} {\bibinfo {author} {\bibfnamefont {G.~M.}\ \bibnamefont {Stocks}}, \bibinfo {author} {\bibfnamefont {W.~M.}\ \bibnamefont {Temmerman}},\ and\ \bibinfo {author} {\bibfnamefont {B.~L.}\ \bibnamefont {Győrffy}},\ }\bibfield  {title} {\bibinfo {title} {Complete {Solution} of the {Korringa}-{Kohn}-{Rostoker} {Coherent}-{Potential}-{Approximation} {Equations}: {Cu}-{Ni} {Alloys}},\ }\href {https://doi.org/10.1103/PhysRevLett.41.339} {\bibfield  {journal} {\bibinfo  {journal} {Physical Review Letters}\ }\textbf {\bibinfo {volume} {41}},\ \bibinfo {pages} {339} (\bibinfo {year} {1978})}\BibitemShut {NoStop}%
\bibitem [{\citenamefont {Redka}\ \emph {et~al.}(2024)\citenamefont {Redka}, \citenamefont {Khan}, \citenamefont {Martino}, \citenamefont {Mettan}, \citenamefont {Ciric}, \citenamefont {Tolj}, \citenamefont {Ivšić}, \citenamefont {Held}, \citenamefont {Caputo}, \citenamefont {Guedes}, \citenamefont {Strocov}, \citenamefont {Di~Marco}, \citenamefont {Ebert}, \citenamefont {Huber}, \citenamefont {Dil}, \citenamefont {Forró},\ and\ \citenamefont {Minár}}]{redka_interplay_2024}%
  \BibitemOpen
  \bibfield  {author} {\bibinfo {author} {\bibfnamefont {D.}~\bibnamefont {Redka}}, \bibinfo {author} {\bibfnamefont {S.~A.}\ \bibnamefont {Khan}}, \bibinfo {author} {\bibfnamefont {E.}~\bibnamefont {Martino}}, \bibinfo {author} {\bibfnamefont {X.}~\bibnamefont {Mettan}}, \bibinfo {author} {\bibfnamefont {L.}~\bibnamefont {Ciric}}, \bibinfo {author} {\bibfnamefont {D.}~\bibnamefont {Tolj}}, \bibinfo {author} {\bibfnamefont {T.}~\bibnamefont {Ivšić}}, \bibinfo {author} {\bibfnamefont {A.}~\bibnamefont {Held}}, \bibinfo {author} {\bibfnamefont {M.}~\bibnamefont {Caputo}}, \bibinfo {author} {\bibfnamefont {E.~B.}\ \bibnamefont {Guedes}}, \bibinfo {author} {\bibfnamefont {V.~N.}\ \bibnamefont {Strocov}}, \bibinfo {author} {\bibfnamefont {I.}~\bibnamefont {Di~Marco}}, \bibinfo {author} {\bibfnamefont {H.}~\bibnamefont {Ebert}}, \bibinfo {author} {\bibfnamefont {H.~P.}\ \bibnamefont {Huber}}, \bibinfo {author} {\bibfnamefont {J.~H.}\ \bibnamefont {Dil}}, \bibinfo {author} {\bibfnamefont {L.}~\bibnamefont
  {Forró}},\ and\ \bibinfo {author} {\bibfnamefont {J.}~\bibnamefont {Minár}},\ }\bibfield  {title} {\bibinfo {title} {Interplay between disorder and electronic correlations in compositionally complex alloys},\ }\href {https://doi.org/10.1038/s41467-024-52349-8} {\bibfield  {journal} {\bibinfo  {journal} {Nature Communications}\ }\textbf {\bibinfo {volume} {15}},\ \bibinfo {pages} {7983} (\bibinfo {year} {2024})}\BibitemShut {NoStop}%
\bibitem [{\citenamefont {Bista}\ \emph {et~al.}(2025)\citenamefont {Bista}, \citenamefont {Beeson}, \citenamefont {Sengupta}, \citenamefont {Jackson}, \citenamefont {Khanna}, \citenamefont {Liu},\ and\ \citenamefont {Yin}}]{bista_fast_2025}%
  \BibitemOpen
  \bibfield  {author} {\bibinfo {author} {\bibfnamefont {D.}~\bibnamefont {Bista}}, \bibinfo {author} {\bibfnamefont {W.~B.}\ \bibnamefont {Beeson}}, \bibinfo {author} {\bibfnamefont {T.}~\bibnamefont {Sengupta}}, \bibinfo {author} {\bibfnamefont {J.}~\bibnamefont {Jackson}}, \bibinfo {author} {\bibfnamefont {S.~N.}\ \bibnamefont {Khanna}}, \bibinfo {author} {\bibfnamefont {K.}~\bibnamefont {Liu}},\ and\ \bibinfo {author} {\bibfnamefont {G.}~\bibnamefont {Yin}},\ }\bibfield  {title} {\bibinfo {title} {Fast \textit{ab initio} design of high-entropy magnetic materials},\ }\href {https://doi.org/10.1103/PhysRevMaterials.9.L031401} {\bibfield  {journal} {\bibinfo  {journal} {Physical Review Materials}\ }\textbf {\bibinfo {volume} {9}},\ \bibinfo {pages} {L031401} (\bibinfo {year} {2025})}\BibitemShut {NoStop}%
\bibitem [{\citenamefont {Samolyuk}\ \emph {et~al.}(2018)\citenamefont {Samolyuk}, \citenamefont {Mu}, \citenamefont {May}, \citenamefont {Sales}, \citenamefont {Wimmer}, \citenamefont {Mankovsky}, \citenamefont {Ebert},\ and\ \citenamefont {Stocks}}]{samolyuk_temperature_2018}%
  \BibitemOpen
  \bibfield  {author} {\bibinfo {author} {\bibfnamefont {G.~D.}\ \bibnamefont {Samolyuk}}, \bibinfo {author} {\bibfnamefont {S.}~\bibnamefont {Mu}}, \bibinfo {author} {\bibfnamefont {A.~F.}\ \bibnamefont {May}}, \bibinfo {author} {\bibfnamefont {B.~C.}\ \bibnamefont {Sales}}, \bibinfo {author} {\bibfnamefont {S.}~\bibnamefont {Wimmer}}, \bibinfo {author} {\bibfnamefont {S.}~\bibnamefont {Mankovsky}}, \bibinfo {author} {\bibfnamefont {H.}~\bibnamefont {Ebert}},\ and\ \bibinfo {author} {\bibfnamefont {G.~M.}\ \bibnamefont {Stocks}},\ }\bibfield  {title} {\bibinfo {title} {Temperature dependent electronic transport in concentrated solid solutions of the $3d$-transition metals {Ni}, {Fe}, {Co} and {Cr} from first principles},\ }\href {https://doi.org/10.1103/PhysRevB.98.165141} {\bibfield  {journal} {\bibinfo  {journal} {Physical Review B}\ }\textbf {\bibinfo {volume} {98}},\ \bibinfo {pages} {165141} (\bibinfo {year} {2018})}\BibitemShut {NoStop}%
\bibitem [{\citenamefont {Tian}\ \emph {et~al.}(2013)\citenamefont {Tian}, \citenamefont {Delczeg}, \citenamefont {Chen}, \citenamefont {Varga}, \citenamefont {Shen},\ and\ \citenamefont {Vitos}}]{tian_structural_2013}%
  \BibitemOpen
  \bibfield  {author} {\bibinfo {author} {\bibfnamefont {F.}~\bibnamefont {Tian}}, \bibinfo {author} {\bibfnamefont {L.}~\bibnamefont {Delczeg}}, \bibinfo {author} {\bibfnamefont {N.}~\bibnamefont {Chen}}, \bibinfo {author} {\bibfnamefont {L.~K.}\ \bibnamefont {Varga}}, \bibinfo {author} {\bibfnamefont {J.}~\bibnamefont {Shen}},\ and\ \bibinfo {author} {\bibfnamefont {L.}~\bibnamefont {Vitos}},\ }\bibfield  {title} {\bibinfo {title} {Structural stability of {NiCoFeCrAl}$_\textrm{x}$ high-entropy alloy from \textit{ab initio} theory},\ }\href {https://doi.org/10.1103/PhysRevB.88.085128} {\bibfield  {journal} {\bibinfo  {journal} {Physical Review B}\ }\textbf {\bibinfo {volume} {88}},\ \bibinfo {pages} {085128} (\bibinfo {year} {2013})}\BibitemShut {NoStop}%
\bibitem [{\citenamefont {Tian}\ \emph {et~al.}(2017)\citenamefont {Tian}, \citenamefont {Wang},\ and\ \citenamefont {Vitos}}]{tian_impact_2017}%
  \BibitemOpen
  \bibfield  {author} {\bibinfo {author} {\bibfnamefont {F.}~\bibnamefont {Tian}}, \bibinfo {author} {\bibfnamefont {Y.}~\bibnamefont {Wang}},\ and\ \bibinfo {author} {\bibfnamefont {L.}~\bibnamefont {Vitos}},\ }\bibfield  {title} {\bibinfo {title} {Impact of aluminum doping on the thermo-physical properties of refractory medium-entropy alloys},\ }\href {https://doi.org/10.1063/1.4973489} {\bibfield  {journal} {\bibinfo  {journal} {Journal of Applied Physics}\ }\textbf {\bibinfo {volume} {121}},\ \bibinfo {pages} {015105} (\bibinfo {year} {2017})}\BibitemShut {NoStop}%
\bibitem [{\citenamefont {Huang}\ \emph {et~al.}(2018)\citenamefont {Huang}, \citenamefont {Tian},\ and\ \citenamefont {Vitos}}]{huang_elasticity_2018}%
  \BibitemOpen
  \bibfield  {author} {\bibinfo {author} {\bibfnamefont {S.}~\bibnamefont {Huang}}, \bibinfo {author} {\bibfnamefont {F.}~\bibnamefont {Tian}},\ and\ \bibinfo {author} {\bibfnamefont {L.}~\bibnamefont {Vitos}},\ }\bibfield  {title} {\bibinfo {title} {Elasticity of high-entropy alloys from \textit{ab initio} theory},\ }\href {https://doi.org/10.1557/jmr.2018.237} {\bibfield  {journal} {\bibinfo  {journal} {Journal of Materials Research}\ }\textbf {\bibinfo {volume} {33}},\ \bibinfo {pages} {2938} (\bibinfo {year} {2018})}\BibitemShut {NoStop}%
\bibitem [{\citenamefont {Győrffy}\ and\ \citenamefont {Stocks}(1983)}]{gyorffy_concentration_1983}%
  \BibitemOpen
  \bibfield  {author} {\bibinfo {author} {\bibfnamefont {B.~L.}\ \bibnamefont {Győrffy}}\ and\ \bibinfo {author} {\bibfnamefont {G.~M.}\ \bibnamefont {Stocks}},\ }\bibfield  {title} {\bibinfo {title} {Concentration {Waves} and {Fermi} {Surfaces} in {Random} {Metallic} {Alloys}},\ }\href {https://doi.org/10.1103/PhysRevLett.50.374} {\bibfield  {journal} {\bibinfo  {journal} {Physical Review Letters}\ }\textbf {\bibinfo {volume} {50}},\ \bibinfo {pages} {374} (\bibinfo {year} {1983})}\BibitemShut {NoStop}%
\bibitem [{\citenamefont {Johnson}\ \emph {et~al.}(1986)\citenamefont {Johnson}, \citenamefont {Nicholson}, \citenamefont {Pinski}, \citenamefont {Gyorffy},\ and\ \citenamefont {Stocks}}]{johnson_density-functional_1986}%
  \BibitemOpen
  \bibfield  {author} {\bibinfo {author} {\bibfnamefont {D.~D.}\ \bibnamefont {Johnson}}, \bibinfo {author} {\bibfnamefont {D.~M.}\ \bibnamefont {Nicholson}}, \bibinfo {author} {\bibfnamefont {F.~J.}\ \bibnamefont {Pinski}}, \bibinfo {author} {\bibfnamefont {B.~L.}\ \bibnamefont {Gyorffy}},\ and\ \bibinfo {author} {\bibfnamefont {G.~M.}\ \bibnamefont {Stocks}},\ }\bibfield  {title} {\bibinfo {title} {Density-{Functional} {Theory} for {Random} {Alloys}: {Total} {Energy} within the {Coherent}-{Potential} {Approximation}},\ }\href {https://doi.org/10.1103/PhysRevLett.56.2088} {\bibfield  {journal} {\bibinfo  {journal} {Physical Review Letters}\ }\textbf {\bibinfo {volume} {56}},\ \bibinfo {pages} {2088} (\bibinfo {year} {1986})}\BibitemShut {NoStop}%
\bibitem [{\citenamefont {Johnson}\ \emph {et~al.}(1990)\citenamefont {Johnson}, \citenamefont {Nicholson}, \citenamefont {Pinski}, \citenamefont {Gy\H{o}rffy},\ and\ \citenamefont {Stocks}}]{johnson_total-energy_1990}%
  \BibitemOpen
  \bibfield  {author} {\bibinfo {author} {\bibfnamefont {D.~D.}\ \bibnamefont {Johnson}}, \bibinfo {author} {\bibfnamefont {D.~M.}\ \bibnamefont {Nicholson}}, \bibinfo {author} {\bibfnamefont {F.~J.}\ \bibnamefont {Pinski}}, \bibinfo {author} {\bibfnamefont {B.~L.}\ \bibnamefont {Gy\H{o}rffy}},\ and\ \bibinfo {author} {\bibfnamefont {G.~M.}\ \bibnamefont {Stocks}},\ }\bibfield  {title} {\bibinfo {title} {Total-energy and pressure calculations for random substitutional alloys},\ }\href {https://doi.org/10.1103/PhysRevB.41.9701} {\bibfield  {journal} {\bibinfo  {journal} {Physical Review B}\ }\textbf {\bibinfo {volume} {41}},\ \bibinfo {pages} {9701} (\bibinfo {year} {1990})}\BibitemShut {NoStop}%
\bibitem [{\citenamefont {Staunton}\ \emph {et~al.}(1994)\citenamefont {Staunton}, \citenamefont {Johnson},\ and\ \citenamefont {Pinski}}]{staunton_compositional_1994}%
  \BibitemOpen
  \bibfield  {author} {\bibinfo {author} {\bibfnamefont {J.~B.}\ \bibnamefont {Staunton}}, \bibinfo {author} {\bibfnamefont {D.~D.}\ \bibnamefont {Johnson}},\ and\ \bibinfo {author} {\bibfnamefont {F.~J.}\ \bibnamefont {Pinski}},\ }\bibfield  {title} {\bibinfo {title} {Compositional short-range ordering in metallic alloys: {Band}-filling, charge-transfer, and size effects from a first-principles all-electron {Landau}-type theory},\ }\href {https://doi.org/10.1103/PhysRevB.50.1450} {\bibfield  {journal} {\bibinfo  {journal} {Physical Review B}\ }\textbf {\bibinfo {volume} {50}},\ \bibinfo {pages} {1450} (\bibinfo {year} {1994})}\BibitemShut {NoStop}%
\bibitem [{\citenamefont {Johnson}\ \emph {et~al.}(1994)\citenamefont {Johnson}, \citenamefont {Staunton},\ and\ \citenamefont {Pinski}}]{johnson_first-principles_1994}%
  \BibitemOpen
  \bibfield  {author} {\bibinfo {author} {\bibfnamefont {D.~D.}\ \bibnamefont {Johnson}}, \bibinfo {author} {\bibfnamefont {J.~B.}\ \bibnamefont {Staunton}},\ and\ \bibinfo {author} {\bibfnamefont {F.~J.}\ \bibnamefont {Pinski}},\ }\bibfield  {title} {\bibinfo {title} {First-principles all-electron theory of atomic short-range ordering in metallic alloys: {D}0\textsubscript{22}- versus {L}1\textsubscript{2}-like correlations},\ }\href {https://doi.org/10.1103/PhysRevB.50.1473} {\bibfield  {journal} {\bibinfo  {journal} {Physical Review B}\ }\textbf {\bibinfo {volume} {50}},\ \bibinfo {pages} {1473} (\bibinfo {year} {1994})}\BibitemShut {NoStop}%
\bibitem [{\citenamefont {Clark}\ \emph {et~al.}(1995)\citenamefont {Clark}, \citenamefont {Pinski}, \citenamefont {Johnson}, \citenamefont {Sterne}, \citenamefont {Staunton},\ and\ \citenamefont {Ginatempo}}]{clark_van_1995}%
  \BibitemOpen
  \bibfield  {author} {\bibinfo {author} {\bibfnamefont {J.~F.}\ \bibnamefont {Clark}}, \bibinfo {author} {\bibfnamefont {F.~J.}\ \bibnamefont {Pinski}}, \bibinfo {author} {\bibfnamefont {D.~D.}\ \bibnamefont {Johnson}}, \bibinfo {author} {\bibfnamefont {P.~A.}\ \bibnamefont {Sterne}}, \bibinfo {author} {\bibfnamefont {J.~B.}\ \bibnamefont {Staunton}},\ and\ \bibinfo {author} {\bibfnamefont {B.}~\bibnamefont {Ginatempo}},\ }\bibfield  {title} {\bibinfo {title} {van {Hove} {Singularity} {Induced} {L1}$_{\textrm{1}}$ {Ordering} in {CuPt}},\ }\href {https://doi.org/10.1103/PhysRevLett.74.3225} {\bibfield  {journal} {\bibinfo  {journal} {Physical Review Letters}\ }\textbf {\bibinfo {volume} {74}},\ \bibinfo {pages} {3225} (\bibinfo {year} {1995})}\BibitemShut {NoStop}%
\bibitem [{\citenamefont {Woodgate}\ \emph {et~al.}(2024)\citenamefont {Woodgate}, \citenamefont {Marchant}, \citenamefont {Pártay},\ and\ \citenamefont {Staunton}}]{woodgate_structure_2024}%
  \BibitemOpen
  \bibfield  {author} {\bibinfo {author} {\bibfnamefont {C.~D.}\ \bibnamefont {Woodgate}}, \bibinfo {author} {\bibfnamefont {G.~A.}\ \bibnamefont {Marchant}}, \bibinfo {author} {\bibfnamefont {L.~B.}\ \bibnamefont {Pártay}},\ and\ \bibinfo {author} {\bibfnamefont {J.~B.}\ \bibnamefont {Staunton}},\ }\bibfield  {title} {\bibinfo {title} {Structure, short-range order, and phase stability of the {Al}$_{\textrm{x}}${CrFeCoNi} high-entropy alloy: insights from a perturbative, {DFT}-based analysis},\ }\href {https://doi.org/10.1038/s41524-024-01445-w} {\bibfield  {journal} {\bibinfo  {journal} {npj Computational Materials}\ }\textbf {\bibinfo {volume} {10}},\ \bibinfo {pages} {271} (\bibinfo {year} {2024})}\BibitemShut {NoStop}%
\bibitem [{\citenamefont {Woodgate}\ \emph {et~al.}()\citenamefont {Woodgate}, \citenamefont {Naguszewski}, \citenamefont {Deacon}, \citenamefont {Potel}, \citenamefont {Min\'ar}, \citenamefont {Quigley},\ and\ \citenamefont {Staunton}}]{woodgate_formation_nodate}%
  \BibitemOpen
  \bibfield  {author} {\bibinfo {author} {\bibfnamefont {C.~D.}\ \bibnamefont {Woodgate}}, \bibinfo {author} {\bibfnamefont {H.~J.}\ \bibnamefont {Naguszewski}}, \bibinfo {author} {\bibfnamefont {S.~L.}\ \bibnamefont {Deacon}}, \bibinfo {author} {\bibfnamefont {M.~P.}\ \bibnamefont {Potel}}, \bibinfo {author} {\bibfnamefont {J.}~\bibnamefont {Min\'ar}}, \bibinfo {author} {\bibfnamefont {D.}~\bibnamefont {Quigley}},\ and\ \bibinfo {author} {\bibfnamefont {J.~B.}\ \bibnamefont {Staunton}},\ }\bibfield  {title} {\bibinfo {title} {Formation of {L1}\textsubscript{2}-ordered $\gamma'$-{Ni}\textsubscript{3}{Al} precipitates in ternary {Cu}--{Ni}--{Al} alloys modelled using an \textit{ab initio} concentration wave theory and atomistic simulations},\ }\href {https://doi.org/10.48550/arXiv.2607.27108} {\bibinfo  {journal} {arXiv:2607.27108}\ }\BibitemShut {NoStop}%
\bibitem [{\citenamefont {Woodgate}\ and\ \citenamefont {Staunton}(2024)}]{woodgate_competition_2024}%
  \BibitemOpen
\bibfield  {journal} {  }\bibfield  {author} {\bibinfo {author} {\bibfnamefont {C.~D.}\ \bibnamefont {Woodgate}}\ and\ \bibinfo {author} {\bibfnamefont {J.~B.}\ \bibnamefont {Staunton}},\ }\bibfield  {title} {\bibinfo {title} {Competition between phase ordering and phase segregation in the {Ti}$_{\textrm{x}}${NbMoTaW} and {Ti}$_{\textrm{x}}${VNbMoTaW} refractory high-entropy alloys},\ }\href {https://doi.org/10.1063/5.0200862} {\bibfield  {journal} {\bibinfo  {journal} {Journal of Applied Physics}\ }\textbf {\bibinfo {volume} {135}},\ \bibinfo {pages} {135106} (\bibinfo {year} {2024})}\BibitemShut {NoStop}%
\bibitem [{\citenamefont {Bragg}\ and\ \citenamefont {Williams}(1934)}]{bragg_effect_1934}%
  \BibitemOpen
  \bibfield  {author} {\bibinfo {author} {\bibfnamefont {W.~L.}\ \bibnamefont {Bragg}}\ and\ \bibinfo {author} {\bibfnamefont {E.~J.}\ \bibnamefont {Williams}},\ }\bibfield  {title} {\bibinfo {title} {The effect of thermal agitation on atomic arrangement in alloys},\ }\href {https://doi.org/10.1098/rspa.1934.0132} {\bibfield  {journal} {\bibinfo  {journal} {Proceedings of the Royal Society of London. Series A, Containing Papers of a Mathematical and Physical Character}\ }\textbf {\bibinfo {volume} {145}},\ \bibinfo {pages} {699} (\bibinfo {year} {1934})}\BibitemShut {NoStop}%
\bibitem [{\citenamefont {Bragg}\ and\ \citenamefont {Williams}(1935)}]{bragg_effect_1935}%
  \BibitemOpen
  \bibfield  {author} {\bibinfo {author} {\bibfnamefont {W.~L.}\ \bibnamefont {Bragg}}\ and\ \bibinfo {author} {\bibfnamefont {E.~J.}\ \bibnamefont {Williams}},\ }\bibfield  {title} {\bibinfo {title} {The effect of thermal agitaion on atomic arrangement in alloys-{II}},\ }\href {https://doi.org/10.1098/rspa.1935.0165} {\bibfield  {journal} {\bibinfo  {journal} {Proceedings of the Royal Society of London. Series A - Mathematical and Physical Sciences}\ }\textbf {\bibinfo {volume} {151}},\ \bibinfo {pages} {540} (\bibinfo {year} {1935})}\BibitemShut {NoStop}%
\bibitem [{\citenamefont {van~de Walle}\ and\ \citenamefont {Ceder}(2002)}]{van_de_walle_effect_2002}%
  \BibitemOpen
  \bibfield  {author} {\bibinfo {author} {\bibfnamefont {A.}~\bibnamefont {van~de Walle}}\ and\ \bibinfo {author} {\bibfnamefont {G.}~\bibnamefont {Ceder}},\ }\bibfield  {title} {\bibinfo {title} {The effect of lattice vibrations on substitutional alloy thermodynamics},\ }\href {https://doi.org/10.1103/RevModPhys.74.11} {\bibfield  {journal} {\bibinfo  {journal} {Rev. Mod. Phys.}\ }\textbf {\bibinfo {volume} {74}},\ \bibinfo {pages} {11} (\bibinfo {year} {2002})}\BibitemShut {NoStop}%
\bibitem [{\citenamefont {Metropolis}\ \emph {et~al.}(1953)\citenamefont {Metropolis}, \citenamefont {Rosenbluth}, \citenamefont {Rosenbluth}, \citenamefont {Teller},\ and\ \citenamefont {Teller}}]{metropolis_equation_1953}%
  \BibitemOpen
  \bibfield  {author} {\bibinfo {author} {\bibfnamefont {N.}~\bibnamefont {Metropolis}}, \bibinfo {author} {\bibfnamefont {A.~W.}\ \bibnamefont {Rosenbluth}}, \bibinfo {author} {\bibfnamefont {M.~N.}\ \bibnamefont {Rosenbluth}}, \bibinfo {author} {\bibfnamefont {A.~H.}\ \bibnamefont {Teller}},\ and\ \bibinfo {author} {\bibfnamefont {E.}~\bibnamefont {Teller}},\ }\bibfield  {title} {\bibinfo {title} {Equation of {State} {Calculations} by {Fast} {Computing} {Machines}},\ }\href {https://doi.org/10.1063/1.1699114} {\bibfield  {journal} {\bibinfo  {journal} {Journal of Chemical Physics}\ }\textbf {\bibinfo {volume} {21}},\ \bibinfo {pages} {1087} (\bibinfo {year} {1953})}\BibitemShut {NoStop}%
\bibitem [{\citenamefont {Landau}\ and\ \citenamefont {Binder}(2014)}]{landau_guide_2014}%
  \BibitemOpen
  \bibfield  {author} {\bibinfo {author} {\bibfnamefont {D.~P.}\ \bibnamefont {Landau}}\ and\ \bibinfo {author} {\bibfnamefont {K.}~\bibnamefont {Binder}},\ }\href {https://doi.org/10.1017/CBO9781139696463} {\emph {\bibinfo {title} {A {Guide} to {Monte} {Carlo} {Simulations} in {Statistical} {Physics}}}},\ \bibinfo {edition} {4th}\ ed.\ (\bibinfo  {publisher} {Cambridge University Press},\ \bibinfo {address} {Cambridge, UK},\ \bibinfo {year} {2014})\BibitemShut {NoStop}%
\bibitem [{\citenamefont {Wang}\ and\ \citenamefont {Landau}(2001)}]{wang_efficient_2001}%
  \BibitemOpen
  \bibfield  {author} {\bibinfo {author} {\bibfnamefont {F.}~\bibnamefont {Wang}}\ and\ \bibinfo {author} {\bibfnamefont {D.~P.}\ \bibnamefont {Landau}},\ }\bibfield  {title} {\bibinfo {title} {Efficient, {Multiple}-{Range} {Random} {Walk} {Algorithm} to {Calculate} the {Density} of {States}},\ }\href {https://doi.org/10.1103/PhysRevLett.86.2050} {\bibfield  {journal} {\bibinfo  {journal} {Physical Review Letters}\ }\textbf {\bibinfo {volume} {86}},\ \bibinfo {pages} {2050} (\bibinfo {year} {2001})}\BibitemShut {NoStop}%
\bibitem [{\citenamefont {Kawasaki}(1966)}]{kawasaki_diffusion_1966}%
  \BibitemOpen
  \bibfield  {author} {\bibinfo {author} {\bibfnamefont {K.}~\bibnamefont {Kawasaki}},\ }\bibfield  {title} {\bibinfo {title} {Diffusion {Constants} near the {Critical} {Point} for {Time}-{Dependent} {Ising} {Models}. {I}},\ }\href {https://doi.org/10.1103/PhysRev.145.224} {\bibfield  {journal} {\bibinfo  {journal} {Physical Review}\ }\textbf {\bibinfo {volume} {145}},\ \bibinfo {pages} {224} (\bibinfo {year} {1966})}\BibitemShut {NoStop}%
\bibitem [{\citenamefont {Naguszewski}\ \emph {et~al.}(2026)\citenamefont {Naguszewski}, \citenamefont {Woodgate},\ and\ \citenamefont {Quigley}}]{naguszewski_optimal_2026}%
  \BibitemOpen
  \bibfield  {author} {\bibinfo {author} {\bibfnamefont {H.~J.}\ \bibnamefont {Naguszewski}}, \bibinfo {author} {\bibfnamefont {C.~D.}\ \bibnamefont {Woodgate}},\ and\ \bibinfo {author} {\bibfnamefont {D.}~\bibnamefont {Quigley}},\ }\bibfield  {title} {\bibinfo {title} {Optimal parallelisation strategies for flat histogram {Monte} {Carlo} sampling},\ }\href {https://doi.org/10.1016/j.cpc.2026.110125} {\bibfield  {journal} {\bibinfo  {journal} {Computer Physics Communications}\ }\textbf {\bibinfo {volume} {324}},\ \bibinfo {pages} {110125} (\bibinfo {year} {2026})}\BibitemShut {NoStop}%
\bibitem [{\citenamefont {Cowley}(1950)}]{cowley_approximate_1950}%
  \BibitemOpen
  \bibfield  {author} {\bibinfo {author} {\bibfnamefont {J.~M.}\ \bibnamefont {Cowley}},\ }\bibfield  {title} {\bibinfo {title} {An {Approximate} {Theory} of {Order} in {Alloys}},\ }\href {https://doi.org/10.1103/PhysRev.77.669} {\bibfield  {journal} {\bibinfo  {journal} {Physical Review}\ }\textbf {\bibinfo {volume} {77}},\ \bibinfo {pages} {669} (\bibinfo {year} {1950})}\BibitemShut {NoStop}%
\bibitem [{\citenamefont {Cowley}(1965)}]{cowley_short-range_1965}%
  \BibitemOpen
  \bibfield  {author} {\bibinfo {author} {\bibfnamefont {J.~M.}\ \bibnamefont {Cowley}},\ }\bibfield  {title} {\bibinfo {title} {Short-{Range} {Order} and {Long}-{Range} {Order} {Parameters}},\ }\href {https://doi.org/10.1103/PhysRev.138.A1384} {\bibfield  {journal} {\bibinfo  {journal} {Physical Review}\ }\textbf {\bibinfo {volume} {138}},\ \bibinfo {pages} {A1384} (\bibinfo {year} {1965})}\BibitemShut {NoStop}%
\bibitem [{\citenamefont {Sheriff}\ \emph {et~al.}(2024)\citenamefont {Sheriff}, \citenamefont {Cao}, \citenamefont {Smidt},\ and\ \citenamefont {Freitas}}]{sheriff_quantifying_2024}%
  \BibitemOpen
  \bibfield  {author} {\bibinfo {author} {\bibfnamefont {K.}~\bibnamefont {Sheriff}}, \bibinfo {author} {\bibfnamefont {Y.}~\bibnamefont {Cao}}, \bibinfo {author} {\bibfnamefont {T.}~\bibnamefont {Smidt}},\ and\ \bibinfo {author} {\bibfnamefont {R.}~\bibnamefont {Freitas}},\ }\bibfield  {title} {\bibinfo {title} {Quantifying chemical short-range order in metallic alloys},\ }\href {https://doi.org/10.1073/pnas.2322962121} {\bibfield  {journal} {\bibinfo  {journal} {Proceedings of the National Academy of Sciences of the United States of America}\ }\textbf {\bibinfo {volume} {121}},\ \bibinfo {pages} {e2322962121} (\bibinfo {year} {2024})}\BibitemShut {NoStop}%
\bibitem [{\citenamefont {Butler}(1985)}]{butler_theory_1985}%
  \BibitemOpen
  \bibfield  {author} {\bibinfo {author} {\bibfnamefont {W.~H.}\ \bibnamefont {Butler}},\ }\bibfield  {title} {\bibinfo {title} {Theory of electronic transport in random alloys: {Korringa}-{Kohn}-{Rostoker} coherent-potential approximation},\ }\href {https://doi.org/10.1103/PhysRevB.31.3260} {\bibfield  {journal} {\bibinfo  {journal} {Physical Review B}\ }\textbf {\bibinfo {volume} {31}},\ \bibinfo {pages} {3260} (\bibinfo {year} {1985})}\BibitemShut {NoStop}%
\bibitem [{\citenamefont {Kubo}(1957)}]{kubo_statistical-mechanical_1957}%
  \BibitemOpen
  \bibfield  {author} {\bibinfo {author} {\bibfnamefont {R.}~\bibnamefont {Kubo}},\ }\bibfield  {title} {\bibinfo {title} {Statistical-{Mechanical} {Theory} of {Irreversible} {Processes}. {I}. {General} {Theory} and {Simple} {Applications} to {Magnetic} and {Conduction} {Problems}},\ }\href {https://doi.org/10.1143/JPSJ.12.570} {\bibfield  {journal} {\bibinfo  {journal} {Journal of the Physical Society of Japan}\ }\textbf {\bibinfo {volume} {12}},\ \bibinfo {pages} {570} (\bibinfo {year} {1957})}\BibitemShut {NoStop}%
\bibitem [{\citenamefont {Greenwood}(1958)}]{greenwood_boltzmann_1958}%
  \BibitemOpen
  \bibfield  {author} {\bibinfo {author} {\bibfnamefont {D.~A.}\ \bibnamefont {Greenwood}},\ }\bibfield  {title} {\bibinfo {title} {The {Boltzmann} {Equation} in the {Theory} of {Electrical} {Conduction} in {Metals}},\ }\href {https://doi.org/10.1088/0370-1328/71/4/306} {\bibfield  {journal} {\bibinfo  {journal} {Proceedings of the Physical Society}\ }\textbf {\bibinfo {volume} {71}},\ \bibinfo {pages} {585} (\bibinfo {year} {1958})}\BibitemShut {NoStop}%
\bibitem [{\citenamefont {Mu}\ \emph {et~al.}(2019{\natexlab{b}})\citenamefont {Mu}, \citenamefont {Wimmer}, \citenamefont {Mankovsky}, \citenamefont {Ebert},\ and\ \citenamefont {Stocks}}]{mu_influence_2019}%
  \BibitemOpen
  \bibfield  {author} {\bibinfo {author} {\bibfnamefont {S.}~\bibnamefont {Mu}}, \bibinfo {author} {\bibfnamefont {S.}~\bibnamefont {Wimmer}}, \bibinfo {author} {\bibfnamefont {S.}~\bibnamefont {Mankovsky}}, \bibinfo {author} {\bibfnamefont {H.}~\bibnamefont {Ebert}},\ and\ \bibinfo {author} {\bibfnamefont {G.}~\bibnamefont {Stocks}},\ }\bibfield  {title} {\bibinfo {title} {Influence of local lattice distortions on electrical transport of refractory high entropy alloys},\ }\href {https://doi.org/https://doi.org/10.1016/j.scriptamat.2019.05.032} {\bibfield  {journal} {\bibinfo  {journal} {Scripta Materialia}\ }\textbf {\bibinfo {volume} {170}},\ \bibinfo {pages} {189} (\bibinfo {year} {2019}{\natexlab{b}})}\BibitemShut {NoStop}%
\bibitem [{\citenamefont {Ebert}\ \emph {et~al.}(2015)\citenamefont {Ebert}, \citenamefont {Mankovsky}, \citenamefont {Chadova}, \citenamefont {Polesya}, \citenamefont {Minár},\ and\ \citenamefont {Ködderitzsch}}]{ebert_calculating_2015}%
  \BibitemOpen
  \bibfield  {author} {\bibinfo {author} {\bibfnamefont {H.}~\bibnamefont {Ebert}}, \bibinfo {author} {\bibfnamefont {S.}~\bibnamefont {Mankovsky}}, \bibinfo {author} {\bibfnamefont {K.}~\bibnamefont {Chadova}}, \bibinfo {author} {\bibfnamefont {S.}~\bibnamefont {Polesya}}, \bibinfo {author} {\bibfnamefont {J.}~\bibnamefont {Minár}},\ and\ \bibinfo {author} {\bibfnamefont {D.}~\bibnamefont {Ködderitzsch}},\ }\bibfield  {title} {\bibinfo {title} {Calculating linear-response functions for finite temperatures on the basis of the alloy analogy model},\ }\href {https://doi.org/10.1103/PhysRevB.91.165132} {\bibfield  {journal} {\bibinfo  {journal} {Physical Review B}\ }\textbf {\bibinfo {volume} {91}},\ \bibinfo {pages} {165132} (\bibinfo {year} {2015})}\BibitemShut {NoStop}%
\bibitem [{\citenamefont {Andersen}(1975)}]{andersen_linear_1975}%
  \BibitemOpen
  \bibfield  {author} {\bibinfo {author} {\bibfnamefont {O.~K.}\ \bibnamefont {Andersen}},\ }\bibfield  {title} {\bibinfo {title} {Linear methods in band theory},\ }\href {https://doi.org/10.1103/PhysRevB.12.3060} {\bibfield  {journal} {\bibinfo  {journal} {Physical Review B}\ }\textbf {\bibinfo {volume} {12}},\ \bibinfo {pages} {3060} (\bibinfo {year} {1975})}\BibitemShut {NoStop}%
\bibitem [{spr()}]{sprkkr_website}%
  \BibitemOpen
  \href@noop {} {\bibinfo {title} {The {SPR-KKR} package}},\ \bibinfo {howpublished} {\url{https://www.sprkkr.org}}\BibitemShut {NoStop}%
\bibitem [{\citenamefont {Vosko}\ \emph {et~al.}(1980)\citenamefont {Vosko}, \citenamefont {Wilk},\ and\ \citenamefont {Nusair}}]{vosko_accurate_1980}%
  \BibitemOpen
  \bibfield  {author} {\bibinfo {author} {\bibfnamefont {S.~H.}\ \bibnamefont {Vosko}}, \bibinfo {author} {\bibfnamefont {L.}~\bibnamefont {Wilk}},\ and\ \bibinfo {author} {\bibfnamefont {M.}~\bibnamefont {Nusair}},\ }\bibfield  {title} {\bibinfo {title} {Accurate spin-dependent electron liquid correlation energies for local spin density calculations: a critical analysis},\ }\href {https://doi.org/10.1139/p80-159} {\bibfield  {journal} {\bibinfo  {journal} {Canadian Journal of Physics}\ }\textbf {\bibinfo {volume} {58}},\ \bibinfo {pages} {1200} (\bibinfo {year} {1980})}\BibitemShut {NoStop}%
\bibitem [{\citenamefont {Bruno}\ and\ \citenamefont {Ginatempo}(1997)}]{bruno_algorithms_1997}%
  \BibitemOpen
  \bibfield  {author} {\bibinfo {author} {\bibfnamefont {E.}~\bibnamefont {Bruno}}\ and\ \bibinfo {author} {\bibfnamefont {B.}~\bibnamefont {Ginatempo}},\ }\bibfield  {title} {\bibinfo {title} {Algorithms for {Korringa}-{Kohn}-{Rostoker} electronic structure calculations in any {Bravais} lattice},\ }\href {https://doi.org/10.1103/physrevb.55.12946} {\bibfield  {journal} {\bibinfo  {journal} {Physical Review B}\ }\textbf {\bibinfo {volume} {55}},\ \bibinfo {pages} {12946} (\bibinfo {year} {1997})}\BibitemShut {NoStop}%
\bibitem [{\citenamefont {Naguszewski}\ \emph {et~al.}(2025)\citenamefont {Naguszewski}, \citenamefont {Pártay}, \citenamefont {Quigley},\ and\ \citenamefont {Woodgate}}]{naguszewski_brawl_2025}%
  \BibitemOpen
  \bibfield  {author} {\bibinfo {author} {\bibfnamefont {H.~J.}\ \bibnamefont {Naguszewski}}, \bibinfo {author} {\bibfnamefont {L.~B.}\ \bibnamefont {Pártay}}, \bibinfo {author} {\bibfnamefont {D.}~\bibnamefont {Quigley}},\ and\ \bibinfo {author} {\bibfnamefont {C.~D.}\ \bibnamefont {Woodgate}},\ }\bibfield  {title} {\bibinfo {title} {{BraWl}: {Simulating} the thermodynamics and phase stability of multicomponent alloys using conventional and enhanced sampling techniques},\ }\href {https://doi.org/10.21105/joss.08346} {\bibfield  {journal} {\bibinfo  {journal} {Journal of Open Source Software}\ }\textbf {\bibinfo {volume} {10}},\ \bibinfo {pages} {8346} (\bibinfo {year} {2025})}\BibitemShut {NoStop}%
\end{thebibliography}
\end{document}